\documentclass[12pt,preprint]{emulateapj}

\begin{document}
\title{The $z\sim6$ Luminosity Function Fainter than $-15$ mag from
  the {\it Hubble} Frontier Fields: The Impact of Magnification
  Uncertainties}
\author{R.J. Bouwens\altaffilmark{1}, P.A. Oesch\altaffilmark{2,3}, G.D. Illingworth\altaffilmark{4}, R.S. Ellis\altaffilmark{5,6}, M. Stefanon\altaffilmark{1}}
\altaffiltext{1}{Leiden Observatory,
  Leiden University, NL-2300 RA Leiden, Netherlands}
\altaffiltext{2}{Department of Astronomy, Yale University, New Haven,
  CT 06520}
\altaffiltext{3}{Observatoire de Gen{\`e}ve, 1290 Versoix, Switzerland}
\altaffiltext{4}{UCO/Lick Observatory, University of California, Santa
  Cruz, CA 95064}
\altaffiltext{5}{European Southern Observatory (ESO), Karl-Schwarzschild-Strasse 2, 85748 Garching, Germany}
\altaffiltext{6}{Department of Physics and Astronomy, University College London, Gower Street, London, WC1E 6BT, UK}
\begin{abstract}
We use the largest sample of $z\sim 6$ galaxies to date from the first
four Hubble Frontier Fields clusters to set constraints on the shape
of the $z\sim6$ luminosity functions (LFs) to fainter than
$M_{UV,AB}=-14$ mag.  We quantify, for the first time, the impact of
magnification uncertainties on LF results and thus provide more
realistic constraints than other recent work.  Our simulations reveal
that for the highly-magnified sources the systematic uncertainties can
become extremely large fainter than $-$14 mag, reaching
\textit{several orders of magnitude} at 95\% confidence at $\sim-$12
mag.  Our new forward-modeling formalism incorporates the impact of
magnification uncertainties into the LF results by exploiting the
availability of many independent magnification models for the same
cluster. One public magnification model is used to construct a mock
high-redshift galaxy sample that is then analyzed using the other
magnification models to construct a LF.  Large systematic errors occur
at high magnifications ($\mu\gtrsim30$) because of differences between
the models.  The volume densities we derive for faint ($\gtrsim-$17
mag) sources are $\sim$3-4$\times$ lower than one recent report and
give a faint-end slope $\alpha=-1.92\pm0.04$, which is 3.0-3.5$\sigma$
shallower (including or not including the size uncertainties,
respectively).  We introduce a new curvature parameter $\delta$ to
model the faint end of the LF and demonstrate that the observations
permit (at 68\% confidence) a turn-over at $z\sim6$ in the range
$-$15.3 to $-$14.2 mag, depending on the assumed lensing model.  The
present consideration of magnification errors and new size
determinations raise doubts about previous reports regarding the form
of the LF at $>-14$ mag.  We discuss the implications of our turn-over
constraints in the context of recent theoretical predictions.
\end{abstract}

\section{Introduction}

One of the most important open question in extragalactic studies
regards cosmic reionization and clarifying which sources drive this
important phase transition in the early universe.  While much evidence
suggests that the process might be driven by galaxies (e.g., Robertson
et al.\ 2013, 2015; Mitra et al.\ 2015; Bouwens et al.\ 2015b), others
have suggested that quasars could provide the dominant contribution
(Giallongo et al.\ 2015; Madau \& Haardt 2015; Mitra et al.\ 2016).

The important issues appear to be whether large numbers of faint
quasars exist at high redshift (e.g., Willott et al.\ 2010; McGreer et
al.\ 2013), whether faint galaxies show an appreciable ($>$5\%) escape
fraction (e.g., Siana et al.\ 2010, 2015; Vanzella et al.\ 2012, 2016;
Nestor et al.\ 2013), and what the total emissivity is in the
rest-frame $UV$ in faint galaxies beyond the limits of current surveys
in the {\it Hubble} Ultra Deep Field (HUDF: Beckwith et al.\ 2006;
Ellis et al.\ 2013; Illingworth et al.\ 2013).  Important issues for
the latter question are the precise values of the faint-end slopes and
the faint-end cut-off to the $UV$ luminosity function.  Depending on
the value of the faint-end slope and the luminosity where a cut-off in
the LF occurs (Kuhlen \& Faucher-Gigu{\`e}re 2012; Bouwens et
al.\ 2012; Robertson et al.\ 2013; Bouwens 2016), the total emissivity
from galaxies in the $UV$ can vary by factors of $\sim$2-10.

One potentially promising way to constrain the total luminosity
density in the rest-frame $UV$ is by taking advantage of the impact of
gravitational lensing by galaxy clusters for magnifying individual
sources.  This can bring extremely faint galaxies into view such that
they can be detected with current telescopes (e.g., Bradac et
al.\ 2009; Maizy et al.\ 2010; Coe et al.\ 2015).  There has been a
significant investment in this approach by {\it HST} in the form of
the {\it Hubble} Frontier Fields program (Coe et al.\ 2015; Lotz et
al.\ 2017), which is investing 840 orbits into reaching $\sim$29-mag
in 7 optical+near-IR bands, as well as two UVIS channels in a
supporting effort (Alavi et al.\ 2016; Siana 2013, 2015).

Already, analyses of sources behind the HFF clusters have resulted in
the identification of $z\sim6$-8 sources first to $-15$ mag (Atek et
al.\ 2014, 2015a,b) and later to $\sim$$-13$ mag (Kawamata et
al.\ 2016; Castellano et al.\ 2016a,b; Livermore et al.\ 2017
[hereinafter, L17]).  At $z\sim2$-3, it has been similarly possible
(Alavi et al.\ 2014, 2016) to probe to $\sim$$-13$ mag taking
advantage of very deep WFC3/UVIS observations over Abell 1689 and
various clusters in the HFF program.  Based on these deep searches,
the volume density of galaxies at $>-$16 mag have been estimated, with
quoted faint-end slopes for $z\sim2$-3 LFs that range from $-1.6$ to
$-1.9$ (Alavi et al.\ 2014, 2016) and from $-1.9$ to $-2.1$ for
$z\sim6$-8 LFs (Atek et al.\ 2015a,b; Ishigaki et al.\ 2015; Laporte
et al.\ 2016; Castellano et al.\ 2016b; L17), respectively.

In spite of the great potential that lensing clusters have for probing
the faint end of the $UV$ LF, successfully making use of data over
these clusters to perform this task in an accurate manner is not
trivial.  The entire enterprise is fraught with sources of systematic
error.  One of these sources of systematic error concerns the assumed
size distribution of extremely faint galaxies (Bouwens et al.\ 2017;
Grazian et al.\ 2011; Oesch et al.\ 2015), an issue that also impacts
LF determinations from blank fields like the HUDF (but to a lesser
degree since the faintest sources asymptote towards being entirely
unresolved).  Small differences in the assumed half-light radii have
the potential to change the inferred faint-end slopes by large
factors, i.e., $\Delta\alpha\gtrsim0.3$ depending on whether the mean
size of extremely faint galaxies is 120 mas, 30 mas, or 7.5 mas (e.g.,
see Figure 2 from Bouwens et al.\ 2017).  Fortunately, we found that
most of the extremely faint sources seem consistent with being almost
unresolved, i.e., with intrinsic sizes of $<$10-30 mas (Bouwens et
al.\ 2017; see also Kawamata et al.\ 2015; Laporte et al.\ 2016),
making this issue much more manageable in terms of its impact; but it
still remains an uncertainty.  A second source of systematic error
arises from errors in the magnification maps, since this can have a
profound impact on the LFs derived.  Finally, there are issues related
to subtraction of the foreground cluster light, contamination from
individual sources in the clusters (e.g., globular clusters), and from
other less important systematic effects that affect determinations of
the volume densities in the field vs. the cluster.\footnote{For
  example, the HFF program does not feature deep observations in the
  $z_{850}$-band, which is useful for discriminating between $z\sim6$
  and $z\sim7$ galaxies, while the HUDF and CANDELS (Grogin et
  al.\ 2011; Koekemoer et al.\ 2011) programs do feature deep
  integrations in this filter.  The availability or not of deep
  observations in the $z_{850}$ band could impact the $z\sim6$ and
  $z\sim7$ samples and LF results derived from these data sets in
  different ways.}

Even without such considerations, it is easy to see that systematics
could be a concern for LF studies from lensing clusters, simply by
comparing several recent LF results from clusters with similar results
based on deep field studies using the HUDF.  To give one recent
example, Alavi et al.\ (2016) reported a faint-end slope $\alpha$ of
$-1.94\pm0.06$ for the $UV$ LF at $z\sim3$ based on an analysis of
sources behind 3 lensing clusters, while Parsa et al.\ (2016) reported
a faint-end slope of $-1.31\pm0.04$ based on a deep $z\sim3$ search
over the HUDF.  These results differ at a significance level of
$\sim$9$\sigma$ taking at face value the quoted statistical errors.
This is but one example of the large differences frequently present
between LF results derived from deep field studies and those derived
on the basis of lensing clusters (see Figure 1 for several other
examples).\footnote{We plan to both investigate and try to resolve
  these large differences in a future work (R.J. Bouwens et al.\ 2017,
  in prep).}

In addition to the clear scientific importance of the faint-end slope
$\alpha$ for computing the total ionizing emissivity from faint
galaxies, the observations also allow us to test for a possible
flattening or turn-over of the $UV$ LF at low luminosities.  Many
cosmological hydrodynamic simulations of galaxy formation predict a
flattening in the $UV$ LF at $\sim-13$ or $\sim-15$ due to less
efficient atomic and molecular hydrogen cooling, respectively
(Mu{\~n}oz \& Loeb 2011; Krumholz \& Dekel 2012; Kuhlen et al.\ 2013;
Jaacks et al.\ 2013; Wise et al.\ 2014; Liu et al.\ 2016; Gnedin
2016; Finlator et al.\ 2016), while other simulations predict a
flattening in the range $\sim-16$ to $\sim-13$ mag due to the impact
of radiative feedback (O'Shea et al.\ 2015; Ocvirk et al.\ 2016).
Meanwhile, by combining abundance matching and detailed studies of the
color-magnitude diagram of low-luminosity dwarfs in the local
universe, evidence for a low-mass turn-over in the luminosity function
has been reported at $-13$ mag (Boylan-Kolchin et al.\ 2015; see also
Boylan-Kolchin et al.\ 2014).  Current observations likely provide us
with some constraints in this regime.  However, given the significant
systematics that appear to be present in current determinations of the
faint-end slope $\alpha$ from lensing clusters (vs. field results), it
is not at all clear that current constraints on the form of the $UV$
LF at $>-$15 mag are reliable, particularly at $z>4$.

\begin{figure}
\epsscale{1.15}
\plotone{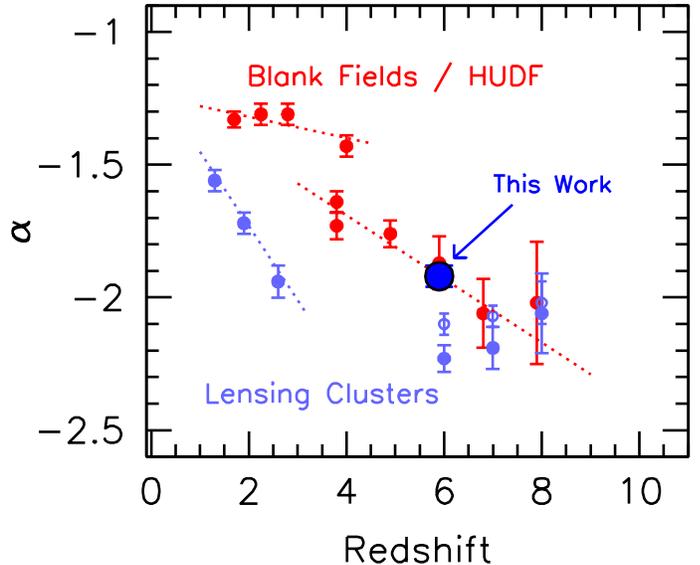}
\caption{Some recent measurements of the faint-end slope $\alpha$
  vs. redshift from the literature using deep fields (\textit{red
    solid circles}) and using lensing clusters (\textit{light blue
    solid circles}).  The field LF results are from Parsa et
  al.\ (2016) at $z\leq4$, Bouwens et al.\ (2007) at $z\sim4$, and
  Bouwens et al.\ (2015a) at $z\geq4$.  The $z\leq3$ cluster LF
  results are from Alavi et al.\ (2016).  The dotted lines shown the
  approximate trends in faint-end slope from each of these studies.
  The $z=6$-8 cluster LF results shown are based on a fit to the L17
  cluster stepwise LFs anchored to one point ($-$20 mag) at the bright
  end of the field LF (see Appendix E).  This ensures that the
  presented faint-end slope $\alpha$ results from L17 are almost
  entirely independent of field constraints; the nominal faint-end
  slope results from L17 (including constraints from the field) are
  shown with the open circles.  The large solid dark blue circle shows
  the faint-end slope $\alpha$ we estimate from our $z\sim6$ HFF
  cluster sample in \S4.  As field and lensed LFs potentially probe
  different luminosity regimes in the $UV$ LF (bright and fainter,
  respectively), it is possible there would be slight differences in
  the derived slopes; however, the differences run in the opposite
  direction normally predicted in simulations (e.g., see right panel
  in Figure 1 from Gnedin 2016).  Given that the differences between
  the derived $\alpha$'s are often much larger than the plotted
  statistical error bars, systematic errors must clearly contribute
  substantially to some of the determinations plotted
  here.\label{fig:systdiff}}
\end{figure}

\begin{figure*}
\epsscale{1.15}
\plotone{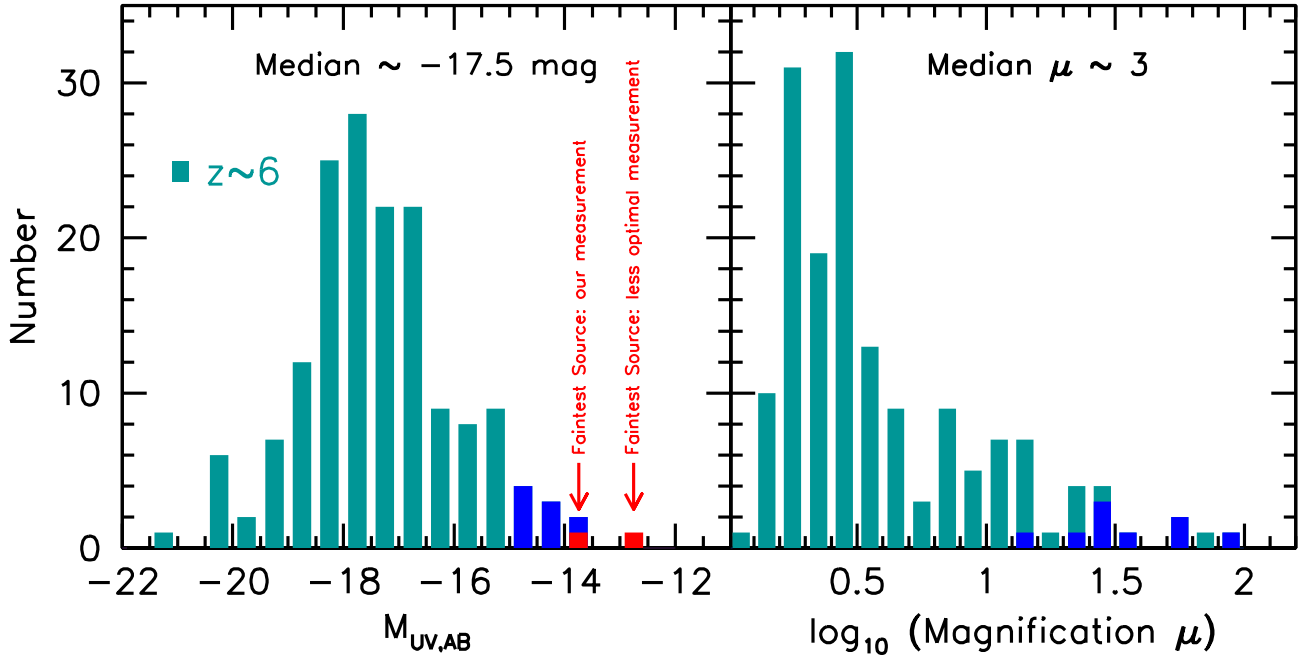}
\caption{Number of galaxies found in our conservative selection of
  $z\sim6$ galaxies behind the first four HFF clusters vs. their
  inferred $M_{UV}$ luminosity (\textit{left panel}) and magnification
  factor (\textit{right panel}).  We take the magnification factor to
  be the median of those derived from the four parametric models
  (\textsc{GLAFIC}, CATS, Sharon/Johnson, and Zitrin-NFW), enforcing a
  maximum value of 100 (due to the much weaker predictive power for
  the models at such high magnification factors: see
  Figure~\ref{fig:predpow}).  The one source over our fields with a
  magnification factor in excess of 100 and that is M0416I-6118103480
  (04:16:11.81, $-$24:03:48.1) with a nominal magnification of 145
  (nominally implying an absolute magnitude of $-13.4$ mag).  The nine
  sources with the faintest intrinsic luminosities are shown in blue
  in each panel.  The faintest source in our probe is sensitive to how
  total magnitude measurements are made and which magnification models
  are used.  The two red squares shows the luminosity of our faintest
  source, as we measure it with our total flux approach (\textit{left
    red square}) and also (\textit{right red square}) consistent with
  the way that L17 measure luminosities for many of the sources in
  their $z\sim6$-8 samples (see \S2 and \S6.1.2).  The luminosity
  shifts $\sim$0.7 mag faintward for these sources in the latter
  approach.\label{fig:sample}}
\end{figure*}

In the present paper, we take the next step in our examination of the
impact of systematic errors on derived LF results from lensing
clusters, after our previous paper on this subject, i.e., Bouwens et
al.\ (2017), where the emphasis was on the uncertain sizes of faint
sources.  Here the focus will be more on the uncertainties in LF
results that arise from errors in the gravitational lensing models.
As we will demonstrate explicitly, the recovered LF from a
straightforward analysis tends to migrate towards a faint-end slope of
$\sim-2$ (or slightly steeper), if uncertainties in the magnification
factor are large.  The impact of these uncertainties is to wash out
features in the LF, particularly at low luminosities.  Given that
magnification factors $\mu$ necessarily become uncertain when these
factors are high, i.e., $\mu>10$ and especially $\mu>50$, accurately
constraining the shape of the LF at extremely low luminosities and
also detecting a turn-over or flattening is very challenging.

The purpose of this paper is to look at the constraints we can set on
the faint end of the $z\sim6$ $UV$ LF with a thorough assessment of
the possible systematic errors.  In doing so, we will look for
possible evidence for a turn-over in the LF at very low luminosities
and if not present, what constraints can be placed on the luminosity
of a turn-over.  Evidence for a turn-over will be evaluated through
the introduction of a {\it curvature parameter} which we constrain
through extensive Markov-Chain Monte-Carlo (MCMC) trials.  The
confidence intervals we obtain on the shape of the $UV$ LF at faint
magnitudes will provide theorists with some important constraints for
comparison with models and cosmological hydrodynamic simulations.
Most importantly, these results provide balance to some discussion in
the literature, where premature claims appear to have possibly been
made regarding the LF's rising steeply to very low luminosities.  To
keep the focus of this paper on our new techniques, we restrict our
analysis to just the $z\sim6$ LF from the first four HFF clusters.

The plan for this paper is as follows.  \S2 summarizes the data sets
we use to select our $z\sim6$ sample and derive constraints on the
$z\sim6$ LF.  \S3 provides some useful context for the issue of errors
in the magnification models and shows the general impact it would have
on LF results.  \S4 and \S5 present new LF results at $z\sim6$ using
our new forward-modeling methodology.  \S6 compares our new results
with previous reported LF results, as well as results from various
theoretical models.  Finally, in \S7, we summarize and conclude.  We
refer to the HST F814W, F850LP, F105W, F125W, F140W, and F160W bands
as $I_{814}$, $z_{850}$, $Y_{105}$, $J_{125}$, $JH_{140}$, and
$H_{160}$, respectively, for simplicity.  Estimates of the $UV$
luminosities are made at $\sim$1800$\AA$ for the typical source in the
sample.  Through this paper, a standard ``concordance'' cosmology with
$H_0=70$ km s$^{-1}$ Mpc$^{-1}$, $\Omega_{\rm m}=0.3$ and
$\Omega_{\Lambda}=0.7$ is assumed.  This is in good agreement with
recent cosmological constraints (Planck Collaboration et al.\ 2015).
Magnitudes are in the AB system (Oke \& Gunn 1983).

\begin{deluxetable*}{ccccccc}
\tablewidth{0pt} \tabletypesize{\footnotesize}
\tablecaption{Magnification Models Used
  Here\tablenotemark{a}\label{tab:lensingmodel}} \tablehead{
  \colhead{} & \colhead{Mass-} & \colhead{} & \colhead{} & \colhead{}
  & \colhead{}\\ \colhead{Model} & \colhead{Traces-} & \colhead{Dark-}
  & \colhead{} & \colhead{} & \colhead{Resolution} &
  \colhead{}\\ \colhead{Name} & \colhead{Light} & \colhead{Matter} &
  \colhead{Code} & \colhead{Parametric\tablenotemark{b}} &
  \colhead{(``)} & \colhead{References}} \startdata
\multicolumn{6}{c}{``Parametric''
  Models\tablenotemark{b}}\\
\textsc{GLAFIC} & Y & Y & \textsc{GLAFIC} & Y & 0.03$''$ & Oguri
(2010); Ishigaki et al.\ 2015; Kawamata et al.\ (2016)\\
CATS & Y & Y & \textsc{lenstool} & Y & 0.1$''$ & Jullo \& Kneib (2009); Richard
et al.\ (2014); Jauzac et\\
& & & & & & al.\ (2015a,b) \\
Sharon/Johnson & Y & Y & \textsc{lenstool} & Y & 0.06$''$ & Johnson et
al.\ (2014) \\
Zitrin-NFW & Y & Y & Zitrin & Y & 0.06$''$ & Zitrin et al.\ (2013, 2015)\\\\
\multicolumn{6}{c}{``Non-Parametric'' Models\tablenotemark{b}}\\
\textsc{Grale} & N & Y & \textsc{Grale} & N & 0.22$''$ & Liesenborgs et
al.\ (2006); Sebesta et al.\ (2016)\\
Bradac & N & Y & Bradac & N & 0.2$''$ & Bradac et al.\ (2009) \\
Zitrin-LTM & Y & N & Zitrin & N & 0.06$''$ & Zitrin et al.\ (2012, 2015).
\enddata
\tablenotetext{a}{This includes all publicly available lensing models
  which have high-resolution mass maps and are generally available for
  the first four HFF clusters.  Our analyses therefore do not include
  the public HFF models of Diego et al.\ (2015) and Merten et
  al.\ (2015).}

\tablenotetext{b}{Parametric models assume that mass in the cluster is
  in the form of one or more dark matter components with an
  ellipsoidal Navarro-Frenk-White (NFW: Navarro et al.\ 1997) form and
  to include a contribution from galaxies following specific
  mass-to-light scalings.  Two well-known parametric modeling codes
  are \textsc{lenstool} (Jullo \& Kneib 2009) and \textsc{GLAFIC}
  (Oguri 2010).  For the non-parametric models, both assumptions are
  typically relaxed, and the mass distributions considered in the
  models typically allow for much more flexibility than with the
  parametric models.}
\end{deluxetable*}

\begin{figure}
\epsscale{1.18}
\plotone{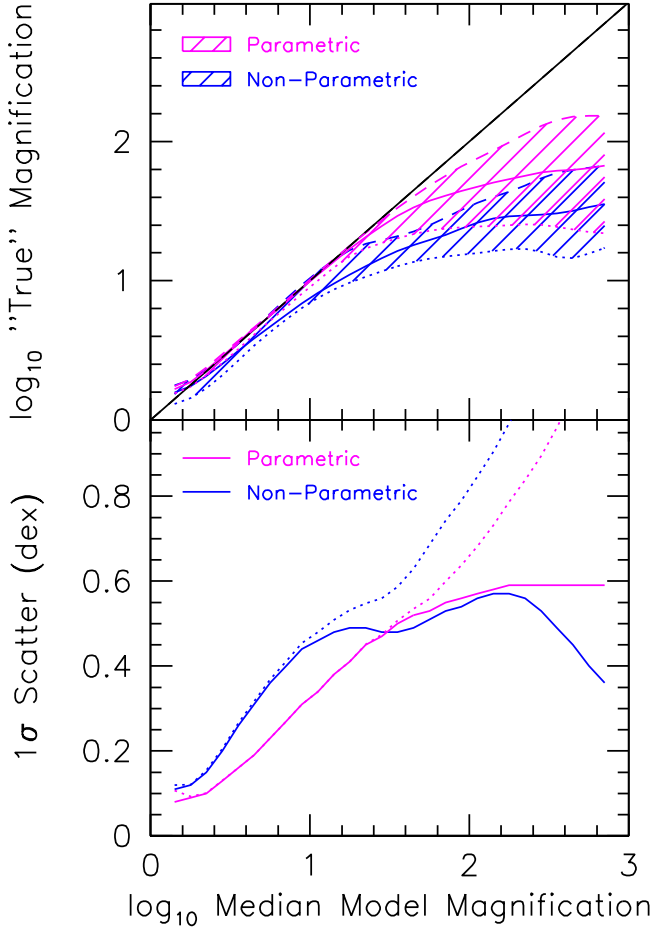}
\caption{An evaluation of the predictive power of the lensing models
  and the median magnification maps.  (\textit{upper}) Illustration of
  how well the median magnification factor from all the magnification
  models but one (variable on the x-axis) predicts the median
  magnification factor for the excluded magnification model, i.e.,
  ``truth'' model (variable on the y-axis).  The plotted magnification
  plotted along the y-axis shows the geometric mean of the results,
  alternatively taking each model to be the truth.  The dashed and
  dotted magenta lines show the recovered magnification factors for
  the parametric magnification models (i.e., \textsc{GLAFIC}, CATS,
  Sharon/Johnson, Zitrin-NFW) from the best and worst performing
  cluster as well.  The solid magenta line shows the geometric mean of
  the recovered magnification factor across all clusters considered
  here.  The blue dashed and dotted lines show the equivalent results
  excluding the non-parametric magnification models (\textsc{Grale},
  Bradac, and Zitrin-LTM) from the process.  Again the solid blue
  shows the geometric mean of the recovered magnification factors for
  all clusters.  For perfectly predictive magnification models, the
  plotted lines would follow the black diagonal line with a slope of
  1. (\textit{lower}) Scatter in the magnification factors vs. median
  magnification factor for the parametric magnification models
  (\textit{magenta solid line}).  The blue solid line gives the
  results for the non-parametric models.  The dotted lines are the
  same as the solid lines but also add in quadrature the logarthmic
  differences between the actual magnification factors in a model and
  that predicted from a median of the other models.  From this figure,
  it is clear that the median magnification model has largely lost its
  predictive power by magnification factors of $\sim$10 and $\sim$30
  assuming that the available non-parametric and parametric models,
  respectively, are representative of reality.  \label{fig:predpow}}
\end{figure}

\section{Data Sets and $z\sim6$ Sample}

In our selection of $z\sim6$ galaxies, we make use of the v1.0
reductions of the deep {\it HST} optical and near-IR {\it HST}
observations available over the first four clusters in the HFF
program: Abell 2744, MACS0416, MACS0717, and MACS1149 (A. Koekemoer et
al.\ 2016, in prep; Lotz et al.\ 2017).  The optical observations
include $\sim$18, $\sim$10, and $\sim$42 orbits of ACS observations in
the F435W, F606W, and F814W bands from 0.4$\mu$m to 0.9$\mu$m.
Near-IR observations over these fields total 34, 12, 10, and 24 orbits
in the F105W, F125W, F140W, and F160W, reaching to roughly a $5\sigma$
limiting magnitude of 28.8 to 29.0 mag.

Subtraction of foreground light from cluster galaxies and cluster
galaxies was performed using galfit (Peng et al.\ 2002) and the
median-filtering algorithm of SExtractor (Bertin \& Arnouts 1996) run
at two different grid scales.  There are many similarities of our
procedure to that from Merlin et al.\ (2016).  The only areas clearly
inaccessible to us in our searches for faint $z\sim6$ galaxies occur
directly under the cores of bright stars or galaxies in the cluster.
Our procedure performs at least as well as any other procedure
currently in use (Merlin et al.\ 2016; L17).  Relative to the
approaches of Merlin et al.\ (2016) or L17, our procedure appears to
perform comparably well.  One measure of this is the number of $z=6$-8
galaxies we identify behind Abell 2744 and MACS0416 (considered in
both previous studies) for the current analysis.  Our samples are
$\gtrsim$10\% larger than that utilized in either previous study and
could be enlarged further by 10-20\% by making use of different
detection images (Appendix A).

A complete description of both our photometric procedure and selection
criteria for identifying $z\sim6$ galaxies is provided in R.J. Bouwens
et al.\ (2017, in prep).  In most respects, our procedures are similar
to that done in many of our previous papers (e.g., Bouwens et
al.\ 2015a), but we do note that we perform our photometric
measurements after subtraction of the intracluster and bright
elliptical galaxy light.  While other procedures report sizeable
differences between the total magnitude measurements on the original
and subtracted images (e.g., L17), with measurements on the original
images giving brighter magnitudes, we only find a 0.03$\pm$0.07 mag
difference for the median source in these measurements.  Further
evidence for the fact that our procedures do not underestimate the
total flux in sources can be seen by comparing our photometry with
other groups (\S6.1.1 and \S6.1.2).  Our magnitude measurements are
typically $\sim$0.1-0.3 mag brighter than other groups for the same
sources.

We briefly summarize our criteria here for selecting a robust and
large sample of $z\sim6$ galaxies.  We select all sources that satisfy
the following $I_{814}$-dropout color criteria
\begin{eqnarray*}
(I_{814}-Y_{105}>0.6)\wedge(Y_{105}-H_{160}<0.45)\wedge\\
(I_{814}-Y_{105}>0.6(Y_{105}-H_{160}))\wedge\\
(Y_{105}-H_{160}<0.52+0.75(J_{125}-H_{160}))
\end{eqnarray*}
and which are detected at $>$6.5$\sigma$ adding in quadrature the S/N
of sources in the $Y_{105}$, $J_{125}$, $JH_{140}$, and $H_{160}$ band
images measured in a 0.35$''$-diameter aperture.  The above color
selection criterion also explicitly excludes the inclusion of $z\sim8$
$Y_{105}$-dropout galaxies.  As the above criteria identify sources at
both $z\sim6$ and $z\sim7$, we compute the redshift likelihood
function $P(z)$ for each source and only include those sources where
the best-fit photometric redshift is less than 6.3.  Sources are
further required to have a cumulative probability $<$35\% at $z<4$ to
keep contamination to a minimum in our high-redshift samples.

Our sample of 160 $z\sim6$ candidate galaxies is the largest
compilation reported to date.  Each of the HFF clusters we examine in
this study have at least seven independent lensing models available,
with both convergence $\kappa$ and shear $\gamma$ maps
(Table~\ref{tab:lensingmodel}).  We estimate the magnification of
sources based on publicly-available models by first multiplying the
$\kappa$ and $\gamma$ maps of each cluster by $D_{ls}/D_{s}$ and then
computing the magnification $\mu$ as follows:
\begin{equation}
\mu = \frac{1}{|(1-\kappa)^2 - \gamma^2)|}
\label{eq:mu}
\end{equation}
where $D_{ls}$ and $D_s$ represent the angular-diameter distances from
the lensing cluster to the magnified galaxy and the angular-diameter
distances to the source, respectively.  For our magnification estimates
for individual sources, we take the median of the model magnifications
from the CATS (Jullo \& Kneib 2009; Richard et al.\ 2014; Jauzac et
al.\ 2015a,b), \textsc{GLAFIC} (Oguri 2010; Ishigaki et al.\ 2015;
Kawamata et al.\ 2016), Sharon/Johnson (Johnson et al.\ 2014), and
Zitrin parametric NFW models (Zitrin-NFW: Zitrin et al.\ 2013, 2015).
The parametric models generally provided the best estimates of the
magnification for individual sources in the HFF comparison project
(Meneghetti et al.\ 2016), but we emphasize that many non-parametric
magnification models also performed very well.

We present in Figure~\ref{fig:sample} the distribution of absolute
magnitudes and magnification factors we estimate for sources in our
$z\sim6$ sample.  Absolute magnitudes for our z$\sim$6 sample are
taken to equal the inverse-weighted mean of the fluxes measured in the
F105W, F125W, F140W, and F160W bands (such that the rest-frame
wavelength for our absolute magnitude measurements is
$\sim$1800$\AA$).  We set an arbitrary maximum magnification factor of
100, given the lack of predictive power for magnification maps at such
high values (see \S3.1).\footnote{Our use of an upper limit on the
  magnification factors only affects one source, i.e.,
  M0416I-6118103480 (04:16:11.81, $-$24:03:48.1) with a nominal
  magnification of 145 (nominally implying an absolute magnitude of
  $-13.4$ mag) and only has a minor impact on the parameters we derive
  for the $z\sim6$ LF in $\S5$ (changing $\alpha$, $\delta$, and
  $\phi^*$ by $\leq 0.01$, $\leq 0.1$, and less than 2\%,
  respectively).}  Our selection includes sources ranging from $-22$
mag to $-13.5$ and with magnification factors ranging from 1.2 to 145,
with the bulk of the sources having absolute magnitudes of $-18$ and
magnification factors of $\sim$2.

We should emphasize that the inferred luminosities and total
magnitudes we report for sources are intended to provide a rather
complete accounting for light in individual sources.  They are based
on scaled-aperture photometry following the Kron (1980) method, with a
correction for flux on the wings of the PSF (e.g., see Bouwens et
al.\ 2015a).  However, in comparing our total magnitude measurements
with the magnitude measurements from other groups (e.g., L17: see
\S6.1.2), we have found some sources being reported to have apparent
magnitude measurements fainter by $\sim$0.3-0.5 mag than what we
measure for the same sources.  In addition, other teams occasionally
report 1.3-1.8$\times$ higher values for the magnification factor of
individual sources than we calculate based on the same models, e.g.,
the faintest source in L17 (Appendix F).

If we quote the luminosities of sources in our study using a similar
procedure as to what L17 appear to use -- where individual sources are
often $\sim$0.4 mag fainter than we find -- and adopt 1.3-1.8$\times$
higher magnification factors, our probe would extend to $-12.6$ mag
(indicated in Figure~\ref{fig:sample} with the red bin), essentially
identical to that claimed by L17 (see also Castellano et al.\ 2016a and
Kawamata et al.\ 2016).  We emphasize, however, that the low
luminosities claimed by measuring magnitudes in this way (and
computing magnification factors in this way: see Appendix F) likely
exaggerate how faint the HFF program probes.  We discuss this further
in \S6.1.2 and \S6.2.  We prefer our photometric scheme for accounting
for the total light in faint sources.

\begin{figure*}
\epsscale{0.9}
\plotone{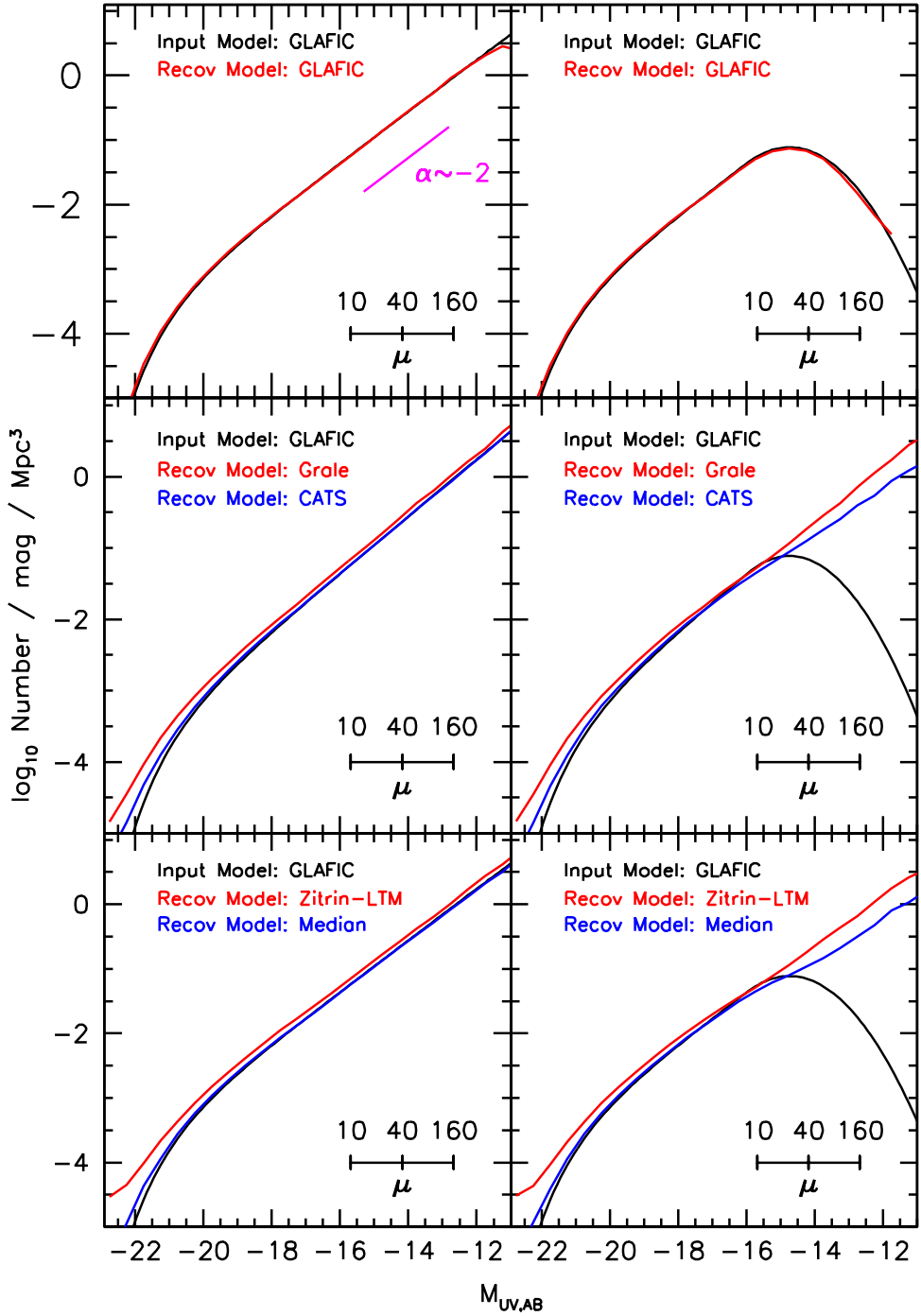}
\caption{Comparison of the input LFs (\textit{black lines}) into our
  forward-modeling simulations and the recovered LFs when using the
  same magnification models (\textit{top panels}) and when using four
  different magnification models, including \textsc{Grale}
  (\textit{red lines: middle panels}), CATS (\textit{blue lines:
  middle panels}), Zitrin-LTM (\textit{red lines: lower panels}), and
  the median parametric model (\textit{blue lines: lower panels}).
  A ticked horizontal bar is added to the panels to indicate the
  approximate luminosities probed by sources of a given magnification
  factor near the faint end of the HFF data set, i.e., 28.5 mag.  Two
  input LFs are considered: one where the LF exhibits a faint-end
  slope of $-2$ with no turn-over at low luminosities (\textit{left
    panels}) and a second also exibiting a faint-end slope of $-2$ but
  with a turn-over at $-15$ mag (\textit{right panels}).  In the first
  case, the recovered LFs show a faint-end slope $\alpha$ of $-2$ to
  very low luminosities, in agreement with the input LF.  However, for
  the second case, the recovered LFs again show a faint-end slope
  $\alpha$ of $-2$ to very low luminosities, in significant contrast
  to the input LF.  As a result, interpreting the LF results from
  lensing clusters can potentially be tricky, as the detection of a
  turn-over in the LF at $>-15$ mag is very challenging (see \S3.2).
  This is due to the weaker predictive power of the magnification
  models at high magnification factors $\mu$ $>$10 and especially
  $\mu$$>$30 (Figure~\ref{fig:predpow}).  See also
  Figures~\ref{fig:illustsize} and \ref{fig:illust15} from Appendices
  B and C.\label{fig:illust}}
\end{figure*}

\section{Impact of Magnification Errors on the Derived LFs}

An important aspect of the present efforts to provide constraints on
the $z\sim6$ LF will be our explicit efforts to include a full
accounting of the uncertainties present in the magnification models we
utilize.  We begin by looking first at the general size of errors in
the magnification models and second at how the errors would impact LFs
derived from lensing clusters.

\subsection{Predictive Value of the Public Magnification Models}

In making use of various gravitational lensing models to derive
constraints on the prevalence of extremely faint galaxies at high
redshift, it is important to obtain an estimate of how predictive the
lensing models are for the true magnification factors.

One way of addressing this issue is the fully end-to-end approach
pursued by Meneghetti et al.\ (2016) and involves constructing
highly-realistic mock data sets, analyzing the mock data sets using
exactly the same approach as are used on the real observations, and
then quantifying the performance of the different methods by comparing
with the actual magnification maps.  While each of the methods did
fairly well in reproducing the magnification maps to magnification
factors of $\sim$10, the best performing methods for reconstructing
the magnification maps of clusters were the parametric models, with
perhaps the best reconstructions achieved by the \textsc{GLAFIC}
models, the Sharon/Johnson models, and the CATS models.

An alternate way of addressing this issue is by comparing the public
lensing models against each other.  Here we pursue such an approach.
We treat one of the models as the truth and then to quantify the
effectiveness of the other magnification models taken as a set for
predicting that model's magnifiction map.  We consider both the case
that the true mass profile of the HFF clusters is considered (1) to
lie among parametric class of models built on NFW-type mass profiles
and (2) to lie among the non-parametric class of models which allow
for more freedom in the modeling process.  We take the former models
to include the \textsc{GLAFIC}, CATS, Sharon/Johnson, and Zitrin-NFW
models, and the latter to include the Bradac et al. [2009],
\textsc{Grale} [Liesenborgs et al.\ 2006; Sebesta et al.\ 2016], and
Zitrin-LTM [Zitrin et al.\ 2012, 2015].\footnote{Zitrin-LTM does not
  technically qualify as parametric or non-parametric, since the mass
  profile is governed by the distribution of light in a cluster.
  However, since the model shows a greater dispersion relative to the
  parametric models, we include it in the non-parametric group.}  A
brief description of the general properties of the public lensing
models can be found in Table~\ref{tab:lensingmodel}.  In performing
this test, we assume that the median of the magnification models
provides our best means for predicting magnifications in the model we
are treating as the truth.  The truth model is always excluded when
constructing the median magnification map for this test.

Alternatively treating each of the magnifictions models as the truth,
we then quantify what the median magnification factor is in the truth
model as a function of the median magnification factors from the other
models.  For perfectly predictive models, the magnification factors in
the truth model would be precisely centered around the median
magnification factors from the other models.  In practice, this is not
true, given the difficulty in predicting the precise locations of the
rare regions around the cluster with the highest magnification
factors.  While one can control for these uncertainties through use of
quantities like the median, even the median will overpredict the true
magnification, due to the impact of model ``noise'' on the medians and
the possibility for chance overlap in the high-magnification regions
across the models.

For the most general results, we take a geometric mean of the median
magnification factors considering each model as the truth and then
plot the results in the upper panel of Figure~\ref{fig:predpow}.
Results on the predictive power of the parametric (\textsc{GLAFIC},
CATS, Sharon/Johnson, Zitrin-NFW) and non-parametric (\textsc{Grale},
Bradac, Zitrin-LTM) models are presented separately with magenta and
blue colored lines.  The dashed and dotted lines give the ``true''
magnifications recovered vs. median magnification factors for best and
worst performing cluster.  Meanwhile, the solid line between the
dashed and dotted lines gives the geometric mean of the ``true''
magnifications across all 4 clusters considered here.  The lower panel
of Figure~\ref{fig:predpow} shows the position-to-position scatter
around the median magnification in the model treated as the truth.
From this exercise, it is clear that the magnification maps have
excellent predictive power to magnification factors of $\sim$10 in all
cases and perhaps to even higher magnification factors assuming that
the magnification profiles of HFF clusters are as well behaved as in
the parametric models.  The scatter, however, is already very large at
magnification factors of 10.  We will extend this exercise in a future
work (R.J. Bouwens et al.\ 2017, in prep).

\begin{figure*}
\epsscale{1.17}
\plotone{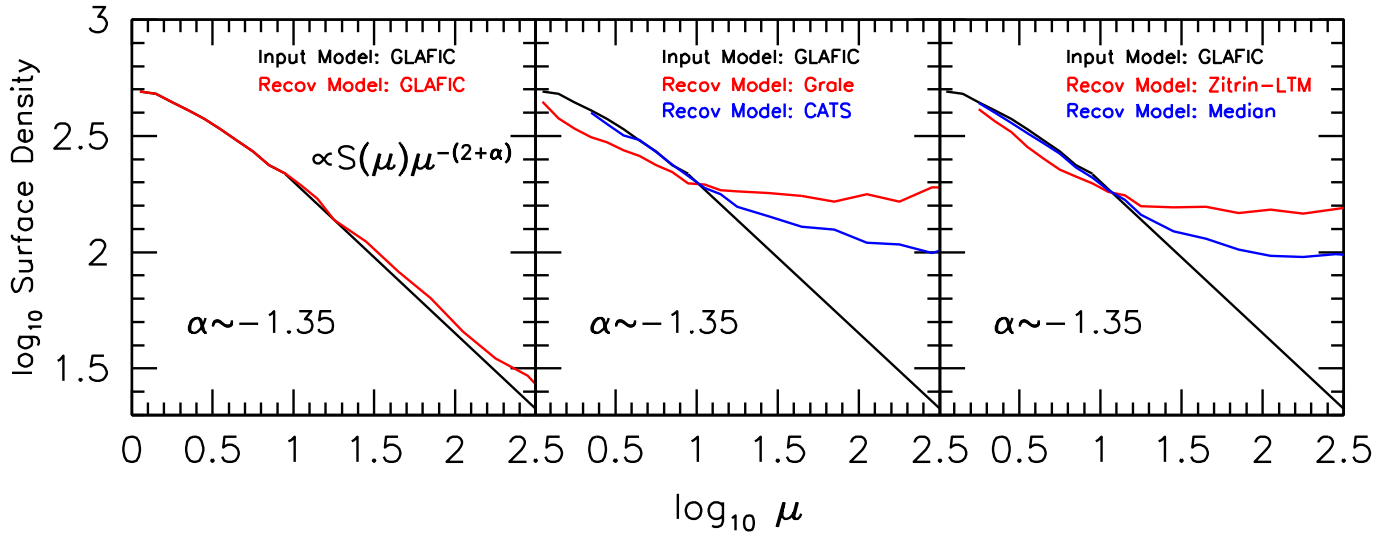}
\caption{Use of forward modeling to demonstrate the expected
  dependence of the recovered surface densities of $z\sim6$ sources on
  the model magnification factor $\mu$.  The input distribution of
  sources around the first four HFF clusters is generated using the
  \textsc{glafic} model and a faint end slope of $-1.35$ and then
  recovered using the \textsc{glafic}, \textsc{grale}, \textsc{CATS},
  Zitrin-LTM, and median parametric model.  In the case of perfect
  magnification maps, the surface density of sources is expected to
  depend on magnification $\mu$ as $S(\mu) \mu^{-(2+\alpha)}$ where
  $S(\mu)$ is the magnification-dependent selection efficiency.  At
  sufficiently high magnifications, the predictive power of the
  lensing models breaks down and one would expect there to be no
  correlation between the surface density of galaxies and the model
  magnification factor, as the present forward model results
  illustrate in the center and right panels.  In such a case, the
  recovered LF has a faint-end slope that asymptotes towards the value
  that implies a fixed surface density of sources above some
  magnification factor.  In the case that the selection efficiency
  does not depend on the magnification factor, this faint-end slope
  would be $-2$.  However, in the more general case presented in
  Appendix C, the faint-end slope asymptotes to $-2+d(\ln
  S(\mu))/d(\ln \mu)$.\label{fig:surf15}}
\end{figure*}

The exercise we perform in this section shows similarities in
philosophy to the analyses that Priewe et al.\ (2017) pursue, in
comparing magnification models over the HFF clusters with each other
to determine the probable errors in the individual magnification maps.
One prominent conclusion from that study was that differences between
the magnification maps was almost always larger than the estimated
errors in the magnification for a given map, pointing to large
systematics in the construction of some subset of the individual maps.
This provides some motivation for the tests we perform here and in
future sections in this paper, and confirmation of the importance of
this study.  Other powerful tests of the predictive power of the
magnification maps, and the challenges, were provided by observations
of SNe Ia (Rodney et al.\ 2015).

\subsection{Impact of Magnification Errors on the Recovered LFs}

The purpose of this subsection is illustrate the impact of
magnification errors on the derived LFs from the HFF clusters.  Two
different example LFs are considered for this exercise: (1) one with a
faint-end slope of $-2$ and a turn-over at $-15$ and (2) another with
a fixed faint-end slope $\alpha$ of $-2$ and no turn-over.

How well can we recover these LFs given uncertainties in the
magnification maps?  We can evaluate this by generating a mock catalog
of sources for each of the first four clusters from the HFF program
using one set of magnification models (``input'' models) and then
attempting to recover the LF using another set of magnification models
(``recovery'' models).  These catalogs include positions and
magnitudes for all the individual sources in each cluster.  In
computing the impact of lensing, the redshifts are fixed to $z=6$ for
all sources.  The input magnification models are taken to be the
\textsc{GLAFIC} models for this exercise.  Following previous work
(e.g. Ishigaki et al.\ 2015; Oesch et al.\ 2015), each galaxy in the
image plane is treated as coming from an independent volume of the
universe, allowing us to construct the input catalogs from the model
magnification maps alone (and therefore not requiring use of the
deflection maps).  This choice does not bias the LF results in our
analysis relative to analyses that account for multiple imaging of the
same galaxies (from the source plane), since both the cosmological
volume and total number of background galaxies is increased in
proportion to the overcounting.  The selection efficiencies of sources
are accounted for when creating the mock catalogs, as estimated in
Appendix B.  In performing this exercise, we ignore errors in our
estimates of the selection efficiencies and small number statistics at
the faint end of the LF.

One can try to recover the input LFs from these mock catalogs, using
various magnification models.  Sources are binned according to
luminosity using the ``recovery'' magnification model.  The selection
volumes available in each luminosity bin are also estimated as
described in Appendix B using this ``recovery'' magnification model.
To demonstrate the overall self-consistency in our approach, we show the
recovered LFs using the same magnification model as we used to
construct the input catalogs in the top two panels in
Figure~\ref{fig:illust}.

What is the impact if different lensing models are used to recover the
LF than those used to construct the mock catalogs?  The lowest two
rows of panels in Figure~\ref{fig:illust} show the results using the
latest magnification models by \textsc{Grale}, CATS, Zitrin-LTM, and
the median of the CATS, Sharon/Johnson, and Zitrin-NFW models where
available.

These simulation results demonstrate that the recovery process appears
to work very well for input LFs with faint-end slopes of $-2$
(\textit{left panels} in Figure~\ref{fig:illust}) independent of the
magnification model, with all recovered LFs showing a very similar
form to the input LFs.

Very different results are, however, obtained in our attempts to
recover input LFs with a turn-over at $-$15 mag (\textit{right panels}
in Figure~\ref{fig:illust}) using magnification models that are
different from the input model.  For all four magnification models we
consider, the recovered LFs look very similar to the LF example we
just considered.  All recovered LFs show a steep faint-end slope to
$-11$ mag.  What is striking is that they do not reproduce the
turn-over present in the input model at $-15$ mag.  There are some
differences in the recovered LFs depending on how similar the input
magnification model is to the recovery model, with effective faint-end
slopes of $-2$, $-1.8$, and $-1.7$ achieved with the \textsc{Grale}
and Zitrin-LTM models, the CATS models, and the median parametric
models, respectively.  The \textsc{GLAFIC} magnification model is not
used when constructing the median parametric magnification model.

Both examples demonstrate that the faint-end slope for the recovered
LFs tend to gravitate towards a value of $-2$.  It is useful to
provide a brief explanation as to why.  For a power-law LF, i.e.,
$L^{\alpha}$, and ignoring any dependence of the selection efficiency
on magnification factor, one expects the surface density of sources on
the sky to depend on magnification factor $\mu$ as $L^{\alpha}/\mu\,
dL |_{L=L_{obs}/\mu} \propto \mu^{-\alpha-2} \propto
\mu^{-(2+\alpha)}$ where $L$ and $L_{obs}$ represent the intrinsic and
observed luminosities, respectively.  For faint-end slopes shallower
than $-2$, one therefore expects a systematic decrease in the surface
density of sources on the sky as the magnification increases; for
faint-end slopes of $-2$, one expects no dependence on source
magnification; and for faint-end slopes steeper than $-2$, one expects
a systematic increase in the surface density of sources as the
magnification increases.  All of the above statements are for the
intrinsic surface densities.  The observed surface densities will be
impacted by the magnification-dependent selection efficiencies
$S(\mu)$.

We illustrate this expected dependence on the magnification factor in
Figure~\ref{fig:surf15} for a LF with a faint-end slope of $-1.35$ by
laying down sources behind the HFF clusters using the \textsc{glafic}
magnification model.  The surface density of the sources
vs. magnification factor can then be recovered using a variety of
other models.  At sufficiently high magnification factors, the
uncertainties in the magnication factors become large, washing out any
dependence on the magnification factor.  This results in a relatively
constant surface density of sources and a faint-end slope of $-2$.

\begin{figure*}
\epsscale{1.17}
\plotone{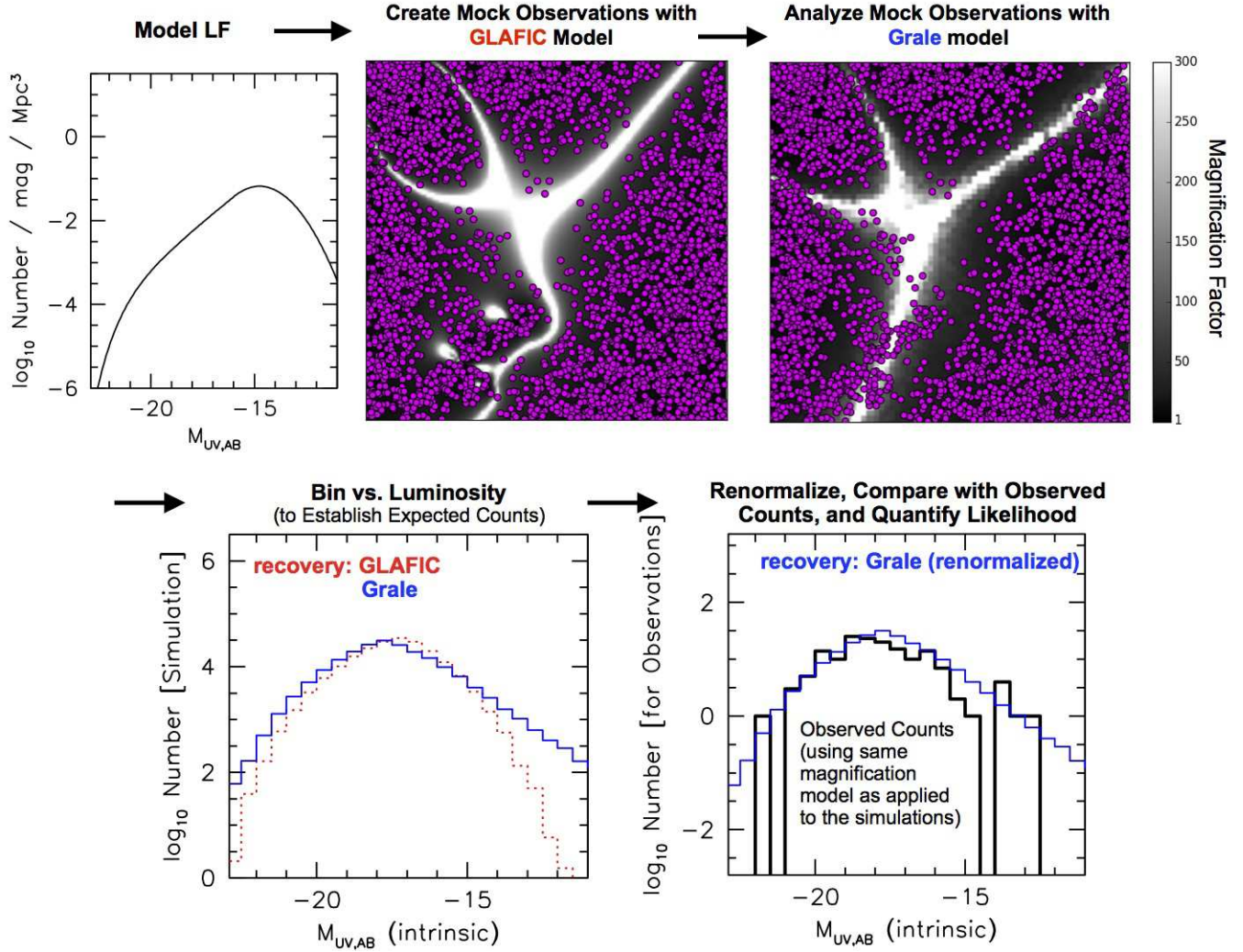}
\caption{Illustration of the steps in our forward-modeling approach to
  determine the impact of errors in the lensing models on the derived
  LF results (\S4.2: see also \S3.2 and \S4.1).  The upper middle
  panel shows the positions of the faint $H_{160,AB}>28$ sources
  (\textit{violet circles}) from a mock catalog created over a
  14$''$$\times$14$''$ region in the image plane near the center of
  Abell 2744 based on a model LF (\textit{shown in the upper left
    panel}) and the \textsc{GLAFIC} lensing model, with the color bar
  at the top right providing the magnification scale for various
  shades of black (low) and white (high).  [Note that sources are
    distributed uniformly over the source plane for the construction
    of the mock catalog.]  The upper right panel shows where this same
  catalog of sources lies in the image plane relative to the critical
  lines in the \textsc{Grale} lensing model over the same region in
  Abell 2744.  The lower left panel shows histograms of the number of
  sources in our mock catalogs vs. luminosity, using both the original
  \textsc{GLAFIC} model used to construct the mock catalogs
  (\textit{dotted red histogram}) and \textsc{Grale} model used for
  recovery (\textit{blue histogram}).  We use these simulations to
  derive the expected number of galaxies per luminosity bin for a
  given LF and compare this with the observed numbers (where the
  intrinsic $M_{UV}$ is calculated using the \textsc{Grale} model) to
  estimate the likelihood of a given LF model (\textit{lower right
    panel}).  In the presented example, the turn-over in the LF at the
  faint end translates into a significant deficit of sources near the
  critical lines using the input magnification model.  However, when
  interpreting this same catalog using a different lensing model,
  i.e., \textsc{Grale} in this case, many sources nevertheless lie
  very close to the critical lines.  As a result of the uncertain
  position of the critical curves, it can be challenging to detect a
  turn-over in the LF at $>-15$ mag.\label{fig:flowchart}}
\end{figure*}

Two other examples of the impact of large magnification errors on LF
results are presented in Figures~\ref{fig:illustsize} and
\ref{fig:illust15} in Appendices C and D, utilizing an input LF with a
faint-end slope of $-1.3$.  For each of these examples, the recovered
LF closely matches the input LF; dramatically, however, faintward of
$-15$ mag (and even $-$16 mag for some models), the recovered LFs
steepen and asymptote again towards a faint-end slope of $\sim$$-2$
(or steeper if sources are resolved), even if the actual slope of the
LF is much shallower (or the LF turns over!).

Each of these examples demonstrate that, regardless of the input LF, a
faint-end slope $\alpha$ of $\sim-2$ will be recovered whenever the
magnification uncertainties are large.  One cannot, therefore, use the
consistent recovery of a steep faint-end slope based on a large suite
of lensing models to argue that the actual LF maintains a steep form
to extremely low luminosities (as was done by L17 using their Figure
11).  The presented examples show this is not a valid argument.

How then can one interpret LF results from lensing clusters when a
steep LF $\alpha\sim-2$ is found?  As we have demonstrated, such a
result could be indicative of the LFs truly being steep or simply an
artifact of large magnification uncertainties.  To determine which is
the case, the safest course of action is to simulate all steps in the
LF recovery process, to determine the impact of magnification
uncertainties on the shape of the LF, and finally to interpret the
recovered LFs from the observations.  While we showed a few examples
here, we formalize the process in the next section.

\section{New Forward-Modeling Methodology to Derive LF Results}

The purpose of the present section is to describe a new methodology
for quantifying the constraints on the $UV$ LF to very low
luminosities, given the uncertainties in the magnification maps.  The
development of such a procedure is useful given the challenges
presented in the previous section.  We will apply this formalism in
\S5.

\subsection{Basic Idea and Utility}

The LF recovery results presented in \S3 (Figure~\ref{fig:illust})
illustrate the impact that errors in the magnification maps can have
on the recovered LFs.  Input LFs, of very different form, can be
driven towards a faint-end slope $\alpha$ of $-2$ at the faint end,
after accounting for the impact of magnification errors.  The results
from \S3.2 demonstrate the importance of forward modeling the entire
LF recovery process to ensure that both the results and uncertainties
are reliable.

We then utilize our forward-modeling approach to derive constraints on
the $z\sim6$ LF.  The basic idea behind our approach follows closely
from the simulations we ran in the previous section and is illustrated
in Figure~\ref{fig:flowchart}.  We begin by treating one of the public
magnification models as providing an exact representation of reality.
In conjunction with an input LF, those models are used to create a
mock data set for the four HFF clusters considered.  The mock data set
is then interpreted using other magnification models for the clusters
to determine the distribution of sources vs. $UV$ luminosity $M_{UV}$
and also to recover the $UV$ LF.  As illustrated by the LF recovery
experiments presented in \S3.2, the recovery could be done with the
models individually or by using some combination of models like the
median.

There are many advantages to using the present procedure to derive
accurate errors on the overall shape of the $UV$ LF.  Probably the
most significant of these is inherent in the end-to-end nature of the
present procedure and our relying significantly on forward modeling to
arrive at accurate errors on the observational results.  Through the
construction of many mock data sets using plausible magnification
models and recovery using other similarly plausible models, the
proposed procedure allows us to determine the full range of allowed
LFs.

In addition, the advocated procedure provides us with a natural means
to account for ``noise'' in the lensing model magnification maps.  The
presence of lensing model ``noise'' is obvious looking at the range in
magnification factors across the various models (e.g., compare the
\textsc{GLAFIC} and \textsc{GRALE} critical lines in
Figure~\ref{fig:flowchart} or see the lower panel in
Figure~\ref{fig:predpow}).  Such noise can even be present in a median
magnification model created from the combination of many individual
models, as there will be regions where the high-magnification regions
of the lensing maps simply overlap due to chance coincidence.
Analogous to considerations of low-significance sources in imaging
observations, the robustness of specific magnification factors in the
median map can be assessed, by considering comparisons with
independent determinations of the same map.  Through the treatment of
one of the public magnification models as the truth, the present
forward-modeling approach effectively formalizes such a technique to
determine the robustness of specific features in the magnification
maps.  The advantage of the current procedure is that this robustness
can be determined using the full magnification maps available for each
cluster (and not just at a limited number of positions where candidate
high-magnification $\mu>10$ sources happen to be found in the real
observations), while also allowing us to derive more reliable
likelihood distributions (with realistic errors).

Another advantage of our forward modeling procedure is that it
explicitly incorporates source selection.  This is important, since
the selection efficiency $S$ could depend on the magnification factor
$\mu$ in the sense that the most magnified sources would also be the
most incomplete, as is likely the case for the brightest and most
extended objects, due to the impact of lensing shear on source
detection (Oesch et al.\ 2015).  If the same situation applied to the
faintest sources in the HFFs (and one did not utilize a procedure that
included forward modeling or an explicit correction), the recovered
LFs would be biased.  This is due to the fact that the actual surface
density of the sources on the sky is proportional to $S(\mu_{true})$,
but it is assumed to be proportional to $S(\mu_{model})$ and
$\mu_{true} < \mu_{model}$ at high magnifications $\mu>10$
(Figure~\ref{fig:predpow}).  The recovered LF would therefore be
higher by the factor $S(\mu_{true})/S(\mu_{model})$.

We would expect such an issue to affect recovered LFs, in all cases
where sources have non-zero size (since the selection efficiency would
then depend on the magnification factor).  For example, if the
completeness of sources shows an inverse correlation with the
magnification factor as Oesch et al.\ (2015) find, i.e.,
$S(\mu)\propto \mu^{-0.3}$ (e.g., as in their Figure 3), a direct
approach would recover a faint-end slope that is $\Delta\alpha \sim
0.3$ steeper than in reality (Figure~\ref{fig:illustsize} from
Appendix C), at very low luminosities where $\mu>10$ (where
$<$$\mu_{true}$$>$ is typically less than $<$$\mu_{model}$$>$).

Remarkably, there is no evidence that this issue is even considered in
many recently derived LFs, which is worrisome given the size
assumptions which were made.  This is most problematic for analyses
pushing to very low luminosities, i.e., $>-15$ mag, while quoting tiny
statistical uncertainties (e.g., L17 who quote statistical
uncertainties on $\alpha$ of $\pm$0.04 vs. this bias which is
$\sim$8$\times$ larger).  

\subsection{Procedure}

We perform our forward modeling simulations at the catalog level, to
ensure that the time requirements on these simulations are manageable.
This involves the construction of catalogs of sources with precise
positions and apparent magnitudes.  Both in the construction of the
mock catalogs and in recovering the LF from these catalogs, the
selection efficiency $S$ must be accounted for, which is in general a
function of the apparent magnitude $m$ and magnification factor $\mu$,
i.e., $S(m,\mu)$.  For the lowest luminosity $z\sim5$-8 galaxies,
there is little evidence to suggest that these sources show
significant spatial extension (Bouwens et al.\ 2017), which implies
that we can credibly treat their selection efficiencies as just a
function of the apparent magnitudes, i.e., $S(m)$.  We describe our
procedure for estimating $S(m)$ in this case in Appendix
B.\footnote{Of course, we also recognize the value in understanding
  the impact on the results if the sizes of sources are larger, and
  this is discussed in Appendix C.  The outcome is similar but leads
  to an even bigger disconnect from the real LF shape.}

In putting together the mock observed catalogs for each LF parameter
set we are considering, i.e., $\phi^*$, $\alpha$, and a third
parameter $\delta$ to be introduced in the next section,
$\sim$2$\times$10$^5$ sources are inserted at random positions (but
yet uniformly in the source plane) across the 4 HFF cluster fields we
are considering (Figure~\ref{fig:flowchart}).  Sources are included in
the catalogs in proportion to their estimated selection efficiencies
$S(m,\mu)$ (Appendix B), their implied volume densities (according to
the model LF), and cosmological volume element (proportional to the
area divided by the magnification factor).  It is the inclusion of
sources in the catalogs in inverse proportion to the magnification
factor that ensures that galaxies are distributed uniformly within the
source plane (since our catalog construction process does not consider
the deflection maps from the lensing models or multiple
imaging).\footnote{We verified that sources in the mock catalogs our
  procedure produced showed a uniform volume density of galaxies (to
  some limiting luminosities) in the source plane, independent of the
  magnification factor tying some region of the image plane to the
  source plane.}  All sources in the input catalogs are assumed to
have the same input redshift $z=6$.

During the recovery process, sources are placed into individual bins
in $UV$ luminosity using a ``recovery'' magnification model, which we
take to be the median of the parametric magnification
models.\footnote{The input magnification model is always excluded when
  constructing the median magnification map (used for recovery) to
  keep the process fair.}  During the recovery process, the redshift
is taken to have the same mean value as assumed in constructing the
mock catalogs, i.e., $z=6.0$, but with a $1\sigma$ uncertainty of 0.3.
This is to account for the impact of uncertainties in the estimated
redshifts of individual sources on the recovered LFs.

We use the results of the simulations we run for each parameter set
(each with $\sim$2$\times$10$^5$ sources) to establish the expectation
values for the number of sources per luminosity bin (in the same
0.5-mag intervals used in the previous section).  We then compute the
likelihood of a given parameterization of the LF by comparing the
observed number of galaxies per bin in luminosity (considering all 4
clusters at the same time) with the expected numbers assuming a
Poissonian distribution, as
\begin{displaymath}
\Pi_{i} e^{-N_{exp,i}} \frac{(N_{exp,i})^{N_{obs,i}}}{(N_{obs,i})!}
\end{displaymath}
where $N_{exp,i}$ is the expected number of sources in a given bin in
intrinsic $UV$ luminosity.  The observed number of sources in a given
bin in $UV$ luminosity $N_{obs,i}$ is computed from the median
parametric magnification maps, as done in \S2
(Figure~\ref{fig:sample}).  A flow chart showing one example of our
forward-modeling approach is provided in Figure~\ref{fig:flowchart}.

\begin{figure}
\epsscale{1.17}
\plotone{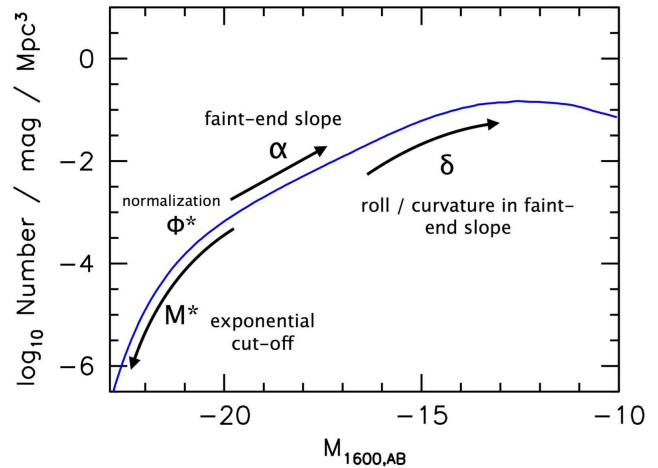}
\caption{Illustration of the parameterization we utilize for the $UV$
  LF in assessing the possibility it may turn over at low luminosities
  (\S4.3).  In addition to the standard Schechter parameters, $M^*$,
  $\phi^*$, $\alpha$, we also allow for curvature in the effective
  slope of the LF using a fourth parameter $\delta$ -- which can be
  used either to represent a roll-over or to indicate a possible
  steepening in the slope towards lower luminosities.  We include such
  a curvature at $>-$16 mag, coincident with the luminosity range
  where magnification uncertainties become larger for individual
  sources.  $-$16 mag also corresponds to that expected for a
  typically-faint source ($\sim$28.5 mag) magnified by a factor of
  10.\label{fig:param}}
\end{figure}

\subsection{Parameterization of LF Model}

Use of a parametric form to the LF is particularly useful for
examining the overall LF constraints on the shape of the $UV$ LF in
lensed fields, due to uncertainties that exist on the magnification
factors, and hence luminosities, of individual sources used in the
construction of the LF.  This makes it difficult to place sources in
specific bins of the $UV$ LF (resulting in each bin showing a larger
error).

For the parametric modeling we do of the LF, we start with a general
Schechter form for the LF:
\begin{displaymath}
\phi^* (\ln(10)/2.5) 10^{-0.4(M-M^{*})(\alpha+1)} e^{-10^{-0.4(M-M^{*})}}
\end{displaymath}
However, we modify the basic form of the Schechter function by
multiplying the general Schechter form by the following expression
faintward of $-16$ mag:
\begin{displaymath}
10^{-0.4\delta (M+16)^2}
\end{displaymath}
Positive values of $\delta$ result in the LF turning over faintward of
$-16$ mag, while negative values of $\delta$ result in the LF becoming
steeper faintward of $-16$ mag.  An illustration of this
parameterization is provided in Figure~\ref{fig:param}.

Our use of $-16$ mag allows us to test for possible curvature in the
shape of the LF at $>-$16 mag, as predicted by some models (e.g.,
Kuhlen et al.\ 2013; Jaacks et al.\ 2013; Ocvirk et al.\ 2016).  $-16$
mag is also just faintward of luminosities probed in field studies
(i.e., $-16.77$ mag: Bouwens et al.\ 2015a).  Finally, since $>-16$ mag
corresponds to the luminosity of $\mu>10$ faint sources in the HFFs,
our fitting for a curvature parameter $\delta$ allows us to
investigate how well the shape of the LF can be recovered in the
regime where the magnification factors are large.

With this parameterization, the turn-over luminosity $M_T$ (i.e.,
where $(d\phi/dM)_{M=M_T} = 0$) can be easily shown to be
\begin{equation}
M_T = -16 - \frac{\alpha+1}{2\delta}
\label{eq:mt}
\end{equation}
assuming that $\delta > 0$.  For sufficiently small values for
$\delta$, i.e, $\delta\sim0.05$, a turn-over in the LF would be so
faint as to be impractical to confirm, and below what would be
expected theoretically (\S6.3).  For typical models (\S6.3), $\delta$
is expected to be $\delta \gtrsim 0.08$, resulting in turnover
magnitudes $<-$10.

\section{LF Results at $z\sim6$}

We now make use of the formalism we presented above and our selection of
160 $z\sim6$ galaxies behind the first four HFF clusters by
R.J. Bouwens et al. (2017, in prep) to set constraints on the form of
the UV LF at extremely faint magnitudes.\\

\begin{figure*}
\epsscale{1.07}
\plotone{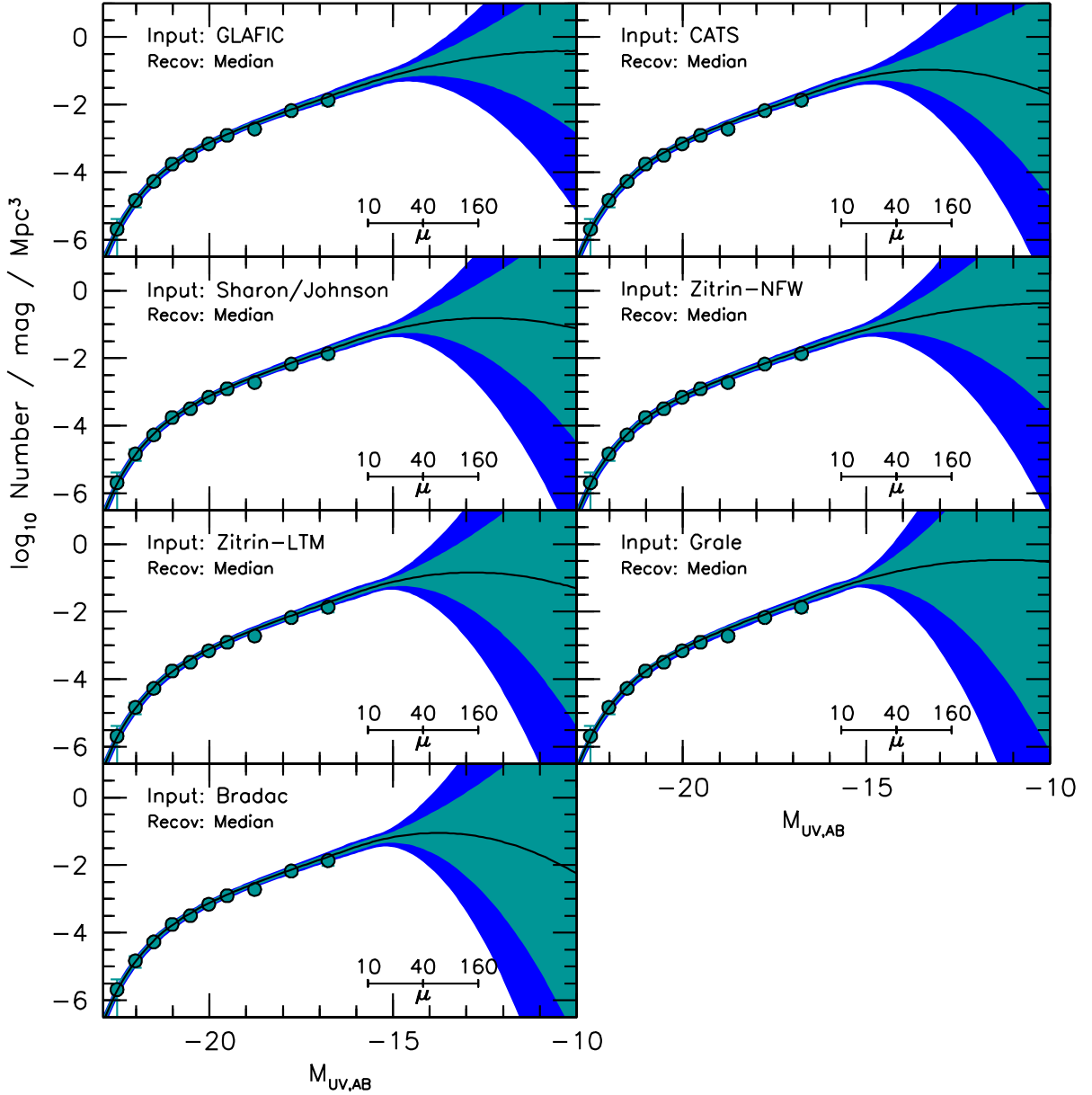}
\caption{Determination of the 68\% and 95\% confidence intervals
  (\textit{shaded in cyan and blue, respectively}: see \S5.2) on the
  overall shape of the $z\sim6$ LF.  The LF here combines constraints
  from the Bouwens et al.\ (2015a) $z\sim6$ study with the HFF
  observations.  We then alternatively assume that the
  \textsc{GLAFIC}, CATS, Sharon/Johnson, Zitrin-NFW, Zitrin-LTM,
  Bradac, and \textsc{Grale} lensing models represent reality and
  recovering the LF using the median of the \textsc{GLAFIC}, CATS,
  Sharon/Johnson, and Zitrin-NFW models (when available and excluding
  models from the median when treated as reality to make the
  assessment fair).  The cyan solid circles show the binned $z\sim6$
  LF from the HUDF, HUDF-parallel fields, and CANDELS (Bouwens et
  al.\ 2015a).  The black line indicates the nominal best-fit LF.  The
  ticked line showing the magnification factors as in
  Figure~\ref{fig:illust}.  The large range of allowed LFs
  (\textit{shaded regions}) is a direct result of the impact of
  magnification uncertainties as illustrated in
  Figure~\ref{fig:illust}.  The plotted confidence intervals are
  tabulated assuming the \textsc{GLAFIC} and Bradac model as inputs in
  Table~\ref{tab:modelswlf}. If differences between this median
  magnification map and the non-parametric magnification maps are
  representative of the actual uncertainties, the present results
  suggest we cannot rule out a turn-over in the LF at $\sim-15$ mag.
  Even taking as the alternate case the assumption that the
  \textsc{GLAFIC} magnification models represent reality, the present
  results suggest a turn-over in the LF is permitted at $\sim-14.2$
  mag within the 68\% confidence intervals.\label{fig:lf6all}}
\end{figure*}

\begin{deluxetable*}{cccccc}
\tablewidth{0pt} \tablecolumns{6} \tabletypesize{\footnotesize}
\tablecaption{ Best-Fit Constraints on the $z\sim6$ $UV$
  LF\label{tab:lfparm}} \tablehead{ \colhead{Input} & \colhead{} &
  \colhead{$\phi^*$ $(10^{-3}$} & \colhead{} & \colhead{} &
  \colhead{}\\ \colhead{Model} & \colhead{$M_{UV}
    ^{*}$\tablenotemark{a}} & \colhead{Mpc$^{-3}$)} &
  \colhead{$\alpha$\tablenotemark{b}} &
  \colhead{$\delta$\tablenotemark{c}} &
  \colhead{$M_{T}$\tablenotemark{$\dagger$}}} 

\startdata
\multicolumn{6}{c}{HFF Observations Alone + $M^*$ from field LF results (\S5.1)}\\
\textsc{GLAFIC} & $-$20.94 & 0.69$\pm$0.04 & $-$1.90$\pm$0.03 & 0.00 & ---\\
CATS & $-$20.94 & 0.66$\pm$0.04 & $-$1.91$\pm$0.02 & 0.00 & ---\\
\textsc{Grale} & $-$20.94 & 0.68$\pm$0.06 & $-$1.98$\pm$0.03 & 0.00 & ---\\
Bradac & $-$20.94 & 0.70$\pm$0.05 & $-$1.89$\pm$0.03 & 0.00 & ---\\
Sharon/Johnson & $-$20.94 & 0.68$\pm$0.04 & $-$1.91$\pm$0.02 & 0.00 & ---\\
Zitrin-NFW & $-$20.94 & 0.58$\pm$0.01 & $-$1.92$\pm$0.03 & 0.00 & ---\\
Zitrin-LTM & $-$20.94 & 0.67$\pm$0.05 & $-$1.95$\pm$0.02 & 0.00 & ---\\
Mean & $-$20.94 & 0.66$\pm$0.06 & $-$1.92$\pm$0.04 & 0.00 & --- \\
Mean Parametric & $-$20.94 & 0.65$\pm$0.06 & $-$1.91$\pm$0.03 & 0.00 & --- \\\\
\multicolumn{6}{c}{Fiducial (\S5.2): HFF Observations + CANDELS/HUDF/HUDF-Parallel (Bouwens et al.\ 2015a)}\\
\textsc{GLAFIC} & $-$20.94 & 0.57$\pm$0.05 & $-$1.92$\pm$0.04 & 0.07$\pm$0.16 & $>$$-$14.2\\
CATS & $-$20.94 & 0.58$\pm$0.05 & $-$1.91$\pm$0.04 & 0.17$\pm$0.20 & $>$$-$14.9\\
\textsc{Grale} & $-$20.94 & 0.63$\pm$0.07 & $-$1.95$\pm$0.03 & 0.16$\pm$0.30 & $>$$-$15.2\\
Bradac & $-$20.94 & 0.57$\pm$0.05 & $-$1.92$\pm$0.04 & 0.21$\pm$0.32 & $>$$-$15.3\\
Sharon/Johnson & $-$20.94 & 0.57$\pm$0.05 & $-$1.92$\pm$0.03 & 0.12$\pm$0.21 & $>$$-$14.9\\
Zitrin-NFW & $-$20.94 & 0.56$\pm$0.06 & $-$1.91$\pm$0.03 & 0.07$\pm$0.20 & $>$$-$14.6\\
Zitrin-LTM & $-$20.94 & 0.58$\pm$0.05 & $-$1.93$\pm$0.03 & 0.14$\pm$0.25 & $>$$-$15.1\\
Mean & $-$20.94 & 0.58$\pm$0.06 & $-$1.92$\pm$0.04 & 0.14$\pm$0.24 & --- \\
Mean Parametric & $-$20.94 & 0.57$\pm$0.05 & $-$1.92$\pm$0.04 & 0.11$\pm$0.20 & ---\\\\
\multicolumn{6}{c}{Idem (but estimating completeness from conventional size-luminosity relations: \S5.4)}\\
\textsc{GLAFIC} & $-$20.94 & 0.54$\pm$0.06 & $-$1.93$\pm$0.04 & $-$0.08$\pm$0.18 & ---\\
CATS & $-$20.94 & 0.54$\pm$0.06 & $-$1.93$\pm$0.04 & $-$0.08$\pm$0.22 & ---\\
\textsc{Grale} & $-$20.94 & 0.58$\pm$0.06 & $-$1.94$\pm$0.03 & $-$0.27$\pm$0.27 & ---\\
Bradac & $-$20.94 & 0.52$\pm$0.06 & $-$1.91$\pm$0.04 & $-$0.25$\pm$0.28 & ---\\
Sharon/Johnson & $-$20.94 & 0.54$\pm$0.06 & $-$1.93$\pm$0.04 & $-$0.03$\pm$0.22 & ---\\
Zitrin-NFW & $-$20.94 & 0.53$\pm$0.06 & $-$1.91$\pm$0.04 & $-$0.21$\pm$0.23 & ---\\
Zitrin-LTM & $-$20.94 & 0.55$\pm$0.05 & $-$1.93$\pm$0.04 & $-$0.25$\pm$0.27 & ---\\
Mean & $-$20.94 & 0.54$\pm$0.06 & $-$1.93$\pm$0.04 & $-$0.17$\pm$0.26 & ---\\
Mean Parametric & $-$20.94 & 0.54$\pm$0.06 & $-$1.92$\pm$0.04 & $-$0.10$\pm$0.22 & ---\\\\
\multicolumn{6}{c}{Literature Including HFF Observations}\\
Atek et al. (2015) & $-20.9\pm0.7$ & 0.28$_{-0.18}^{+0.59}$ & $-2.04_{-0.13}^{+0.17}$\tablenotemark{*} \\
Livermore et al. (2017) & $-20.82_{-0.05}^{+0.04}$ & 0.23$_{-0.02}^{+0.02}$ & $-$2.10$_{-0.03}^{+0.08}$ & --- & $>-11.1_{-0.8}^{+0.4}$\tablenotemark{$\ddagger$} \\\\
\multicolumn{6}{c}{Literature Before HFF Observations}\\
Bouwens et al. (2015a) & $-20.94\pm0.20$ & 0.50$_{-0.16}^{+0.22}$ & $-1.87\pm0.10$ \\
Finkelstein et al. (2015) & $-21.13_{-0.31}^{+0.25}$ & $0.19_{-0.08}^{+0.09}$ & $-2.02\pm0.10$ 
\enddata
\tablenotetext{a}{Fixed} 
\tablenotetext{b}{The faint-end slopes we derive are moderately
  dependent on the assumptions we make about the intrinsic size
  distribution of very low luminosity galaxies.  Nevertheless,
  motivated by the results from a companion paper (Bouwens et
  al.\ 2017) where extremely faint $z\sim5$-8 galaxies were found to
  have a size distribution consistent with point sources, we used this
  assumption in deriving results for the faint end of the $UV$ LF.
  However, since obtaining direct constraints on the size distribution
  and hence completeness of extremely faint galaxies over the HFF
  clusters is difficult, we could underestimate the volume density of
  faint sources.  This could result in faint end slopes that are
  steeper by $\Delta \alpha \sim 0.08$ (if we adopt 20\% larger sizes
  than the Shibuya et al.\ (2015) size luminosity relation instead of
  assuming galaxies to be point sources).}

\tablenotetext{c}{Best-fit curvature in the shape of the $UV$ LF faintward of $-16$ mag (see Figure~\ref{fig:param}).  For HFF-only determinations, the curvature is fixed to 0 for simplicity.}
\tablenotetext{$\dagger$}{Brightest luminosity at which the current constraints from the HFF permit a turn-over in the $z\sim6$ LF (within the 68\% confidence intervals).}
\tablenotetext{$\ddagger$}{This is the luminosity where according to
  Figure 12 of L17, L17 find $\Delta (\textrm{BIC}) = 2$, where BIC denotes the
  Bayesian Information Criteria (analogous to $\Delta\chi^2$ for their
  usage).  Strictly speaking, it is closer to a 84\% confidence limit
  than a 68\% confidence limit.}
\tablenotetext{*}{LF constraints obtained for a combined $z\sim6$-7 sample}
\end{deluxetable*}

\subsection{Using the HFF Observations Alone} 

We begin by looking at the constraints we can set on the shape of the
$z\sim6$ LF by restricting our analysis to $z\sim6$ samples found
behind the first four HFF clusters.  Such an exercise is useful, since
it allows us to examine the LF constraints we obtain from our HFF
search results, entirely independent of results in the field.

In deriving best-fit LF results, we use the following approach.  We
fix $M^*$ to the same value Bouwens et al.\ (2016) found at $z\sim6$,
i.e., $-20.94$ mag, set the curvature $\delta$ to be zero, and then
fit for $\phi^*$ and $\alpha$.  We use a Markov Chain Monte Carlo
(MCMC) algorithm where we start with the field LF results from Bouwens
et al.\ (2015), i.e., $\phi^* = 0.0005$ Mpc$^{-3}$ and $\alpha =
-1.87$ to determine how the likelihood of various parameter
combinations varies as a function of $\phi^*$ and $\alpha$.  For each
set of parameters $\alpha$ and $\phi^*$, we repeat the simulations
described in \S4.2 to calculate the likelihood of those parameters.
For those simulation, we consistently use magnification maps from one
team to create the mock observations and then recover the results
using the median magnification maps formed from the parametric models.
On the basis of the grid of likelihoods we derive, we determine the
most likely values for $\phi^*$ and $\alpha$, while also determining
covariance matrix which best fits the same likelihood grid.  From the
covariance matrix, we estimate errors on $\phi^*$ and $\alpha$.

To determine the impact that errors in the magnification maps can have
on the derived values for $\phi^*$ and $\alpha$, we repeat the
exercise from the above paragraph seven times.  In each case, we treat
the magnification maps from a different team as the truth and proceed
to derive constraints on $\phi^*$ and $\alpha$ using the results from
the first four HFF clusters.  As two of the HFF clusters we examine do
not have Zitrin-NFW magnification maps available, i.e., MACS0717 and
MACS1149, we make use of the Zitrin-LTM-Gauss models instead.  The
results are presented in Table~\ref{tab:lfparm}.

The faint-end slope $\alpha$ results we obtain from the HFF clusters
alone inhabit the range $-1.89$ to $-1.98$ depending on which
magnification model we treat as reality.  The faint-end slope we
estimate averaging the results from all of the models is
$-1.92\pm0.04$, while the faint-end slope $\alpha$ we find using the
parametric models is $-1.91\pm0.03$.  The quoted uncertainty includes
median statistical error added in quadrature with the standard
deviation among the faint-end slope determinations for the different
magnification models.

These results are interesting in that they are consistent with our own
results over the field, i.e., Bouwens et al.\ (2015), where $\alpha =
-1.87\pm0.10$, as well as other estimates in the literature (Yan \&
Windhorst 2004; Bouwens et al.\ 2007; Calvi et al.\ 2013; Bowler et
al.\ 2015; Finkelstein et al.\ 2015) which generally lie in the range
$\sim-1.8$ to $\sim-2.0$.

\begin{deluxetable}{ccccc}
\tabletypesize{\footnotesize}
\tablecaption{68\% and 95\% Confidence Intervals on the $UV$ LF at $z\sim6$ (\S5.2) adopting the functional form in \S4.3\label{tab:modelswlf}}
\tablehead{
\colhead{} & \multicolumn{4}{c}{$\phi(M)$ [$\log_{10}$(\#/Mpc$^3$/mag])}\\
\colhead{} & \multicolumn{2}{c}{Lower Bound} & \multicolumn{2}{c}{Upper Bound} \\
\colhead{$M_{UV,AB}$} & \colhead{95\%} & \colhead{68\%} & \colhead{68\%\tablenotemark{c}} & \colhead{95\%\tablenotemark{c}}}
\startdata
\multicolumn{5}{c}{Case 1 (\textsc{GLAFIC})\tablenotemark{a}}\\
$-$16.75 & $-$1.88 & $-$1.82 & $-$1.69 & $-$1.63\\
$-$16.25 & $-$1.71 & $-$1.64 & $-$1.50 & $-$1.42\\
$-$15.75 & $-$1.54 & $-$1.47 & $-$1.30 & $-$1.23\\
$-$15.25 & $-$1.38 & $-$1.30 & $-$1.13 & $-$1.05\\
$-$14.75 & $-$1.29 & $-$1.18 & $-$0.95 & $-$0.84\\
$-$14.25 & $-$1.31 & $-$1.12 & $-$0.73 & $-$0.54\\
$-$13.75 & $-$1.43 & $-$1.13 & $-$0.49 & $-$0.18\\
$-$13.25 & $-$1.65 & $-$1.18 & $-$0.22 & 0.25\\
$-$12.75 & $-$1.95 & $-$1.29 & 0.07 & 0.73\\
$-$12.25 & $-$2.34 & $-$1.45 & 0.37 & 1.26\\
$-$11.75 & $-$2.82 & $-$1.66 & 0.70 & 1.84\\
$-$11.25 & $-$3.37 & $-$1.93 & 1.04 & 2.48\\
$-$10.75 & $-$4.01 & $-$2.24 & 1.40 & 3.17\\
$-$10.25 & $-$4.74 & $-$2.60 & 1.79 & 3.91\\\\
\multicolumn{5}{c}{Case 2 (Bradac)\tablenotemark{b}}\\
$-$16.75 & $-$1.85 & $-$1.79 & $-$1.66 & $-$1.60\\
$-$16.25 & $-$1.67 & $-$1.60 & $-$1.46 & $-$1.39\\
$-$15.75 & $-$1.50 & $-$1.42 & $-$1.27 & $-$1.20\\
$-$15.25 & $-$1.34 & $-$1.27 & $-$1.11 & $-$1.04\\
$-$14.75 & $-$1.36 & $-$1.21 & $-$0.90 & $-$0.75\\
$-$14.25 & $-$1.57 & $-$1.27 & $-$0.65 & $-$0.35\\
$-$13.75 & $-$1.93 & $-$1.42 & $-$0.36 & 0.16\\
$-$13.25 & $-$2.43 & $-$1.66 & $-$0.04 & 0.75\\
$-$12.75 & $-$3.08 & $-$1.99 & 0.31 & 1.42\\
$-$12.25 & $-$3.88 & $-$2.40 & 0.68 & 2.18\\
$-$11.75 & $-$4.81 & $-$2.90 & 1.09 & 3.02\\
$-$11.25 & $-$5.88 & $-$3.48 & 1.52 & 3.95\\
$-$10.75 & $-$7.10 & $-$4.17 & 1.98 & 4.95\\
$-$10.25 & $-$8.46 & $-$4.93 & 2.47 & 6.04
\enddata
\tablenotetext{a}{For case 1, we assume that
  differences between the magnifications in the median parametric
  model and the \textsc{GLAFIC} model are a good representation of the
  typical errors in the magnification models we utilize.}
\tablenotetext{b}{For case 2, we assume that differences between the
  magnifications in the median parametric model and the Bradac model
  are a good representation of the typical errors in the magnification
  models we utilize.  Similar confidence regions are obtained if one
  uses the \textsc{Grale} model instead of the Bradac model.}
\tablenotetext{c}{If 50\% of faint sources at $z\sim6$-8 are
  significantly spatially extended (intrinsic half-light radii $>$30
  mas), the 68\%-likelihood upper bounds on the implied LF constraints
  rould increase by $\sim$0.3 dex (Bouwens et al.\ 2016).  The actual
  upper bound on the volume density could be much higher if the
  completeness is substantially less than 50\%.}
\end{deluxetable}

It is worthwhile emphasizing the value of the test we perform in the
previous paragraph.  As we consider the use of searches behind lensing
clusters for constraining the faint end of the $UV$ LFs, it is
essential that we derive constraints from the lensing clusters in
isolation of those obtained from field searches to verify that no
major systematics appear to be present in the LF results from the
lensing clusters.  This is relevant, since recent determinations of
the faint-end slope $\alpha$ to the LFs from field and cluster search
results seem to show a substantial discrepancy (Figure 1).

\subsection{Using results from the HFF clusters and the field}

We now proceed to derive constraints on the overall form of the $UV$
LF at $z\sim6$ combining constraints from the field with those
available from the HFF clusters.  For simplicity, we keep the
characteristic luminosity $M^*$ fixed to the value $-$20.94 mag that
we found in our earlier wide-area field study (Bouwens et al.\ 2015a),
given the lack of substantial information in the HFF cluster program
for constraining this parameter due to the small volume probed.

In combining constraints from the field and from the HFF clusters, we
need to allow for some error in the normalization of both the field
and HFF cluster results, as a result of large-scale structure
variations (``cosmic variance'': Robertson et al.\ 2014; see also
Somerville et al.\ 2004 and Trenti \& Stiavelli 2008) and also small
systematic errors in the estimates of the volume densities of galaxies
in each of our probes.  We assume an uncertainty of $\sim$20\% in the
volume density in both the field and HFF LF results.

The 20\% uncertainty we assume to be present in the normalization of
the LF results at both the bright and faint ends includes a $\sim$10\%
uncertainty in our estimates of the selection volume and $\sim$10\%
systematic error due to uncertainties in the total magnitude
measurements (reflecting a $\sim$0.1 mag systematic error in the
magnitude measurements).  The inclusion of such an error is relevant
given the existence of real errors in the estimated volume densities
of galaxies using photometric criteria.  Small differences appear to
be guaranteed, given that the filters available for the selection of
galaxies from the field are different (in particular including a deep
``z''-band filter) from those available over the HFF clusters (which
do not include a deep ``z'' band filter).  Important factors
contributing to these uncertainties are (1) likely differences between
the assumed sizes and SEDs of galaxies vs. redshift in the
observations vs. those in the simulations and (2) uncertainties in the
contamination rate of observed samples.\footnote{Measurements of the
  total magnitudes typically differ at $\sim$0.1 mag level, as evident
  looking at the broad range of magnituede measurements in Skelton et
  al.\ (2014).  See e.g. their Figures 35-36.  The situation is likely
  even more challenging for galaxies behind lensing clusters due to
  the substantial foreground light from the clusters themselves, and
  in fact the total magnitudes measured with different procedures and
  by different groups are found to differ at the $\sim$0.2 mag to
  0.25-mag level.  See \S6.1.2.  This translates into normalization
  differences of $\sim$20\% to $\sim$25\% assuming a faint-end slope
  of $\sim-2$ for LF results.}

Similar to our LF results using the HFF observations alone, we derive
confidence regions on the $z\sim6$ LF results using a MCMC-type
procedure where we explore a limited region in the
$\phi^*$-$\alpha$-$\delta$ parameter space and calculate the
likelihood of each point in parameter space using the forward modeling
simulations we describe in \S4.1.  Our calculated likelihoods
explicitly include a marginalization across the 20\% volume density
uncertainties we assume.  These likelihoods are then multiplied by the
likelihoods on the same Schechter parameters derived by Bouwens et
al.\ (2015a) at $z\sim6$ using results from the full CANDELS program,
the HUDF, and the HUDF parallels (again after marginalizing the
Bouwens et al.\ 2015a over $\phi^*$ to account for a 20\% uncertainty
in the volume density of sources).  Finally, this 3D likelihood grid
is fit to derive the most likely values for $\phi^*$, $\alpha$, and
$\delta$ and also the covariance matrix.

As in our determinations of the LF parameters from the HFF programs
alone, we repeat this exercise seven different times, treating each of
the magnification models from different teams as reality and
recovering the LF results using the median magnification map.  We
present our constraints on each of the LF parameters in
Table~\ref{tab:lfparm}.  The faint-end slope $\alpha$ we estimate
averaging all of our models and just the parametric models are
$-1.92\pm0.04$ and $-1.92\pm0.04$, respectively.

As in our determinations using only the HFF observations themselves,
the faint-end slopes $\alpha$ we derive at $z\sim6$ are fairly
consistent with LF results in the field.  Our obtaining consistent
results for all seven of the magnification model families we consider
is not especially surprising, given the fact that individual sources
would be expected to scatter in almost the same direction as the
dominant slope of the LF, i.e., $\sim-2$ (the expected slope of the LF
from scatter) vs. $\sim-1.9$ (the actual slope of the LF).

\begin{figure}
\epsscale{1.15}
\plotone{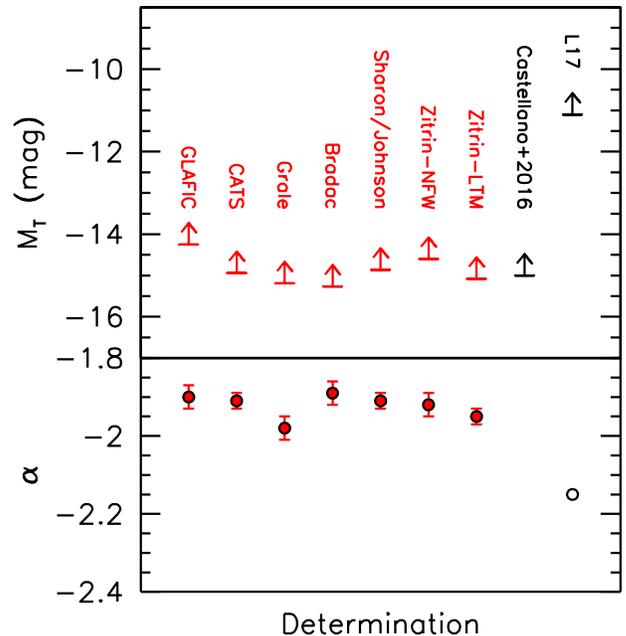}
\caption{(\textit{upper}) Brightest luminosity allowed for a potential
  turn-over in the $z\sim6$ $UV$ LF (within our 68\% confidence
  intervals), using faint $z\sim6$ galaxies identified behind the
  first four HFF clusters and assuming different magnification models
  represent reality.  The reported constraints from Castellano et
  al.\ (2016b) and L17 are also presented, with the L17 constraints
  plotted at a $\Delta(\textrm{BIC})$ value of 2.  The substantially
  fainter allowed turn-over luminosity reported by L17 (significantly
  discrepant with the other estimates) is likely an artifact of the
  very large sizes L17 assume (see Figure~\ref{fig:lf6comp}, \S6.1.2,
  \S6.2, and also Bouwens et al.\ 2017) and a large number of sources
  close to the detection limit of their selection
  (Figure~\ref{fig:l17apm}).  (\textit{lower}) Faint-end slope
  $\alpha$ determinations using only the HFF cluster search results.
  See Table~\ref{tab:lfparm} for a tabulation of these
  results.  \label{fig:turnover}}
\end{figure}

Despite general agreement on the most likely value for $\alpha$ at
$z\sim6$, we find a broad range of values for the curvature parameter
$\delta$, from 0.07 to 0.21, with $1\sigma$ uncertainties ranging from
0.16 to 0.32.  None of the magnification models we considered point
towards our having even modest evidence, i.e., $\delta>0$ at
$>2\sigma$, for the $z\sim6$ LF showing a turn-over at the faint end.

\begin{figure*}
\epsscale{1.15}
\plotone{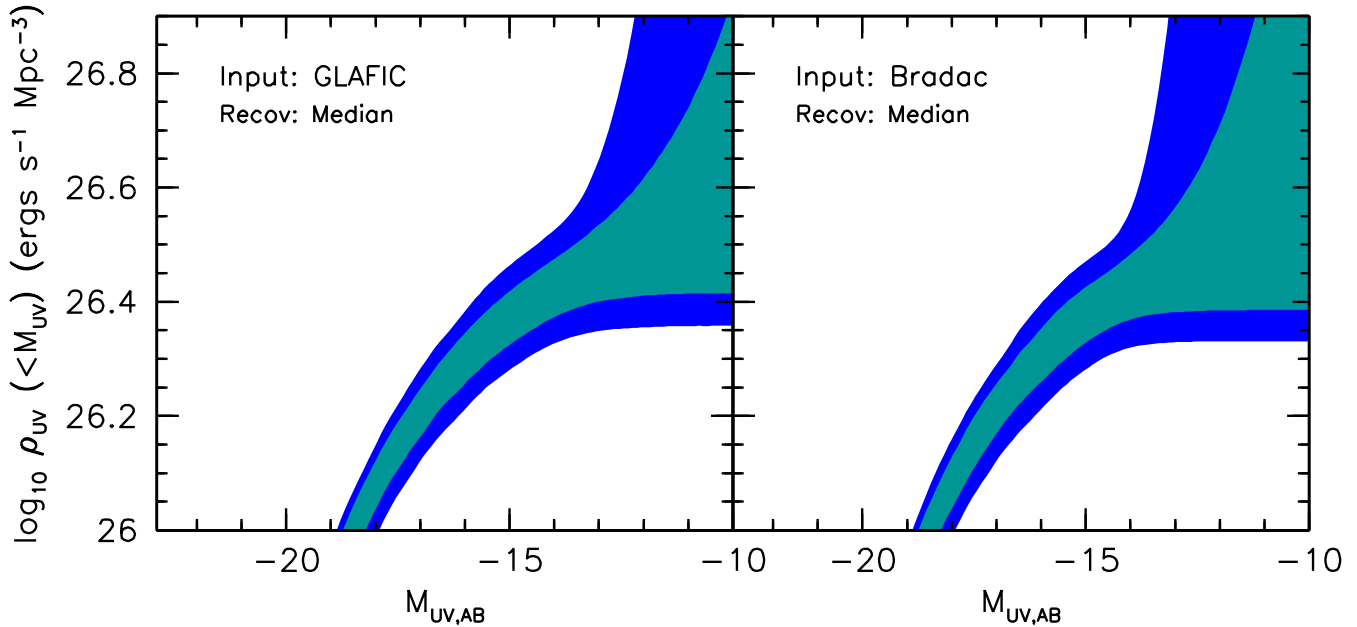}
\caption{68\% and 95\% confidence intervals (\textit{shaded in cyan
    and blue, respectively}) on the estimated $UV$ luminosity density
  for all sources brighter than some $M_{UV}$ (\S5.5).  Shown are the results
  for two different assumptions about the size of the magnification
  errors in the lensing models.  As we integrate further down the $UV$
  LF to derive the luminosity densities, the lower boundaries allowed
  by our analysis show a clear monotonic increase.  However, faintward
  of $\sim-15$ mag, the lower boundaries cease to show an increase
  that is highly significant (i.e., $>$1.5$\sigma$).  This
  demonstrates that it is not yet possible to make strong claims that
  $>-15$ mag galaxies provide an additional reservoir of photons to
  reionize the universe.\label{fig:lumd}}
\end{figure*}

We find the smallest uncertainties on $\delta$ assuming that the
\textsc{GLAFIC} magnification models represent reality.  Slightly larger
uncertainties on $\delta$ are found assuming that the CATS,
Sharon/Johnson, and Zitrin-NFW models represent reality, while the
largest uncertainties on $\delta$ are found, assuming that the
Zitrin-LTM, Bradac, and \textsc{Grale} models represent reality.  The size of
the uncertainty on $\delta$ is a function of how similar the various
magnification models are to the median magnification model formed from
the parametric models.  Given the similarity of assumptions utilized
in the \textsc{GLAFIC}, CATS, Sharon/Johnson, and Zitrin-NFW models, it is not
surprising that their magnification maps agree best with median
magnification maps constructed using similar assumptions.

The results in Table~\ref{tab:lfparm} for the different magnification
models indicate the general range of constraints one could obtain on
the form of the $z\sim6$ LF: the results for the non-parametric models
indicate the errorbars on the LFs if the mass distribution in clusters
do not strictly follow the assumptions made in the parametric models,
while results for the parametric models indicate the likely error
bars, if the mass profiles in the HFF clusters generally do adhere to
those assumptions.

\begin{deluxetable}{ccccc}
\tabletypesize{\footnotesize}
\tablecaption{68\% and 95\% Confidence Intervals on the $UV$ Luminosity Density
at $z\sim6$ to Various Limiting Luminosities (\S5.5)\label{tab:lumd}}
\tablehead{
\colhead{} & \multicolumn{4}{c}{$\log_{10} \rho_{UV}$ ($UV$ Luminosity Density)} \\
\colhead{} & \multicolumn{4}{c}{(ergs s$^{-1}$ Hz$^{-1}$ Mpc$^{-3}$)}\\
\colhead{} & \multicolumn{2}{c}{Lower Bound} & \multicolumn{2}{c}{Upper Bound} \\
\colhead{Faint-End Limit} & \colhead{95\%} & \colhead{68\%} & \colhead{68\%} & \colhead{95\%}}
\startdata
\multicolumn{5}{c}{Case 1 (\textsc{GLAFIC})\tablenotemark{a}}\\
$M_{UV}<-17$ & 26.13 & 26.17 & 26.25 & 26.28\\
$M_{UV}<-15$ & 26.28 & 26.33 & 26.42 & 26.47\\
$M_{UV}<-13$ & 26.35 & 26.40 & 26.54 & 26.65\\
$M_{UV}<-10$ & 26.36 & 26.42 & 26.93 & 28.65\\
$M_{UV}<-3$\tablenotemark{b} & 26.36 & 26.42 & 31.10 & 41.23\\\\
\multicolumn{5}{c}{Case 2 (Bradac)\tablenotemark{a}}\\
$M_{UV}<-17$ & 26.13 & 26.18 & 26.26 & 26.29\\
$M_{UV}<-15$ & 26.29 & 26.33 & 26.43 & 26.47\\
$M_{UV}<-13$ & 26.33 & 26.39 & 26.57 & 27.00\\
$M_{UV}<-10$ & 26.33 & 26.39 & 27.39 & 31.62\\
$M_{UV}<-3$\tablenotemark{b} & 26.33 & 26.39 & 34.35 & 55.95
\enddata
\tablenotetext{a}{Same assumptions as in Table~\ref{tab:modelswlf}.}
\tablenotetext{b}{The $-3$ mag limit is included here for illustrative
  value and takes as its inspiration results from O'Shea et
  al.\ (2015) and Ocvirk et al.\ (2016) which predict sources to such
  faint magnitudes.}
\end{deluxetable}

Uncertainties in the redshifts of the lensed $z\sim6$ galaxies also
contribute to the error in $\delta$.  To estimate the impact, we kept
the redshifts of lensed background souces fixed while rerunning the
forward-modeling simulations from our MCMC chain.  Comparing the
uncertainties we derive in $\delta$ to the uncertainties we derive
including errors in the redshift, we find a typical increase of 0.01
in the uncertainty on $\delta$, i.e., from 0.15 to 0.16 in the case of
the \textsc{GLAFIC} simulations.  If we assume that uncertainties in
the deflection maps and redshifts both add in quadrature, this
suggests that errors in the photometric redshift errors contribute
$\sim$12\% of the fractional error in $\delta$, i.e., (0.16$^2$ -
0.15$^2$)/0.16$^2$$~\sim 0.12$.

To help visualize what our present LF results mean, we have made use
of our parametrized constraints to derive 68\% and 95\% confidence
intervals on the volume density of galaxies as a function of the $UV$
luminosity $M_{UV}$.  These results are presented both in
Figure~\ref{fig:lf6all} and also in Table~\ref{tab:modelswlf}.\\

\subsection{Constraints on a Possible Turn-Over in the $z\sim6$ LF}

One question that has recently been of significant interest in the
literature regards whether there is a flattening or turn-over in the
$UV$ LF at the faint end.  This question is important, since the
answer could indicate to us whether there is a sufficient number of
extremely faint galaxies to produce the photons necessary for driving
the reionization of the universe.

Fortunately, using the likelihood contours for
$\delta$-$\alpha$-$\phi^*$ and Eq.~\ref{eq:mt}, we can directly
determine the brightest point in the LF where a turn-over is permitted
(with the 68\% confidence intervals).  The results do depend somewhat
on which magnification model we assume to be representative of reality
(and therefore which of the panels we consider from
Figure~\ref{fig:lf6all}).  Nevertheless, we find that the HFF
observations allow for a turn-over in the LF as bright as $-14.2$ mag
to $-15.3$ mag (within the 68\% confidence intervals).  The allowed
turn-over luminosities we estimate assuming different magnification
models are presented in Table~\ref{tab:lfparm} and also in
Figure~\ref{fig:turnover}.\\

\subsection{LF Results: The Impact of Galaxy Size}

In a companion paper (Bouwens et al.\ 2017), we showed that the slope
of the luminosity function at low luminosities is strongly dependent
on the size of very faint galaxies (see Figure 2 from that paper).  We
constrained the sizes of faint galaxies ($>-16$ mag) by taking
advantage of the large samples of such galaxies available behind the
first four HFF clusters at $z\sim2$-3 and $z\sim5$-8.  We found no
evidence to indicate that these galaxies were significantly resolved,
looking at (1) the prevalence of high magnification sources as a
function of the predicted shear and (2) their stacked profile along
the expected shear axis.  The slope varied dramatically from
$\alpha\sim-1.9$ for the small sizes we found (taking 7.5 mas to be
representative of the half-light radius) to $\alpha\sim-2.7$ (taking
120 mas as representative).  The resulting luminosity density from
integrating the LF to $-$14 changed by a factor of 40!  Clearly the
size assumed for faint galaxies is a key parameter that is central to
a reliable determination of the LF.

On the basis of our findings in a companion study, can we assume that
$>-$16 mag galaxies are all entirely unresolved?  Unfortunately, we
cannot due to the impact of surface brightness selection effects on
the recovered samples.  The impact is sufficient that we cannot rule
out (86\% confidence) sources having half-light radii of 30 mas, which
is approximately the size we would predict for faint sources
extrapolating conventional size-luminosity relations.

Given this fact, it is therefore quite logical to repeat the present
determination of the LF, but this time assuming a conventional
size-luminosity relation.  We will then treat the results as providing
an upper bound on the $z\sim6$ LF results, given uncertainties in the
size distribution.  To the end, we suppose that the median half-light
radii of galaxies follow the following correlation with luminosity
$r_{hl} = (0.14'') (L/L_{z=3}^*)^{0.27}$, which is in good agreement
with the size-luminosity relation found by Shibuya et al.\ (2015).  In
addition, we adopt the sizes of galaxies exhibit a log-normal
distribution with 0.3 dex $1\sigma$ scatter.  We assume galaxies to
have a Sersic profile and for the Sersic indices to have a log-normal
distribution with a median of 1.5 and scatter of 0.3 dex.  The axial
ratio is also assumed to be log-normal with a median value of 1.8 and
a scatter of 0.3 dex.  A random position angle is adopted for sources.
Finally, the pixel-by-pixel profiles for all sources in the
Monte-Carlo catalogs are computed.  The impact of gravitational
lensing is included using the latest deflection maps from the
\textsc{CATS} team.

The simulated galaxies are then added to the real observations, and we
utilize the same procedure for catalog creation and source selection
as we use on the real data.  We then rederive the selection volumes in
the same way as before (i.e., see Appendix B) and repeat the
determination of the LF results using the outlined forward-modeling
procedure.  Compared to the situation where point-source sizes are
assumed, the selection volumes we derive are lower, increasing the
overall volume density of sources inferred at lower luminosities
$>-$18 mag.

The approximate impact of this use of larger sizes for faint sources
is to increase the volume density of sources at $\sim-17$, $\sim-15$
and $\sim-14$ by factors of $\sim$1.6, $\sim$2, and $\sim$3.3,
respectively.  The amplitude of the correction increases towards lower
luminosities due to the correlation between surface brightness and
luminosity implied by conventional size-luminosity relations (where
$R\propto L^{0.27}$: e.g., Shibuya et al.\ 2015), i.e., $L/R^2 \propto
L/(L^{0.27})^2 \propto L^{0.46}$.\footnote{In Bouwens et al.\ (2017),
  we presented evidence that the scaling may be steeper than this,
  i.e., $R\propto L^{0.5\pm0.07}$, at lower luminosities based on the
  sizes and luminosities of $z\sim6$ sources in the HFFs.  We,
  however, caution that correcting for completeness successfully in
  the HFF data is sufficiently challenging that the true relation
  could be shallower than what we found.}

Based on this scaling, one would expect 0.01 $L^*$ and 0.001 $L^*$
galaxies to have $\sim$8$\times$ and $\sim$24$\times$ lower surface
brightnesses, respectively, than more luminous L$^*$ galaxies.  Since
gravitational lensing preserves surface brightness, it should not be
easy to select extremely low luminosity galaxies behind lensing
clusters, if conventional relations held.  We should emphasize,
however, that it is not clear however that conventional relations hold
down to such low luminosities, i.e., $M_{UV}>-16$.  The light profile
in galaxies may be dominated by just a single super star cluster or
two in this regime, as suggested by our results in Bouwens et
al.\ (2017).

The derived LF results we derive for the stated size assumptions are
presented in Table~\ref{tab:lfparm}.  The results are similar to our
fiducial results.  However, they do nevertheless give faint-end slopes
$\alpha$ that are $\Delta\alpha\sim 0.01$ steeper and curvature
parameters $\delta$ which are approximately 0.2 lower.  With the
present size assumptions, $\delta$ inferred for the $z\sim6$ LF is
formally negative for all of the magnification models we consider.  As
in \S5.3, we emphasize that a possible upturn in the LF (i.e.,
$\delta<0$) is not a statistically robust result.  If we force
$\delta$ to be 0, the faint-end slope we derive is $\Delta\alpha\sim
0.03$ steeper for the typical lensing model.  If we allow for such a
change in $\alpha$, the present tension in faint-end slope $\alpha$
vs. L17 would decrease to $3\sigma$.

The exercise in this section demonstrates the sensitivity of the
curvature parameter in the LF $\delta$ -- and in fact the whole
question as to where or if the $UV$ LF turns over -- to the form of
the size-luminosity relation.  We caution that the conclusions here
are based on an extrapolation of sizes seen at significantly higher
luminosities and that the indications from our recent work on sizes in
the HFF clusters (Bouwens et al. 2017) suggest that galaxies at
luminosities $>-$16 may be very small.  In such a case the
completeness corrections will be much smaller.

\begin{deluxetable}{lc}
\tablewidth{0pt} \tabletypesize{\footnotesize} \tablecaption{Binned
  Determination of the rest-frame $UV$ LF at $z\sim6$
  (\S5.6)\tablenotemark{$\dagger$}\label{tab:swlf}} \tablehead{
  \colhead{$M_{UV,AB}$\tablenotemark{a}} & \colhead{$\phi_k$
    (10$^{-3}$ Mpc$^{-3}$ mag$^{-1}$)}} \startdata
$-$20.75 & 0.0002$_{-0.0002}^{+0.0002}$\\
$-$20.25 & 0.0009$_{-0.0004}^{+0.0004}$\\
$-$19.75 & 0.0007$_{-0.0004}^{+0.0004}$\\
$-$19.25 & 0.0018$_{-0.0006}^{+0.0006}$\\
$-$18.75 & 0.0036$_{-0.0009}^{+0.0009}$\\
$-$18.25 & 0.0060$_{-0.0012}^{+0.0012}$\\
$-$17.75 & 0.0071$_{-0.0014}^{+0.0066}$\tablenotemark{b}\\
$-$17.25 & 0.0111$_{-0.0022}^{+0.0102}$\tablenotemark{b}\\
$-$16.75 & 0.0170$_{-0.0039}^{+0.0165}$\tablenotemark{b}\\
$-$16.25 & 0.0142$_{-0.0054}^{+0.0171}$\tablenotemark{b}\\
$-$15.75 & 0.0415$_{-0.0069}^{+0.0354}$\tablenotemark{b}\\
$-$15.25 & 0.0599$_{-0.0106}^{+0.0757}$\tablenotemark{c}\\
$-$14.75 & 0.0817$_{-0.0210}^{+0.1902}$\tablenotemark{c}\\
$-$14.25 & 0.1052$_{-0.0434}^{+0.5414}$\tablenotemark{c}\\
$-$13.75 & 0.1275$_{-0.0747}^{+1.6479}$\tablenotemark{c}\\
$-$13.25 & 0.1464$_{-0.1077}^{+5.4369}$\tablenotemark{c}\\
$-$12.75 & 0.1584$_{-0.1343}^{+19.8047}$\tablenotemark{c}\\
\enddata
\tablenotetext{$\dagger$}{These LF results are simple estimates,
  representing the \# of sources at a given luminosity (using the
  median magnification maps) after division by the selection volumes.
  No account is made for scatter resulting from errors in the
  magnification maps.  Errors in the magnification maps can be best
  handled using the forward modeling simulations and methodology we
  consider in \S4.2, leading to the results presented in
  Table~\ref{tab:modelswlf}.}
\tablenotetext{a}{Upper limits are $1\sigma$.}
\tablenotetext{b}{The $1\sigma$ upper limits indicate the upper limits
  if one adopts the larger size for sources assumed in \S5.4.}
\tablenotetext{c}{68\% confidence intervals on the $z\sim6$ $UV$ LF at
  $>-16$ mag we achieve using forward modeling and observations of the
  first four HFF clusters in \S4.3.  The quoted constraints give the
  geometric mean of our results using the \textsc{GLAFIC}, CATS, and
  Sharon/Johnson parametric models as inputs.  The $1\sigma$ upper
  limits indicate the upper limits if one adopts the larger size for
  sources assumed in \S5.4 (resulting a $\sim$0.01 and $\sim$0.2-0.3 more
  negative values for $\alpha$ and $\delta$).  The error bars are not
  independent.}
\end{deluxetable}

\subsection{Implied Constraints on the $z\sim6$ UV Luminosity Density}

To determine if galaxies produce enough ionizing photons to drive the
reionization of the universe, we require constraints on the luminosity
density in the rest-frame $UV$ that include the contribution from all
galaxies.

We can compute confidence intervals on the luminosity densities in the
rest-frame $UV$ by using the constraints from the analyses performed
in the previous section and then marginalizing over $\phi^*$,
$\alpha$, and $\delta$.  We compute the results to a number of
different limiting luminosities $M_{UV}$, i.e., $-17$ mag, $-15$ mag,
$-13$ mag, $-10$ mag, and $-3$ mag,\footnote{The faintest limit here,
  $-3$ mag, is only included for illustrative value and takes as its
  inspiration results from O'Shea et al.\ (2015) and Ocvirk et
  al.\ (2016) which predict sources to such faint magnitudes.}  the
first four of which commonly appear in the literature, in considering
whether galaxies might drive cosmic reionization, particularly
including the contribution at very low luminosities.  We have
presented the results we obtain in Figure~\ref{fig:lumd} and also in
Table~\ref{tab:lumd} to these faint-end limits.  Our results imply a
luminosity density of $10^{26.38\pm0.05}$ ergs s$^{-1}$ Hz$^{-1}$
Mpc$^{-3}$ to $-15$ mag.

Not surprisingly, our LF results allow for essentially an arbitrarily
high contribution from very faint galaxies to the $UV$ luminosity
density, particularly including a contribution from galaxies to $-3$
mag.  These results also provide fairly firm $1\sigma$ and $2\sigma$
lower bounds on the luminosity density.  We find $1\sigma$ and
$2\sigma$ lower limits on luminosity density in the rest-frame $UV$ of
$\sim10^{26.40}$ ergs s$^{-1}$ Hz$^{-1}$ Mpc$^{-3}$ and
$\sim10^{26.35}$ ergs s$^{-1}$ Hz$^{-1}$ Mpc$^{-3}$, respectively.

The $1\sigma$ and $2\sigma$ lower bounds we find on the luminosity
density integrating to arbitrarily faint luminosities is not
especially higher than what we find integrating to $-15$ mag.  These
results indicate that it is not yet possible to argue for the
discovery of a significant additional reservoir of photons from
galaxies faintward of $-15$ mag (Atek et al.\ 2015a,b), as has been
the claim in one recent study (L17).

\subsection{Binned Determinations of $z\sim6$ LF Using Direct Method}

Finally, to conclude this section, we derive a binned representation
of the results from the first four HFF clusters.  Binned
representations of the LF results from the HFFs should be very
reliable at high luminosities, where errors in the magnification maps
are smaller.  Binned representations retain the advantage that they
are a much more model independent probes of the LF shape as a function
of luminosity.

In our binned representation of the LF, we adopt bins of width 0.5 mag
and determine the binned LF results $\phi_m$ to be as follows:
\begin{equation}
\phi_m = \frac{N_m}{V_m}
\end{equation}
where $N_m$ is the number of sources in absolute magnitude bin $m$
after correcting for the estimated magnification.  

We derive the selection volumes $V_m$ in a given magnitude bin using
the following equation:
\begin{equation}
V_m = \int_{A} \int_{dz} C(z,m,\mu) \frac{1}{\mu(A)} \frac{dV(z)}{dA} dz dA
\end{equation}
where $A$ denotes the area, $V$ denotes the volume, $C$ denotes the
estimated completeness, and $\mu$ denotes the magnification factor.
Our estimates of the completeness are provided in Appendix B; the
completeness $C$ appears not to be a strong function of the
magnification factor $\mu$ in data sets as deep as the HFFs, if we
take the results of Bouwens et al.\ (2017) to be indicative.

\begin{figure}
\epsscale{1.17}
\plotone{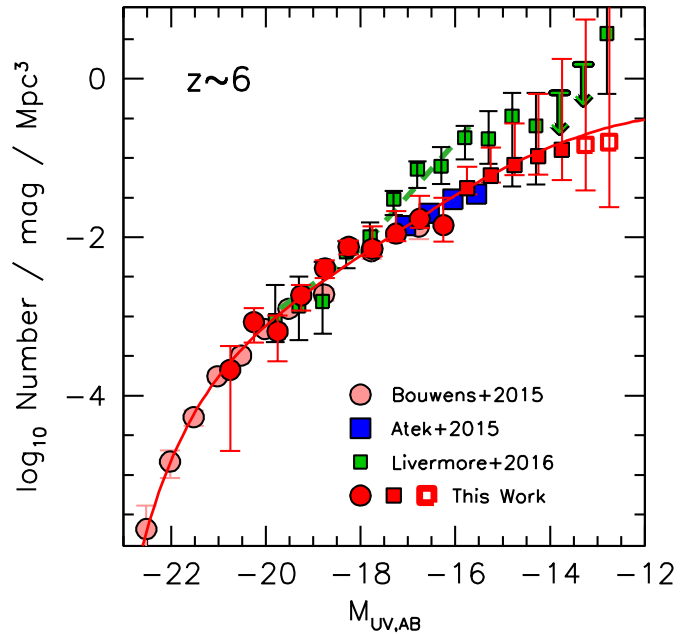}
\caption{Comparison of the present stepwise $UV$ LF from the HFF
  program at $z\sim6$ (\textit{dark red circles}: see \S5.6) with
  previous determinations by Atek et al.\ (2015: \textit{blue
    squares}), L17 (\textit{green squares}), and Bouwens et
  al.\ (2015a) which just come from using the HUDF, HUDF-parallel, and
  CANDELS fields (\textit{light red circles}).  All error bars and
  upper limits are $1\sigma$.  The dark red squares and open red
  squares give the results from our full forward-modeling procedure,
  as given in \S4 (but where the error bars are not independent: see
  \S5.6).  We use open red squares at $>-$13.5 mag where there are no
  $z\sim6$ sources in our selection to indicate a greater uncertainty.
  The upper $1\sigma$ error bars include uncertainties in the size
  distribution (\S5.4).  See Table~\ref{tab:swlf} for a tabulation of
  the present constraints shown here.  The red line shows our best-fit
  LF that we derive by doing a forward-modeling analysis using the
  \textsc{GLAFIC} magnification models as inputs.  The luminosities of
  the individual points in the L17 and Atek et al.\ (2015) LFs have
  been corrected brightward by $\sim$0.3 to ensure better consistency
  with the luminosities (and total magnitudes) measured in our own
  study (see \S6.1.2).  The dashed green line schematically indicates
  the upward break in the $z\sim6$ LF results of L17 (i.e., an
  apparent steepening) that likely impacts their interpretation of the
  shape of the LF at lower luminosities, i.e., their claim that there
  is no turn-over in the $z\sim6$ LF until $\sim-$12 mag (see
  \S6.2).\label{fig:lf6comp}}
\end{figure}

We derived simple stepwise constraints on the $UV$ LF brightward of
$-16$ mag, by dividing the number of sources in each absolute
magnitude bin by the computed selection volume.  The results are
presented in Table~\ref{tab:swlf} and Figure~\ref{fig:lf6comp}.  The
upper limits we quote on the volume densities of sources include a
possible $1.6\times$ underestimate, if the sizes of sources follow the
size-luminosity relation presented in \S5.4.

Faintward of $-16$ mag, we include results in Table~\ref{tab:swlf} and
Figure~\ref{fig:lf6comp} by taking the geometric mean of the best-fit
LF results using the \textsc{GLAFIC}, CATS, and Sharon/Johnson models.
We present those results to luminosities plausibly probed by the
present study.  $1\sigma$ errors on the results at the faint end of
the $z\sim6$ LF are similarly taken to be the geometric mean of the
confidence intervals on the LF fits for the same parametric
magnification models.  The upper limits we quote on the LF results
include the impact of the larger sizes quoted in \S5.4, which we find
to result in a 0.01-0.03 steeper value of the faint-end slope $\alpha$
and to lower the inferred $\delta$ by $\sim$0.2-0.3.

\section{Discussion}

The purpose of this paper is to present new constraints on the form of
the $z\sim6$ LF to low luminosities utilizing new constraints from the
first four clusters available from the HFF program.

\subsection{Comparison with Previous Observational Constraints}

Before looking into comparisons of our new observational constraints
with theory, it is important first to compare the present results with
previous observational results where available to evaluate and
understand any differences.  Doing so helps clarify the gains that are
made with our newer analysis and enables others to also understand
what new factors led to the changes from prior work.

\subsubsection{Atek et al.\ (2015)}

We first consider a comparison with the most recent results of Atek et
al.\ (2015) who make use of observations available over the first
three HFF clusters Abell 2744, MACS0416, and MACS0717 and selected
galaxies using an $I_{814}$-dropout selection criteria which would
identify galaxies from $z\sim6$ to $z\sim7$.  

A comparison with the most recent determination of the $z\sim6$-7 LF
from Atek et al.\ (2015) is provided in Figure~\ref{fig:lf6comp}.  In
comparing against the Atek et al.\ (2015) LF determinations, we
incorporate a $\sim$0.3-mag brightward shift of the Atek et
al.\ (2015) LF to correct for differences in our apparent magnitude
measurements for individual sources.  As already noted in one of the
companion papers to this study (Bouwens et al.\ 2017), overall the
agreement appears to be quite good, at least insofar as the stepwise
points are concerned.

The best-fit $\phi^*$ and luminosity density that Atek et al.\ (2015)
estimate to $-15$ mag, i.e., $\sim10^{26.20\pm0.13}$ ergs s$^{-1}$
Hz$^{-1}$ Mpc$^{-3}$, is $\sim$0.18 dex lower than what we find.  This
is a small but readily understandable difference that arises because
Atek et al.\ (2015b) provide a constraint on the LF at a higher mean
redshift than we do, i.e., $z\sim6.5$ vs. $z\sim6$, and also include
in their determinations results from field surveys, i.e., CANDELS or
the HUDF, which probe $z\sim7$ vs. our $z\sim6$ probe.  Given that the
integrated luminosity density to a limit of $-17$ mag changes by
$\sim$0.2 dex per unit redshift, our larger luminosity density
estimate is entirely expected.

\subsubsection{L17}

As can be seen from Figure~\ref{fig:lf6comp}, the differences between
our results and those from L17 are more substantial than those with
Atek et al.\ (2015) discussed above. These differences with L17 are
particularly large for the LFs at the fainter magnitudes $>-$17 where
much of the current interest lies since this is a region that is
uniquely accessible using the HFF clusters.  As a result, the
discussion of the reasons for these differences is necessarily much
more extensive than that for Atek et al.\ (2015).

To ensure that comparisons with the LF results from L17 were made
using a consistent luminosity scheme, we carefully cross-matched
sources from our catalogs with those from L17.  We also computed
apparent magnitudes for individual sources using the tabulated
absolute magnitudes, redshifts, and magnification factors L17 provide
in their Tables 7-9.  It is to these derived apparent magnitudes we
compare to our own photometry and that of other groups.

Comparing the total magnitudes we derive for sources using our scaled
aperture scheme to the L17 apparent magnitudes, we find a 0.43-mag
median difference, with the L17 apparent magnitudes being fainter than
ours, both for relatively bright $H_{160,AB}<28$ sources and also for
the fainter $H_{160,AB}>28$ sources.  If we instead estimate total
magnitudes for sources by taking the flux in fixed apertures that
would enclose 70\% of the flux for point sources, as performed by the
HUDF12 team (Schenker et al.\ 2013; McLure et al.\ 2013) and derive an
inverse variance-weighted total magnitude from the $Y_{105}$,
$J_{125}$, $JH_{140}$, and $H_{160}$ bands, we find differences of
0.2-mag in the median, with the L17-inferred magnitudes being fainter,
comparing magnitudes for the faintest sources (i.e., $>$28 mag).  The
L17 magnitudes show a similar offset relative to the published
photometry of Atek et al.\ (2015a).

Given that the HUDF12 apparent magnitude measurement scheme should
give a fairly conservative lower limit on the total fluxes for
individual candidates, these comparisons suggest that L17
systematically underestimate the luminosity of individual sources in
their catalog by at least $\sim$0.2 mag, if not more (taking our
scaled-aperture magnitudes as the baseline).

\begin{figure}
\epsscale{1.17}
\plotone{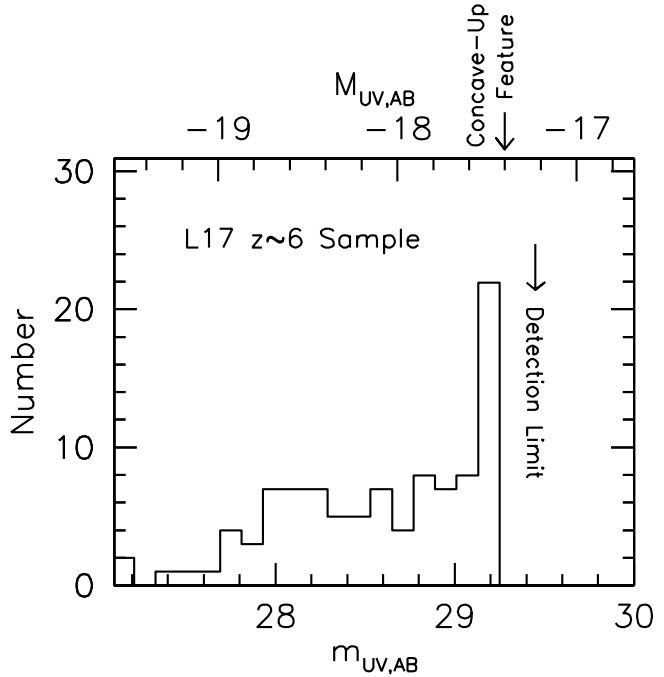}
\caption{Number of sources per apparent magnitude bin ($\Delta m =
  0.12$ mag) in the L17 $z\sim6$ sample shown with respect to the
  approximate detection limit in the HFF data.  The upper horizontal
  axis shows the equivalent absolute magnitude for a $z\sim6$ source
  assuming a magnification of 1.  Apparent magnitudes are derived from
  the absolute magnitudes, redshifts, and magnification factors given
  in Table 7 of L17.  We note 22 sources in the $m_{AB}=29.13$-29.25
  bin just brightward of the detection limit vs. $\sim$7 sources in
  the typical $\sim$0.12 mag bin.  This large pile-up of sources at
  the $z\sim 6$ magnitude limit is not apparent in Figure 9 of L17,
  since L17 set the upper vertical axis to 30 -- even though there are
  actually 45 sources in their faintest bin.  Given that this is also
  where their recovery fraction (and sample completeness) is
  approaching zero (Figure 4 of L17), one would expect their LF to
  show a substantial increase in the volume density of sources over
  the entire absolute magnitude range where sources in their
  $m_{AB}=29.13$-29.25 mag bin impacts the LF, i.e., from $-17.5$ mag
  (where the magnification factor is $\sim$1) to $-12.5$ mag (where
  the magnification factor would be 100).  This apparent upturn in the
  L17 LF is illustrated with the dashed green line in
  Figure~\ref{fig:lf6comp}.  The impact of the 22 sources on the
  $z\sim6$ LF would be exacerbated by L17's assuming larger sizes than
  are found in most observational studies (Kawamata et al.\ 2015;
  Laporte et al.\ 2016; Bouwens et al.\ 2017: see \S6.1.2).
  \label{fig:l17apm}}
\end{figure}

Given this range in values, we adopt a shift of the binned $z\sim6$ LF
of L17 brightward by 0.3 mag to compare volume density measurements at
luminosities closer to what we measure.  After doing so, we find that
the L17 stepwise results appear to be a factor of $\sim3$-4$\times$
higher than our own results in the luminosity range $-17$ to $-14.5$
mag (see Figure~\ref{fig:lf6comp}).  After considering various
explanations for these differences, it seems likely that the sizeable
discrepancies arise due to the excess of sources L17 at the
completeness limit, i.e., there are 22 sources in the
$m_{AB}=29.13$-29.25 bin vs. $\sim$7 sources per bin brightward of
29.1 mag (this can be clearly seen in Figure~\ref{fig:l17apm} which is
adapted from Figure 13 of Bouwens et al.\ 2017).  When one combines
such an excess with the large intrinsic half-light radii that L17
assume (median of 0.09$''$) results, it seems clear that L17 may
significantly overestimate the volume density implied by galaxies at
the faint end of their probe.  In Bouwens et al.\ (2017), we
demonstrated through extensive simulations that the assumptions of
such large sizes for faint galaxies would result in much higher
inferred volume densities for sources (by factors of $>$5) than if
smaller sizes (i.e., $<$10 mas) were used.  The effect of using large
sizes for faint galaxies can be seen in Figure 2 from Bouwens et
al.\ (2017).

Based on recent results from a number of papers (Bouwens et al.\ 2017;
see also Kawamata al.\ 2014; Laporte et al.\ 2016), it is now clear
that the use of such large sizes for very faint galaxies is not
realistic.  Several lines of evidence indicate that the intrinsic
half-light radii of very low luminosity sources (i.e., $>-$16 mag) are
very small, i.e., $\lesssim$0.03$''$, with many sources appearing to
have sizes $\lesssim$0.01'' (Bouwens et al.\ 2017).  Small sizes imply
small completeness corrections and therefore much lower volume
densities for faint galaxies.  It is important to verify that the
sizes assumed in selection volume simulations are realistic, given how
sensitive the selection volume estimates (and hence LFs) for faint
galaxies are to the sizes assumed for faint galaxies (see Figure 13 of
Bouwens et al.\ 2017).

\begin{figure*}
\epsscale{1.17} \plotone{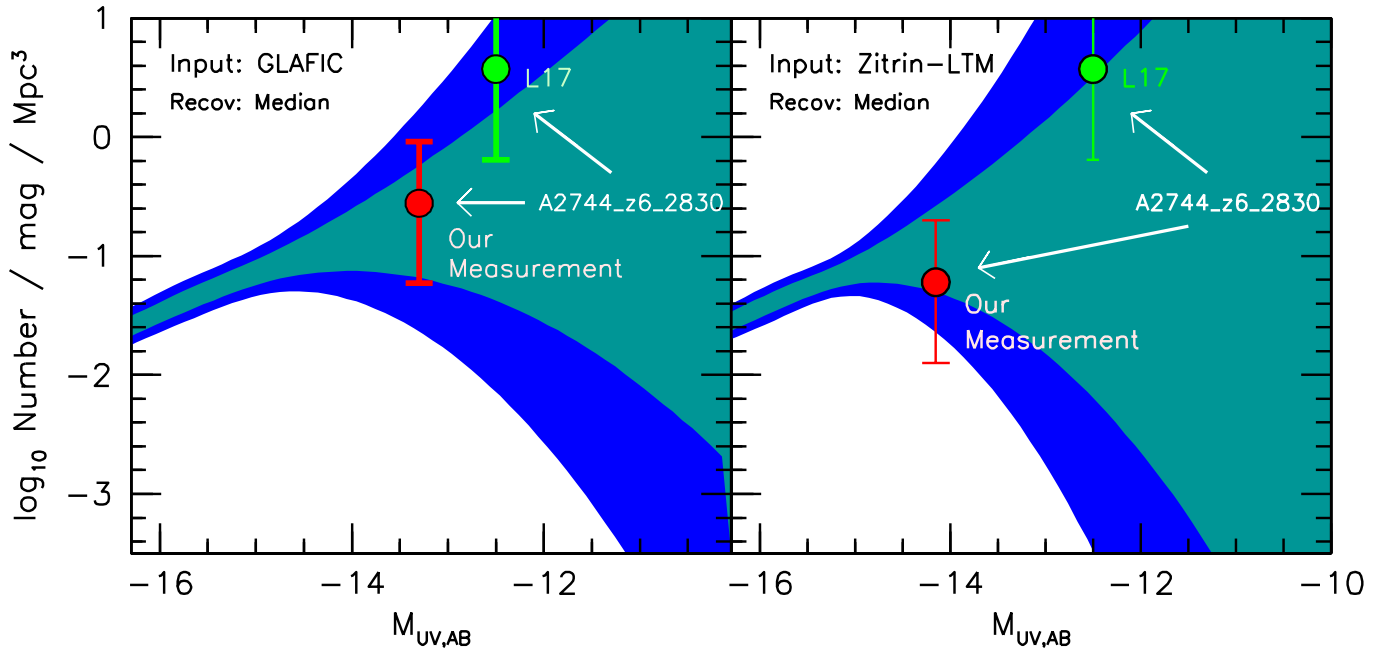}
\caption{Comparison of the faintest point in the $z\sim6$ LF from L17
  (\textit{large green point} with $1\sigma$ error bars) with the 68\%
  and 95\% likelihood intervals implied by our $z\sim6$ LF results
  (\textit{shaded in cyan and blue, respectively}) assuming that the
  \textsc{GLAFIC} and Zitrin-LTM magnification models represent
  reality (\textit{left and right panels, respectively}).  This point
  originates from just a single $z\sim6$ candidate in the L17 catalog,
  but is important because it provides significant leverage in their
  discussion regarding a turn-over.  While this source does not
  satisfy the criteria for our own $z\sim6$ selection due to its
  having an estimated probability of $\sim$50\% of lying at $z<4$, we
  can nevertheless determine the LF constraint we would obtain if we
  had included it in our $z\sim6$ sample.  This is shown with the
  large red point in each panel.  The absolute magnitude we estimate
  for this source is 1.0 mag brighter than what L17 estimate.  The
  Zitrin-LTM magnification model implies that this source is another
  factor of 3 brighter yet than in the \textsc{GLAFIC} models.  For
  either lensing model, we find no significant tension between our
  68\% and 95\% likelihood contours and the volume densities we
  estimate for this candidate using our own photometry and selection
  volume constraints.  While this point plays a significant role in
  L17's discussion regarding a turn-over, a reassessment of its
  luminosity and volume density indicates that it is consistent with
  other forms for the $z\sim6$ LF, including one with a turn-over at
  $\sim-$15 mag.\label{fig:faintz6}}
\end{figure*}

The very high volume densities L17 find for $>-17.5$ mag galaxies in
their $z\sim6$ LF seem likely to have impacted the analyses they
performed regarding a possible turn-over at the faint end, as we
explain in \S6.2.  For similar reasons, the present constraints on the
faint-end slope $\alpha$ that we determine, i.e.,
$\alpha=-1.92\pm0.04$ are shallower than obtained by L17.  Given that
the constraints can also depend on the gravitational lensing model
assumed, it is best to compare faint-end slopes assuming the same
gravitational lensing model.  If we take their formal constraints as
measured assuming the \textsc{GLAFIC} gravitational lensing model, our
estimated faint-end slope $\alpha$ is $-1.92\pm0.04$ vs. the L17
faint-end slope at $z\sim6$ of $-2.10\pm0.03$.  The difference in the
derived slope is 3.5$\sigma$, combining the errors from both
measurements of the slope.

The actual differences between our faint-end slope $\alpha$ estimates
and L17's estimates are likely even larger than 3.5$\sigma$, if we
compare like measurements with like.  L17 combine their HFF
measurements with the Finkelstein et al.\ (2015) field constraints,
which prefer $-2.02\pm0.10$.  Re-estimating the faint-end slope
$\alpha$ from the L17 HFF results alone, we derive their faint-end
slopes to be $-2.15$ to $-2.3$ (see Appendix E here or \S6.2 of
Bouwens et al.\ 2017).  This is significantly ($\geq$4.5$\sigma$)
steeper than our mean estimate using the HFF data alone of
$-1.92\pm0.04$ (\S5.2; Table~\ref{tab:lfparm}).

\subsection{Observational Constraints on a Possible Turn-over at Very Low Luminosities}

One issue we examined in this study concerned the existence of a
possible turn-over in the $z\sim6$ $UV$ LF at the faint end.  This
question is of great interest for the theoretical models, as we shall
see in the next section, and so any observational claims regarding
where a turnover occurs -- or does not occur -- require a very high
degree of careful analysis and credibility if they are to be of real
value to the theoretical modelers.

We noted in (\S5.3) that we were not able to find clear and compelling
evidence for a turn-over at the faint end of the LF.  Using our
likelihood contours for $\delta$-$\alpha$-$\phi^*$, we concluded that
a turn-over in the LF is allowed faintward of $-$14.2 mag or $-$15.3
mag (at 68\% confidence), depending on whether we assumed smaller or
larger uncertainties in the magnification maps.

Since we could not find compelling evidence for a turn-over at the
faint end of the LF (\S5.3), we proceeded to examine what
constraints we can place on the presence of a turn-over as well as the
luminosity at which a turn-over could occur.  Using our search
results, we concluded that any possible turn-over in the LF would need
to occur at $\gtrsim-14.2$ mag or $\gtrsim-15.3$ mag (within the 68\%
confidence intervals), depending on the assumptions we made about
errors in the magnification maps.

The present conclusions parallel those of Castellano et al.\ (2016b),
who also concluded that the $UV$ LF at $z\sim6$ appeared unlikely to
show a turn-over at $<-15$ mag (68\% confidence), after factoring in
the uncertainties in the magnification maps.  We now proceed to
examine in more detail what constraints we can place on the presence
of a turn-over as well as the luminosity at which a turn-over could
occur.  We also look in more detail at the L17 result since they claim
strong evidence against a turn-over to very low luminosities.

L17 had concluded based on their analysis of $z\sim6$ galaxies behind
the first two HFF clusters that they have ``positive'' and ``strong''
evidence that any turn-over in the LF must be fainter than $-11.1$ mag
and $-12$ mag using the criteria $\Delta (\textrm{BIC})=2$ and $\Delta
(\textrm{BIC})=6$, respectively (where we draw these numbers from
their Figure 12).  BIC denotes the Bayesian Information Criterion
(Schwarz 1978).  In the way L17 apply BIC, BIC is effectively just
equal to $\Delta\chi^2$, and the above limits on the turn-over
luminosity translate to nominal confidence levels of 84\% and 98.5\%,
respectively.  On the basis of the L17 fit results assuming different
magnification models, L17 reported an uncertainty in their allowed
turn-over luminosity of $_{-0.8}^{+0.4}$ mag and $_{-0.6}^{+0.3}$ mag,
respectively.  While our conclusions regarding the lack of a bright
turnover (at $\sim-15$ or $\sim-14$ mag) are consistent with the much
lower limit claimed by L17, there are a number of reasons for being
concerned about the validity of their constraints on the luminosity of
a possible turn-over and their strong statements that it cannot occur
at brighter magnitudes.

To be more specific, L17 use their LF results to claim evidence
against a turn-over in the $z\sim6$ LF 3.1 mag fainter than what we
do, despite their examining just a half the HFF data set and with a
resulting smaller $z\sim6$ sample (we use four clusters vs. their
two).  How could they claim stronger constraints?  Part of this could
be because we have recognized, and applied, the very large
uncertainties that are inherent in using very highly-magnified
sources, as indicated in Figure 8, whereas their analysis does not
include this very large source of systematic error.  Remarkably,
however, it appears that the primary reason for the difference lies
elsewhere, and not with the faintest sources they report.  It appears
to arise from the very high volume densities they estimate for the
$UV$ LF in the range $-17$ to $-15$ (which lie in significant excess
of our own determinations and those of Atek et al.\ 2015, by factors
of $\sim3$-4: see Figure~\ref{fig:lf6comp}).

When compared to brighter points at $<-18$ in the LF, the volume
density of sources L17 report over the luminosity range $-17.5$ to
$-15$ is sufficiently high (with small error bars) as to suggest a LF
form which steepens further at lower luminosities, i.e., $\delta < 0$,
rather than one which retains a fixed faint-end slope and then
flattens towards lower luminosities, i.e., $\delta > 0$ (using our
formalism).  Such a shape, with a ``concave up'' feature around
$\sim-17.5$ mag is quite unusual (\textit{this region is indicated in
  Figure~\ref{fig:lf6comp} with the dashed green line}).  Given this,
we suspect that it would be very difficult indeed for L17 to find
evidence for a turn-over at intermediate luminosities (since their LF
is suggesting the opposite curvature).  Given the statistical weight
of this upward change of slope at $\sim-17.5$ to $-15$ mag, L17 would
find that they needed to probe very faint indeed to find a luminosity
where a turn-over was allowed.

A probable explanation for this derives from what we found in the L17
apparent magnitude distribution in Figure~\ref{fig:l17apm}. The
faintest sources are subject to very large completeness corrections.
Normally, such sources would only contribute to the faintest bin in
the LF, but by virtue of a diversity of magnification factors relevant
to sources in this bin, they impact all the fainter bins in the LF.
The large number of sources in the L17 $m_{AB}=29.1$-29.25 bin
exacerbates this effect, and results in large contributions to the LF
over a broad range.  We will discuss this further in a future paper
(R.J. Bouwens et al.\ 2017, in prep), but the current result suggests
the need for particular conservatism in selections near the detection
limit, when taking advantage of gravitational lensing.

There is a second piece of evidence L17 present which could argue
against a possible turn-over in the LF at $\sim-$15 to $\sim-$14 mag.
This involves their $z\sim6$ candidate at $\sim-12.4$ mag
(A2744\_z6\_2830).  Based on this candidate, L17 estimate a volume
density of $\sim$6 galaxies per Mpc$^3$ at $\sim-12.5$ mag, which
would disfavor a turn-over in the LF at any luminosity down to this
limit.  This candidate galaxy thus assumes a critically-important role
in their conclusions, and it is therefore very important both to
consider and examine, as to its robustness.  Interestingly, this
$z\sim6$ candidate from L17 is also detected in our catalogs, though
we do not include it in our $z\sim6$ sample, since the integrated
likelihood of the candidate lying at $z<4$ is $\sim$50\% and hence it
does not meet our selection criteria.  Nonetheless, given its
importance, we need to give consideration to this object.

On the basis of our own photometry, we estimate the source to have a
total apparent magnitude of 28.9 mag.  Using the median magnification
factor 61$_{-15}^{+44}$ we derive for the source based on the latest
publicly-available magnification models (weighting each type equally)
and a redshift of $z=6.11_{-1.22}^{+1.05}$ (L17's estimate), we
calculate an absolute magnitude of $-13.4$ mag for the source.  This
is $\sim$1 mag brighter than what L17 estimate for the same source.
L17's reported luminosity is based on a magnification factor of
110.0$_{-22.2}^{+129.0}$.  This high magnification factor appears to
be at the high end of the publicly available models, as we discuss in
Appendix F, even though L17's estimate is purportedly a median of
those same model results.  By contrast, our magnification estimates
agree very well ($<$2\% difference) with L17's initial estimate of
60.4$_{-22.2}^{+129.0}$.  It is unclear why L17 changed their median
magnification estimate from 60.4 (their version 1 value) to 110 (their
published value) since the median magnification is clearly much closer
to 60 than it is to 110 (as is obvious from Figure 19 and the entire
discussion in Appendix F).  We have verified that this is the case
both from our own calculations and using the public magnification
calculator.\footnote{https://archive.stsci.edu/prepds/frontier/lensmodels/\#magcalc}
Despite this described tension and this faint candidate not being in
our $z\sim 6$ selection, we can examine the implications this source
would have for our UV LF results if it is indeed at $z\sim 6$.

Taking this single source and dividing by the selection volume in the
range $-13.5<M_{UV,AB}<-13$, we estimate a volume density of
$0.28_{-0.22}^{+0.64}$ Mpc$^{-3}$ mag$^{-1}$.  Interestingly, this is
completely consistent, as illustrated in Figure~\ref{fig:faintz6}
(\textit{left panel}), with the LF constraints we find at $\sim-13.4$
mag treating the \textsc{GLAFIC} magnification model as reality and
recovering the LF results using the median magnification model.  It is
also consistent (\textit{right panel} to Figure~\ref{fig:faintz6})
with the LF results we derive using the other magnification models,
e.g., Zitrin-LTM, which suggest a magnification factor of 19, with the
associated volume density of $\sim$0.06 Mpc$^{-3}$ mag$^{-1}$, instead
of the magnification factor of $\sim$110 adopted by L17.  It is not
surprising that we recover lower volume densities than recovered by
L17, given our use of smaller sizes for the faint $>-16$ mag
population as now appears to be appropriate (Bouwens et al.\ 2017).
Smaller sizes translate into a higher completeness and selection
volume.

Since this source is not included in our LF results, we also checked
the impact it would have on our results if it had been included.  We
found that it only had a moderate impact on the allowed luminosity of
a turn-over.  We find our constraints on a possible turn-over in the
$UV$ LF change by 0.8 mag (becoming fainter) assuming that
\textsc{GLAFIC}-vs.-median model are typical of the true magnification
errors and by $\sim$0.4 mag (becoming fainter) assuming that the
Bradac-vs.-median or \textsc{Grale}-vs.-median models are more typical
of the errors.  If we ask which luminosities we can exclude for a
turn-over at 95\% confidence, the excluded luminosities are $<-14.7$
mag for the \textsc{GLAFIC}-vs.-median case (2.7 mag brighter than
what L17 report for this limit).

The considerations discussed in this section suggest that it is not
possible using current observational data to definitively rule out the
presence of a turn-over in the LF as bright as $-15$ mag and
especially at $-14$ mag.  L17 (despite a smaller $z\sim6$ sample) had
previously claimed strong evidence against a turn-over brightward of
$\sim-12$ mag.  From the present discussion, it appears their
conclusions were impacted by the large numbers of faint sources
incorporated into their LF results near the completion limit
(Figure~\ref{fig:l17apm}) and their size assumptions.  This resulted
in an apparent upturn in their $z\sim6$ LF at $\gtrsim-17.5$ mag,
strongly disfavoring a turn-over in their LF results until very low
luminosities.  A reassessment of L17's faintest source using the
public magnification models shows a wide range of estimated
magnifications, mostly lower than their value and many consistent with
a turn-over at higher luminosities.

\subsection{Comparison with Theoretical Expectations}

The observational constraints we have obtained here are obviously of
great value for comparing against the predictions for the form of the
LF at the faint end, as provided by many different teams using
simulations, theoretical models, and on the basis of observations of
the nearby universe.

We compare our LF constraints with the following cosmological
simulation or theoretical model results:\\

\begin{figure*}
\epsscale{1.15}
\plotone{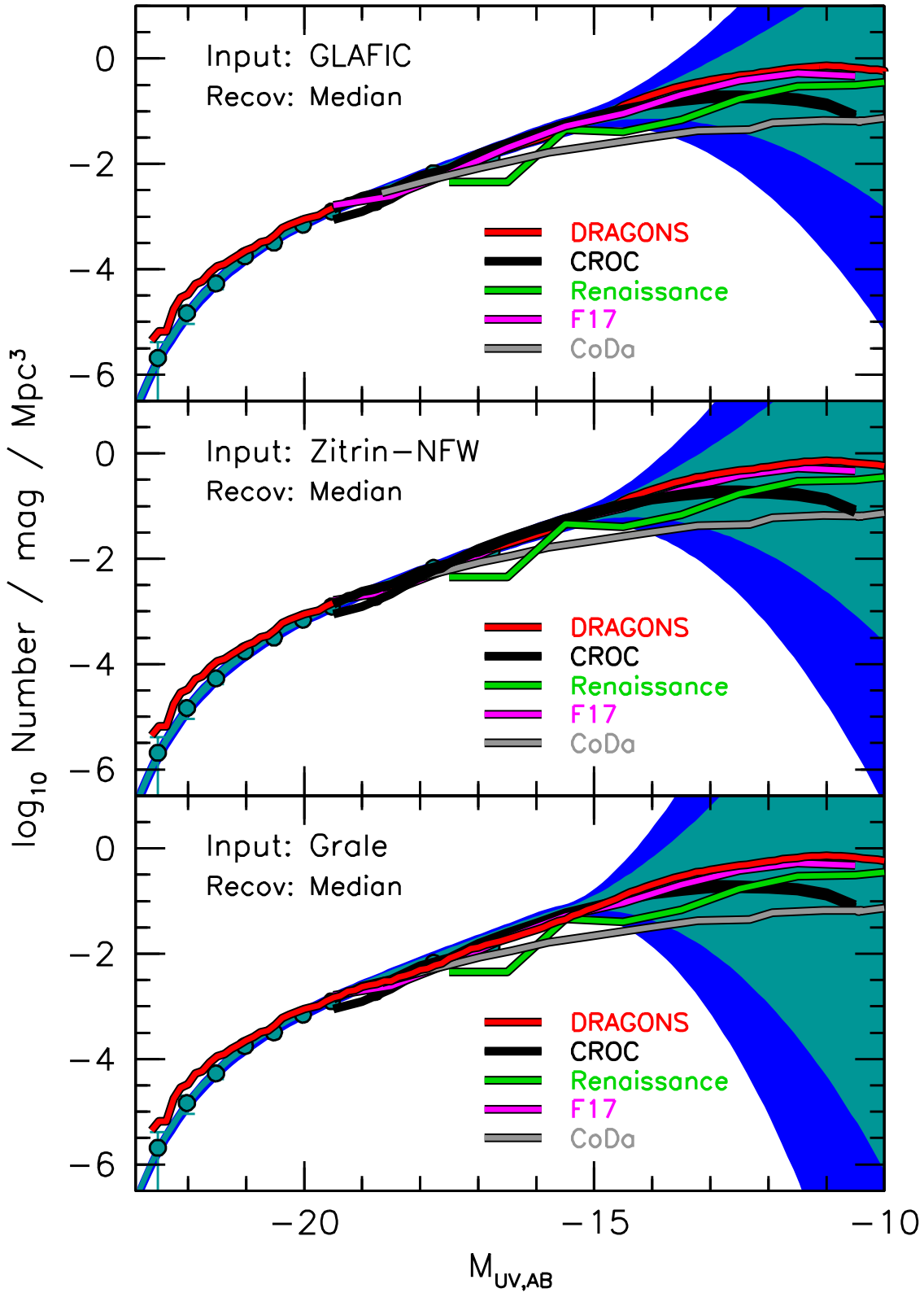}
\caption{Comparison of the 68\% and 95\% confidence intervals we have
  derived on the shape of the $z\sim6$ $UV$ LF with the predictions
  for this LF.  Confidence intervals are shown making different
  assumptions about the typical size of errors in the lensing models,
  assuming these errors to typically be as large as the differences
  between the median parametric model and the \textsc{GLAFIC} model,
  Zitrin-NFW models, and \textsc{Grale} models.  The plotted
  theoretical models include DRAGONS (\textit{red lines}: Liu et
  al.\ 2016), CROC (\textit{black lines}: Gnedin 2016), ENZO
  (\textit{green lines}: O'Shea et al.\ 2016), CoDa (\textit{gray
    lines}: Ocvirk et al.\ 2016), and Finlator et al.\ (2015, 2016,
  2017 [F17]: \textit{purple lines}).  The LF results from O'Shea et
  al.\ (2016) rely on their $z\sim12$ LF, since those simulations have
  not yet run down to $z\sim6$.  Overall, we find good agreement
  between the predicted LF results and the present observational
  constraints.\label{fig:theory}}
\end{figure*}

\begin{figure*}
\epsscale{1.15}
\plotone{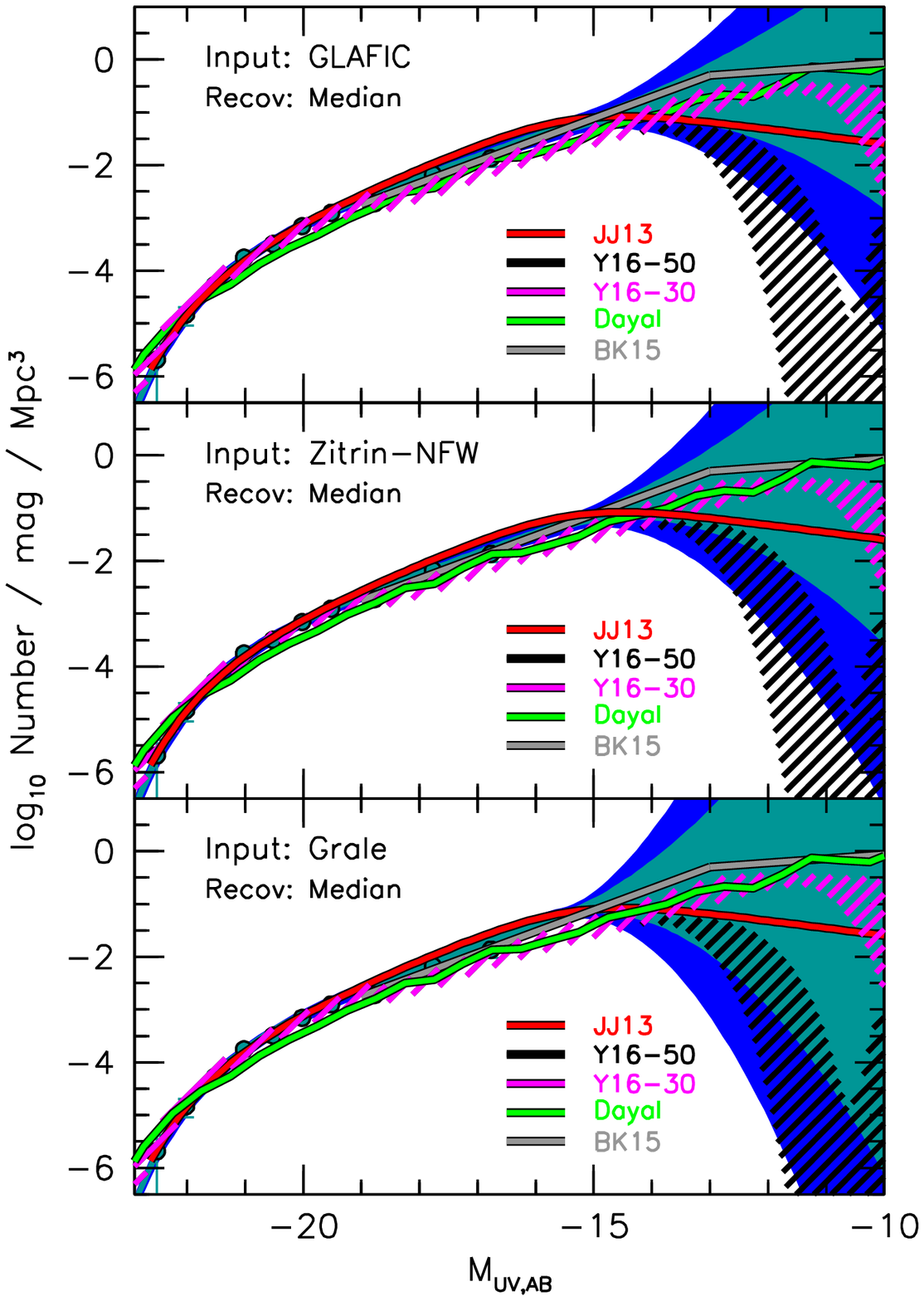}
\caption{Identical to Figure~\ref{fig:theory} but showing the results
  for the Jaacks et al.\ (2013: JJ13) model, two different models from
  Yue et al.\ (2016) where radiative feedback becomes important at
  circular velocities of 30 km s$^{-1}$ and 50 km s$^{-1}$, and the
  Dayal et al.\ (2014) model.  The dip at $-11$ mag in the 50 km
  s$^{-1}$ Yue et al.\ (2016) model is due to the quenching of star
  formation in low-mass halos from radiative feedback.  Also included
  among the presented results are the LF constraints implied from the
  abundance matching analysis Boylan-Kolchin et al.\ (2015: BK15)
  perform using dwarfs in the nearby universe.\label{fig:theory2}}
\end{figure*}

\noindent \textit{Renaissance [O'Shea et al.\ 2015]:} O'Shea et
al.\ (2015) report some of the first results from the ``Renaissance''
simulations.  The ``Renaissance'' simulations are zoom-in simulations
of a $(28.4 \textrm{Mpc}/h)^3$ volume of the universe, powered by the
Enzo code (Bryan et al.\ 2014), and self-consistently following the
evolution of gas and dark matter, including $H_2$ formation and
destruction from photodissociation.  Star formation and supernovae
physics are included and ionizing and UV radiation are produced as
predicted by Starburst99 (Leitherer et al.\ 1999).  Individual
dark-matter particles in the simulations have masses of $2.9\times
10^4$ $M_{\odot}$, meaning that the smallest halos that are resolved
in the simulation are $2\times10^6$ $M_{odot}$ ($\sim$70
particles/halo).  Many details of the physical implementation of the
implementation of the physics and also sub-grid recipes are provided
in Xu et al. (2013, 2014) and Chen et al.\ (2014).  In the
``Renaissance'' simulations, flattening in the UV LF is a direct
result of the decreasing fraction of baryons converted to stars in the
lowest mass halos, due to the impact of radiative feedback and less
efficient cooling processes.  While it is not yet possible to follow
the results of these simulations to $z\sim6$, results are available at
$z\sim12$ and this is the redshift we use for
comparisons.\\\vspace{-0.2cm}

\noindent \textit{CoDa [Ocvirk et al.\ 2016]:} The Cosmic Dawn (CoDa)
simulations are full gravity + hydrodynamic simulations of a large
$\sim (100 \textrm{Mpc})^3$ volume of the universe using the RAMSES
code (Teyssier 2002).  The simulations include standard prescriptions
for star formation and supernovae explosions following standard
recipes (Ocvirk et al.\ 2008; Governato et al.\ 2009, 2010).  One new
feature of the CoDa simulations is the inclusion of radiative transfer
into the simulations, in the sense that hydrodynamics and radiative
transfer are now fully coupled.  As a result, the effects of
photoionization heating on low-mass galaxies are fully included in the
CoDa simulations.  Ocvirk et al.\ (2016) report that radiative
feedback plays a big role in suppressing star formation in low mass
galaxies and modulating the faint-end of the LF.\\\vspace{-0.2cm}

\noindent \textit{CROC [Gnedin 2014, 2016]:} The LF results for the
Cosmic Reionization On Computers (CROC) are based on gravity +
hydrodynamical simulations using Adaptive Refinement Treement (ART)
code (Kravtsov 1999, 2002; Rudd et al.\ 2008).  The CROC
simulations include a wide variety of physical processes, including gas
cooling and heating processes, molecular hydrogen chemistry, star
formation, stellar feedback, radiative transfer of ionizing and UV
light from stars.  These simulations are conducted in 20h$^{-1}$ Mpc
boxes at a variety of resolutions.  The effective slope of CROC LFs
continue to flatten towards fainter magnitudes and reach a peak at
$\sim-12$ mag.  However, the peak at $\sim-12$ mag is reported not to
be a robust prediction of the simulation and to depend on the minimum
particle size in the simulations.  The impact of radiative feedback is
less important in the CROC simulations than in CoDa.\\\vspace{-0.2cm}

\noindent \textit{Finlator et al.\ (2015, 2016, 2017):} The Finlator
et al.\ (2015, 2016, 2017) LF results are based on a cosmological
simulation of galaxy formation in a $(7.5h^{-1})^3$ Mpc$^3$ volume of
the universe including both gravity and hydrodynamics as implemented
in the GADGET-3 code (Springel 2005).  To this code, gas cooling is
added through collisional excitation of hydrogen and helium as in Katz
et al.\ (1996), and metal line cooling is implemented using the
collisional ionization equilibrium tables of Sutherland \& Dopita
(1993).  Star formation is added using the Kennicutt-Schmidt law, with
supernovae feedback included following the ''ezw'' prescription from
Dav{\'e} et al.\ (2013) and metal enrichment from supernovae as
implemented as in Oppenheimer \& Dav{\'e} (2008).  Flattening in the
Finlator et al.\ (2015, 2016, 2017) LFs occurs mostly due to less
efficient gas cooling at lower halo masses.\\\vspace{-0.2cm}

\noindent \textit{DRAGONS [Liu et al.\ 2016]:} The LF results from
Liu et al.\ (2016) are based on the Dark-ages Reionization And
Galaxy-formation Observables from Numerical Simulations
(DRAGONS)\footnote{http://dragons.ph.unimelb.edu.au} project which
build semi-numerical models of galaxy formation on top of halo trees
derived from N-body simulations done over different box sizes to probe
a large dynamical range.  The semi-numerical models include gas
cooling physics, star formation prescriptions, feedback and merging
prescriptions, among other components of the model.  The turn-over in
the LF results of Liu et al.\ (2016) at $\sim-12$ mag correspond to
the approximate halo masses $\sim10^8$ $M_{\odot}$ where the gas
temperature is $10^4$ K.  Above this temperature, atomic cooling
processes become efficient.  In earlier data sets, Mu{\~n}oz \& Loeb
(2011) had looked at what constraints could be placed on this mass
using earlier LFs of Bouwens et al.\ (2007).\\\vspace{-0.2cm}

\noindent \textit{Jaacks et al.\ (2013):} The simulation results in
Jaacks et al.\ (2013) are powered by the GADGET-3 (Springel 2005)
gravity+hydrodynamics code run in three box sizes (10, 33.75, and 100
$h^{-1}$ Mpc) and three different particle sizes (9$\times$10$^5$,
2$\times$10$^7$, and 3$\times$10$^8$ h$^{-1}$ $M_{\odot}$).  Radiation
cooling is included in the simulations by H, He, and metals (Choi \&
Nagamine 2009), a UV background self-shielding effect, and heating by
a uniform UV background (Faucher-Gigu{\`e}re et al.\ 2009).
Supernovae feedback is implemeneted by a momentum-driven wind model
(Choi \& Nagamine 2011).  Star formation in the simulations is
governed by SFR-vs-$H_2$ model of Krumholz et al.\ (2009) rather than
in terms of the total density in cold gas.  The implementation of this
in the GADGET-3 code is as described in Thompson et al.\ (2014) and is
similar to the implementation of the same recipe by Kuhlen et
al.\ (2013) in the Enzo code.  As a result of the lower gas density in
molecular hydrogen $H_2$ in fainter, lower-mass galaxies, the LFs
predicted by Jaacks et al.\ (2013) show a turn-over at
$\sim-$15.4$\pm$0.6 mag.\\\vspace{-0.2cm}

\noindent \textit{Dayal et al.\ (2014):} The LF results from Dayal et
al.\ (2014) are based on a semi-analytic model which follows the
evolution of galaxies in merger tree constructed from extended
Press-Schechter theory (Lacey \& Cole 1993).  The star formation rate
in individual galaxies proceed at such a rate as to balance the impact
of supernovae feedback in expelling all the gas from a galaxy.
Flattening in the $UV$ LF is partially the result of a similar
flattening in the halo mass function, as well as lower efficiency for
star formation in the lower-mass halos that contribute to the low
luminosity end of the LF.\\\vspace{-0.2cm}

\noindent \textit{Yue et al.\ (2016):} Yue et al.\ (2016) derive their
LF results assuming a non-evolving stellar mass-halo mass relation.
Yue et al.\ (2016) adopt a very similar approach to what Mason et
al.\ (2015) employ in predicting the galaxy LF (see also Trenti et
al.\ 2010 and Tacchella et al.\ 2013).  Yue et al.\ (2016) start with
the halo mass function, break up the star formation history of each
halo into segments according to which the halo grows in mass by a
factor of two, and then assume that the SFR must be such to maintain a
constant stellar mass-halo mass relation which they calibrate to the
$z\sim5$ LF of Bouwens et al.\ (2015a).  Yue et al.\ (2016) then look
into the impact that radiative feedback could have during the epoch of
reionization.  Yue et al.\ (2016) derive their LF results assuming
that halos below some fixed circular velocity would have their star
formation quenched.\\\vspace{-0.2cm}

Finally, we also include a comparison with the empirical results on
the faint end of the LF at high redshift:\\\vspace{-0.2cm}

\noindent \textit{Boylan-Kolchin et al.\ (2015):} Boylan-Kolchin et
al.\ (2015) arrive at constraints on the faint-end of the LF at
$z\sim7$ by leveraging deep probes of the color-magnitude relationship
of nearby dwarf galaxies which allow one to estimate the luminosity of
these sources at $z\sim7$.  By comparing the distribution of inferred
luminosities of these dwarfs with the expected numbers extrapolating
$z\sim7$ LFs to $-10$ mag, Boylan-Kolchin et al.\ (2015) infer a break
in the LF at $\sim-13$ mag from a faint-end slope of $\sim-2$ to
$\sim-1.2$.\\\vspace{-0.2cm}

We present comparisons of the predicted LFs from both sophisticated
hydrodynamical simulations, various semi-analytic models, and the
empirical results of Boylan-Kolchin et al.\ (2015) in
Figures~\ref{fig:theory} and \ref{fig:theory2}.

Overall, we find reasonable agreement between our observational
results and the predicted LFs from both hydrodynamic simulations and
various semi-empirical theoretical models.  The predicted LFs from the
CoDa simulation (Ocvirk et al.\ 2016) and from two semi-analytical
models Yue et al.\ (2016) and Dayal et al.\ (2014) against which we
compare fall slightly below our observational constraints at $-15.5$
mag by $\sim$0.2-0.3 dex, but otherwise are in reasonable agreement
with our results.

In particular, we find that our observational constraints allow for
the existence of a flattening or turn-over in the $z\sim6$ LF at
$>-$15 mag as predicted in the theoretical models, due to a variety of
physical processes, including a greater role for radiative feedback
(O'Shea et al.\ 2015; Ocvirk et al.\ 2016; Yue et al.\ 2016) and less
efficient cooling in lower mass halos where atomic cooling processes
would be less important (Wise et al.\ 2014; Gnedin 2016; Liu et
al.\ 2016).  The present results suggest that these physical processes
can impact the shape of the LF at $>-15$ mag, as is predicted, and
there is no fundamental disagreement with observational results to
$>-14$ mag (contrary to reports from L17).

Our observational results are also fully consistent with the abundance
matching constraints obtained by Boylan-Kolchin et al.\ (2015) which
suggest a break in the faint end slope of the LF at $-$13 mag.  The
toy LF from Boylan-Kolchin et al.\ (2015) is almost entirely contained
with our 68\% confidence intervals, suggesting that the flattening
they infer from analyses of nearby dwarf galaxies is fully consistent
with the HFF observations of faint $z\sim6$ galaxies.

\section{Summary}

We have combined a large sample of 160 lensed $z\sim6$ galaxies from
the first four HFF clusters with a first-ever determination of the
systematic uncertainties at high magnification in the massive lensing
clusters.  In so doing, we provide the most realistic determination
yet of the shape of the $z\sim6$ LF to very low luminositis.  This
sample of lensed $z\sim6$ galaxies represents the most comprehensive
sample to date.  The sample reaches to low luminosities comparable to
others when different magnitude and magnification measurement
apporaches are considered (see \S2 and \S6.1-\S6.2).  Our analysis
provides a much more realistic assessment of the impact of the large
magnification uncertainties inherent at high magnifications.  This
allows us to set improved constraints on the faint-end slope $\alpha$
and also to investigate whether the $UV$ LF shows a turn-over at very
low luminosities.

One particular emphasis of this analysis was to include a full account
of systematic errors in deriving accurate constraints on the shape of
the $z\sim6$ $UV$ LF.  We looked especially at the impact of errors in
the magnification maps, but we also considered the impact of
uncertainties in the estimated completeness based on the size
distribution building upon results in a companion paper (Bouwens et
al.\ 2017).

To explore the impact of errors in the magnification map on LF
results, we have developed a new forward modeling approach which
involves using one set of magnification models and a candidate LF to
create mock catalogs over each of the HFF clusters.  These catalogs
are then analyzed using the same type of magnification models as are
used to interpret the real observations (Figure~\ref{fig:flowchart}).
The likelihood of a given LF can then be assessed by comparing the
observed counts with the expected counts derived from the simulations.
Our quantification of the general form of the $z\sim6$ LF includes not
only the normal Schechter parameters, but also a curvature parameter
$\delta$ which we apply faintward of $-$16 mag to characterize the
form of the LF to very low luminosities.

Our new simulation results using forward modeling demonstrate the
substantial impact of magnification errors on the LF results.  We show
that scatter due to magnification errors results in the LF asymptoting
to a faint-end slope of $\sim$$-2$ or steeper in the very low
luminosity regime when the magnification factors are high $\mu>10$-30.
This occurs regardless of the true slope.  This effect is so pervasive
that it can eliminate any indication of a turn-over (even if present
in reality) at the faint end of the LF (see Figure~\ref{fig:illust}).
At lower magnification factors, i.e., $\mu<10$, where the predictive
power of the magnification models is best (e.g., see
Figure~\ref{fig:predpow} and also Meneghetti et al.\ 2016), the impact
appears to be most manageable in terms of the overall impact on the LF
results.

For higher magnification factors, i.e., at $\mu>10$ and especially at
$\mu>30$, the predictive power of the magnification models is much
poorer, resulting in large uncertainties in the magnification factors.
As a result, it can be difficult to determine whether the LF shows a
turn-over at $\sim-$15 mag, whether it steepens further at $\sim-$15
mag, or whether it continues with a fixed faint-end slope to $-12$ mag
(see Figure~\ref{fig:illust}).

Taking advantage of our forward modeling procedure, we derive new
constraints on the faint-end slope of the LF and arrive at a value of
$-1.92\pm0.04$ using the HFF observations alone
(Table~\ref{tab:lfparm}) and, rather coincidentally, also
$-1.92\pm0.04$ combining our constraints with the field results of
Bouwens et al.\ (2015a).  Both constraints are consistent with our
previous determination $\alpha=-1.87\pm0.10$ (Bouwens et al.\ 2015)
using the HUDF+HUDF-parallel+CANDELS data alone.  We nevertheless
caution that the faint-end slope could be steeper (by $\Delta\alpha
\sim 0.01$-0.03) with a potentially more negative $\delta$ (less
consistent with a possible turn-over at the faint end of the LF) if
the sizes of faint sources are not so small as to be essentially point
sources (Bouwens et al.\ 2017) and closer to conventional
size-luminosity relations (e.g., Shibuya et al.\ 2015).  In this case,
when the galaxy sizes are substantially larger, the completeness of
galaxies faintward of $-14.5$ mag could become quite large.

We use our new constraints to derive 68\% and 95\% confidence regions
on the faint-end form of the $z\sim6$ LF, presenting our results in
Figure~\ref{fig:lf6all}.  We find no evidence for a turn-over in the
LF at the faint end.  Nevertheless, we can place constraints on how
faint it must be, though the result does depend on the assumed size of
the errors in the magnification models.  If the true errors in the
models are similar to the differences between the \textsc{GLAFIC}
model and the median parametric model, our results strongly indicate
that a turn-over cannot occur brightward of $-14.2$ mag (68\%
confidence).  However, if differences between the non-parametric
models and the median parametric models are typical, then a turn-over
cannot occur brightward of $-15.3$ mag (68\% confidence).  Our results
are fully consistent with recent observational results from
Boylan-Kolchin et al.\ (2015) and theoretical models (O'Shea et
al.\ 2015; Gnedin 2016; Liu et al.\ 2016; Ocvirk et al.\ 2016)
predicting some flattening in the $UV$ LFs at $>-$15 mag.

The faint-end slope $\alpha$ we derive at $z\sim6$ is $-1.92\pm0.04$
and 3.5$\sigma$ shallower than the Livermore et al.\ (2017) faint-end
slope $\alpha=-2.10$$\pm0.03$$_{-0.01}^{+0.02}$.  The tension with the
faint-end slope result of L17 decreases to $3\sigma$, if we allow for
larger source sizes (\S5.4) and hence a steeper $\alpha$ by 0.03.
Meanwhile, our constraints on the turn-over are consistent with the
findings by Atek et al.\ (2015) and Castellano et al.\ (2016b), but
occur at much higher luminosities than what L17 report.  Despite
having larger samples than L17 and considering twice as many HFF
clusters (while probing to comparably low luminosities), we only find
evidence against a turn-over brightward of $\sim-15.3$ and $\sim-14.2$
mag at 68\% confidence, vs. the $\sim-11.1$ mag reported by L17 at
nominally slightly higher confidence.  We speculate that L17's
stronger claims against a turn-over (and steeper faint-end slope
results) arose as a result of artifacts in their determinations of the
LFs resulting from likely inaccurate size assumptions (see \S6.1-6.2
and Bouwens et al.\ 2017) and a large number of sources near their
completeness limit (\S6.1.2, Figure~\ref{fig:l17apm}, and Figure 13 of
Bouwens et al.\ 2017).  We show that these limitations likely led L17
to set constraints that we cannot reproduce through the analysis of
current data sets.

The new formalism we have developed to derive LF results in the
presence of errors in the magnification map has significant utility
and can be applied to other {\it HST} observations that have been obtained
with the HFF program.  In the immediate future, we plan to make use of
our new forward-modeling methodology to derive LF results at $z\sim6$,
$z\sim7$, $z\sim8$, and $z\sim9$ from the full HFF program.  These
results would provide us perhaps our most complete information on the
faint-end form of the LFs before JWST and provide us with clues as to
how the overall ionizing emissivity evolves with cosmic time.

\acknowledgements

We acknowledge many useful conversations on the algorithms presented
in this paper with Marijn Franx.  We are greatly appreciative to
Pratika Dayal, Kristian Finlator, Nicholas Gnedin, Chuanwu Liu, Brian
O'Shea, Pierre Ocvirk, and Bin Yue for sending us the predictions they
derive for the LF results at high redshift.  We thank Kristian
Finlator for discussing with us at length different turn-over
mechanisms that would impact the $UV$ LF at low luminosities.  Austin
Hoag kindly sent us the v2 Bradac models for HFF cluster MACS0416.
RSE acknowledges financial support from the European Research Council
under Advanced Grant FP7/669253.

\begin{figure*}
\epsscale{1.13}
\plotone{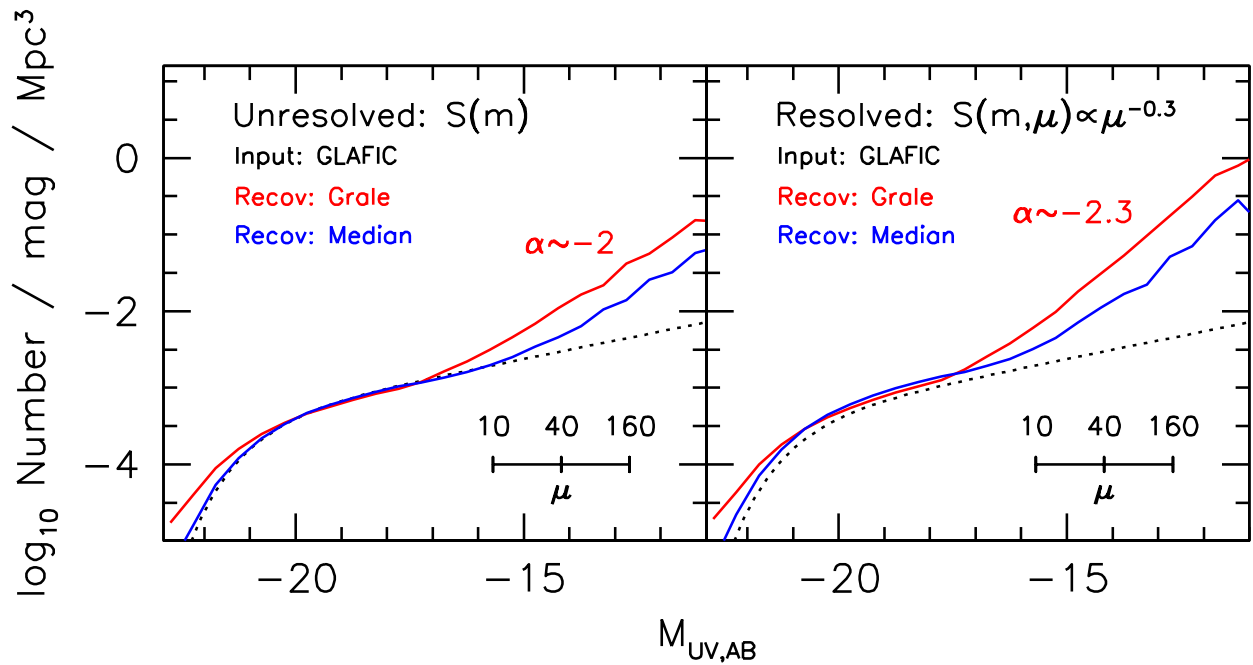}
\caption{Illustration of how the impact of magnification errors on the
  recovered LFs depends on the sizes of sources (Appendix C).  The
  magnification scale in the corner is as in Figure~\ref{fig:illust}.
  An input LF with a faint-end slope of $-1.3$ and no turn-over at the
  faint end is assumed.  In our forward modeling procedure, the
  \textsc{GLAFIC} magnification model is used to create the mock
  catalogs, while the \textsc{Grale} and median magnification maps are
  used for recovering the LFs (\textit{red and blue lines,
    respectively}).  The left panel shows the recovered LFs assuming
  that point-source spatial profiles for all galaxies, while the right
  panel shows the recovered LFs assuming more extended sources and
  where the selection efficiency $S(\mu)$ decreases towards large
  magnification factors as $\mu^{-0.3}$ (as in Figure 3 from Oesch et
  al.\ 2015).  The recovered LFs shows a departure from the input LF
  at $\sim-15.5$ mag, asymptoting towards a slope of $-2$ for
  unresolved sources (\textit{left panel}) and $-2.3$ for resolved
  sources (\textit{right panel}).  Previous studies appear to have
  completely ignored the issue raised in this
  figure.\label{fig:illustsize}}
\end{figure*}

\appendix

\section{A.  Performance of Our Intra-Cluster Light Subtraction Technique}

In this section, we briefly quantify the performance of our procedure
for subtracting the intra-cluster light in galaxy clusters relative to
that obtained by other groups.

One measure of performance regards the total number of $z=6$-8
candidate galaxies that it is possible to recover from the
observations, after subtraction of the intra-cluster light.  We begin
by comparison of the number of $z=6-7$ galaxies in our own samples.
When extracting these samples without our foreground cluster
subtraction procedure, we find 61 $z=6$-7 galaxies over Abell 2744,
but 71 $z=6$-7 galaxies when making use of images where the foreground
light from the cluster has already been removed.  This illustrates the
basic increase in numbers one can achieve from a subtraction of the
foreground light.

The published results of Merlin et al.\ (2016) and L17 provide us with
a separate benchmark.  Merlin et al.\ (2016) report 138 $z\sim7$-8
galaxy candidates over the two clusters, while L17 report 161
$z\sim6$, 7, and 8 candidates.  With our procedure, we recover 176
candidates over the two clusters, which is slightly larger than what
L17 obtain.

However, the numbers of $z=6$-8 candidates L17 report are likely
boosted by their considering selections generated from coadditions of
the $Y_{105}$, $J_{125}$, $JH_{140}$, and $H_{160}$ data in 14
different combinations (e.g., $Y_{105}$, $J_{125}$, $Y_{105}+J_{125}$,
$Y_{105}+J_{125}+JH_{140}$).  By considering selections from many
different combinations of such images, it is possible to increase the
completeness of one's selections.  This occurs since SExtractor often
defines the apertures of specific sources in ways which are not
entirely ideal, and by selecting candidates off many different
detection images, one can improve the overall completeness of a
selection.

An alternative way to achieve similar gains in sample size is by
perturbing the detection image multiple times, rerunning the
selection, and adding in to the main sample any new sources that are
found.  For this test, we perturb the detection image by adding to it
a smoothed noise image the same RMS as the data itself.  The smoothing
is a gaussian kernel with FWHM of 0.28$''$.  Repeating the selection
process 4 additional times and removing redundant sources over
MACS0416, we find that we can recover a 30\% higher surface density of
z=6-8 sources than running the selection on just a single detection
image.  Assuming similar gains in numbers over Abell 2744, we estimate
a total sample size of 228 $z\sim6$-8 galaxies behind Abell 2744 and
MACS0416, 65\% and 40\% larger than claimed by Merlin et al.\ (2016)
and L17, respectively.  

Despite the demonstrated potential to make use of a larger sample of
$z=6$-8 sources, we only make use of 87 $z\sim6$ galaxies (176 $z=6$-8
galaxies) behind Abell 2744 and MACS0416 in the LF analyses we conduct
in this paper.

\section{B.  Selection Volume Estimates}

Here we describe our procedure for estimating the effective selection
volumes for faint galaxies behind the HFF clusters we examine.

Our baseline treatment is to model faint galaxies as point sources in
estimating their selection volumes.  This choice is motivated by our
finding in Bouwens et al.\ (2017) that the faintest $z\sim2$-8
galaxies behind the HFF clusters had properties consistent with
point-source spatial profiles, with no discernible extension along the
expected shear axes.  Also, no discernible dependence was found for
the measured surface densities as a function of the predicted shear in
the high-magnification regions.  Nonetheless, as such small sizes for
galaxies are unexpected (e.g., Liu et al.\ 2017; Kravtsov 2013), we
also consider the impact of larger sizes (and a larger incompleteness)
on the LF results throughout the paper.

As point sources, the only quantity of importance from the lensing
model is the magnification factor; the form of the deflection (or
shear) map has no impact on the results.  This simplifies the
selection volume simulations enormously, since it means we can
estimate the selection volumes for extremely faint sources in the
presence of lensing in exactly the same way we would estimate the
selection volumes in the absence of lensing.  The only quantity of
importance in estimating the selection volumes is the apparent
magnitudes of the sources.  Bouwens et al.\ (2017) discuss this in
their \S6.3.

We adopt a median value of $-2.2$ for the $UV$-continuum slope for
galaxies in our simulations to match that found in the observations
(e.g., Bouwens et al.\ 2014: see also Wilkins et al.\ 2011; Bouwens et
al.\ 2012; Finkelstein et al.\ 2012; Dunlop et al.\ 2013; Rogers et
al.\ 2014; Duncan \& Conselice 2015).

Adopting these assumptions for the color of the sources and a
point-source assumption for the size, we create artificial images for
each source over the full suite of passbands and insert these images
into the real observations.  We then do object detection and
photometry using the same procedure as we use in constructing our
catalogs (\S3) and then apply our selection criteria.  In this way, we
derive the completeness for sources in different regimes.

Selection volumes are computed by multiplying the cosmological volume
element by the estimated completeness and integrating over redshift.
Following previous work (e.g., Ishigaki et al.\ 2015; Oesch et
al.\ 2015), we treat different multiple images of the same source as
entirely independent for the purposes of our analysis.

\section{C.  Source Size Modulates the Impact Magnification Errors Have on the Inferred Shape of the LF}

The impact of magnification errors on the derived LF can also depend
on source size, as discussed in \S4.1.  This can occur as a result of
the fact that higher magnification sources are more generally more
difficult to detect, if they are spatially extended, than if their
magnification is lower.  In other words, the selection efficiency $S$
is a function of the magnification factor $\mu$.

The issue is that while the actual surface density of sources in our
catalogs is proportional to $S(\mu_{true})$, the selection volumes we
estimate for these sources is $S(\mu_{model})$.  This results in the
recovered volume density for these sources being higher than the true
surface density by the factor
$<$$S(\mu_{true})$$>$$/$$<$$S(\mu_{model})$$>$.  When
$<$$\mu_{true}$$>$ $\sim$ $<$$\mu_{model}$$>$, no bias is present in
the recovery volume density of sources.  However, when
$<$$\mu_{true}$$>$ is less than $<$$\mu_{model}$$>$, as is often the
case at higher magnifications $\mu>10$ (Figure~\ref{fig:predpow}), the
recovered surface density of sources can be significantly higher than
reality.  For example, assuming that $S(\mu) \propto \mu^{-0.3}$ as in
Figure 3 of Oesch et al.\ (2015), the LF asymptotes towards a
faint-end slope of $-2.3$.

To illustrate the impact of source size and different assumptions
about $S(\mu)$, we consider two different cases: the first involving
point sources where $S$ is independent of $\mu$ and the second
involving extended sources where $S(\mu)$ is proportional to
$\mu^{-0.3}$.  Similar to the simulations run in \S3.2, we use the
\textsc{GLAFIC} model to construct mock catalogs and then recover the
LF using either the \textsc{Grale} model or the median parametric
magnification model.  In the two cases, we incorporate the different
dependencies of $S$ on the magnification factor $\mu$ for both the
catalog construction and recovery of the LF.  The input LF for the
simulations has a faint-end slope of $-1.3$, with no turn-over at the
faint end.

The results are presented in Figure~\ref{fig:illustsize} with the red
and blue lines indicating recovery by the \textbf{Grale} and median
parametric magnification models, respectively.  Differences between
the two size cases are immediately obvious.  In the point source case
(where $S(\mu)$ is independent of $\mu$), the faint-end slope
asymptotes to $-2$.  However, in the case of extended sources (where
$S(\mu)$ scales as e.g. $\mu^{-0.3}$), the faint-end slope instead
asymptotes to $-2.3$.  In general, one expects the slope to equal
$-2+d(\ln S(\mu))/d(\ln \mu)$, as Figure~\ref{fig:surf15} illustrates
in \S3.2.  In either case, the LFs asymptotically approach these
slopes faintward of $-16$ mag assuming the \textsc{Grale} model and
faintward of $-15$ assuming the median parametric magnification model.

\begin{figure*}
\epsscale{1.17}
\plotone{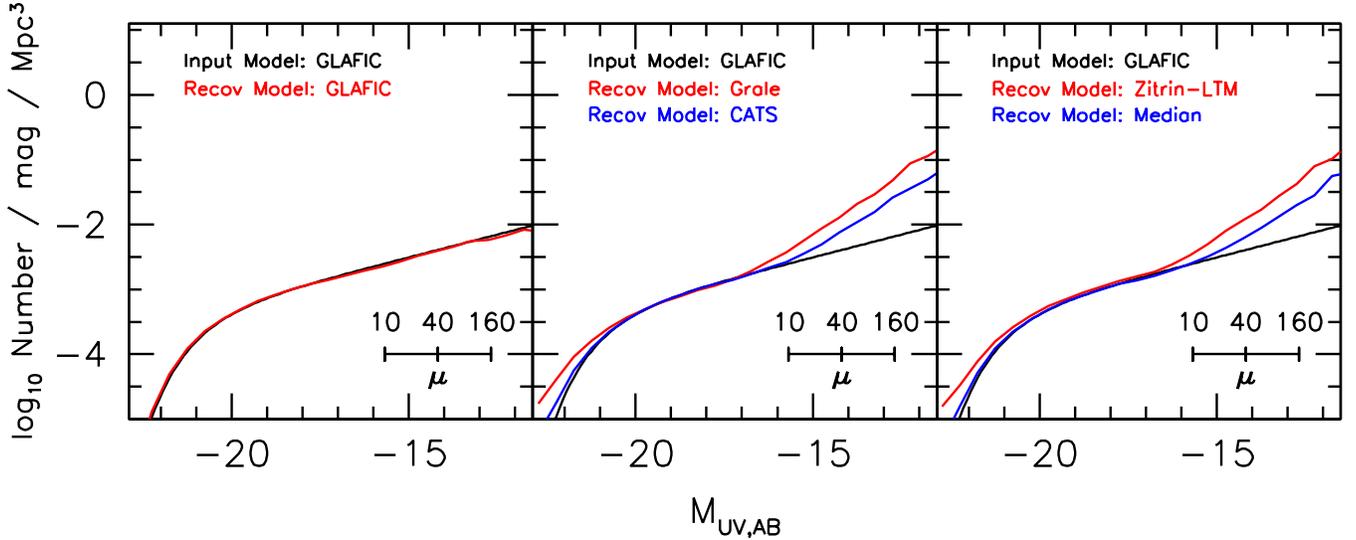}
\caption{Comparison of an input LF with a shallower faint-end slope of
  $-1.3$ with the recovered LFs using a forward-modeling procedure
  where we create mock catalogs using the \textsc{GLAFIC}
  magnification maps and recover the LF results using the CATS,
  \textsc{Grale}, Zitrin-LTM, and median parametric magnification maps
  (see Appendix D).  The typical magnification levels of sources
  probing a given luminosity range are indicated by magnification
  scale in the corner.  The recovered LFs show excellent agreement
  with the input LFs to $-16.5$ mag, but show a departure at $-15$ mag
  and rapidly asymptote towards a faint-end slope of $-2$.
  Interestingly, the absolute magnitudes $M_{UV}$ where this departure
  occurs correspond to magnification factors where the models lose
  their predictive power
  (Figure~\ref{fig:predpow}).\label{fig:illust15}}
\end{figure*}

This example should reinforce how difficult it is to obtain accurate
constraints on the shape of the LF at $>-15$ mag and thus to detect
the existence of a flattening or turn-over in the LF.  Not only do the
results depend on the magnification level to which magnification maps
retain their predictive power (e.g., see Figure~\ref{fig:predpow}),
but the results also depend significantly (i.e., $\Delta\alpha \sim
0.3$) on the size distribution for faint sources.  

We emphasize that the impact this has on the LF shape is distinct from
the effect already discussed in the companion study to the present one
(Bouwens et al.\ 2017), where the faint-end slope $\alpha$ of the LF
could be biased if the sizes and hence selection volumes were
improperly estimated.  This bias explicitly arises \textit{because of
  errors in the magnification map} and due to mismatches between
$<$$S(\mu_{true})$$>$ and $<$$S(\mu_{model})$$>$.

If faint sources are slightly resolved (after magnification) -- as
assumed in many recent studies of faint galaxies -- this bias has the
potential to be quite significant at absolute magnitudes $M_{UV}$ of
$>-15$ mag where $\mu > 20$.  Amazingly, however, previous work appear
to have neither recognized the importance of such a bias, nor made use
of procedures that would allow for its correction, even though given
the size assumptions in e.g. L17, this bias would constitute an
important effect.

\section{D.  Recovery of LFs with shallower faint-end slopes}

Errors in the magnification maps can have a substantial impact on the
shape of the $z\sim6$ LF faintward of $-15$ mag.  We already
illustrated this in \S3.2 of this paper using a LF which turned over
at $-15$ mag (Figure~\ref{fig:illust}).

Here we show the impact of these magnification errors using a $z\sim6$
LF with a faint-end slope $-1.3$, which we intentionally take to be
substantially shallower than $-2$ (the direction in which
magnification errors drive the apparent faint-end slope).  Again, we
use a forward modeling procedure where we create the mock catalogs
using the \textsc{GLAFIC} magnification maps and then alternatively
recover the LFs with the CATS, \textsc{Grale}, Zitrin-LTM, and median
parametric magnification maps.

The result is shown in Figure~\ref{fig:illust15}, and excellent
recovery of the LF is observed brightward of $-16.5$ mag for all
magnification maps.  The best performance is achieved using the median
magnification map; however, we note that faintward of $-15$ mag, the
recovered LF still diverges from the input LF, rapidly transitioning
faintward of $-15$ to a faint-end slope of $-2$.

\section{E.  Estimates of the Faint-end Slopes in Previous Work using only the HFF data}

In utilizing the data from the HFF clusters to map out the faint-end
of the UV LF and derive faint-end slope results, it is valuable to
perform this exercise using only the HFF samples to preserve the
independence of the faint-end slope determinations from those derived
from the field (i.e., the HUDF).  By doing so, one can conduct fair
comparisons between faint-end slope results $\alpha$ derived using
lensing clusters and from the field to test for consistency.

\begin{figure}
\epsscale{1.05}
\plotone{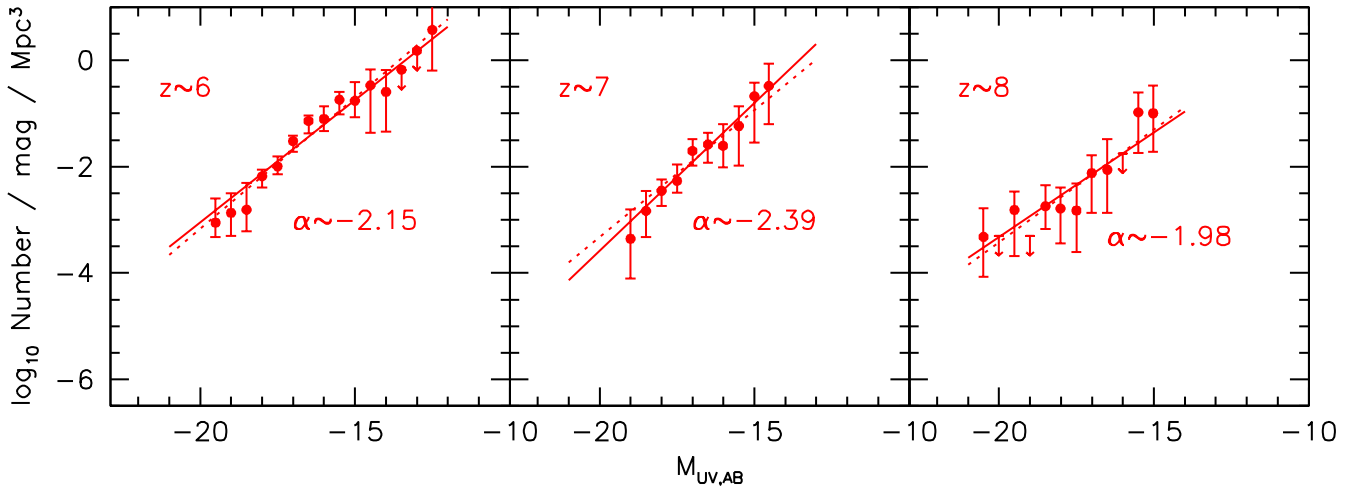}
\caption{Power-law fits (\textit{solid lines}) to the binned $z\sim6$,
  7, and 8 LF results from L17 in an effort to estimate the faint-end
  slope estimates from the HFF results alone.  Also presented
  (\textit{dotted lines}) are fits done to the L17 HFF LF results
  anchored to the field LF results from the same group (Finkelstein et
  al.\ 2015) at $M_{UV,AB} = -20$.  The motivation for deriving the
  faint-end slopes $\alpha$ from the HFF results alone is to keep the
  derived results independent of those derived for the field.  This
  makes it possible to compare the lensed LF results and field LF
  results in a fair way, as we do in Figure~1.  \label{fig:livfit}}
\end{figure}

Towards this end, we have taken the binned LF results from L17 on the
HFF clusters and fit the results to a power law to estimate a
faint-end slope $\alpha$.  The results are presented in
Figure~\ref{fig:livfit} as the solid lines and give slopes of
$-2.15\pm0.09$, $-2.39\pm0.24$, and $-1.98\pm0.27$ at $z\sim6$,
$z\sim7$, and $z\sim8$, respectively.  Interestingly, these results
are mostly steeper than the faint-end slopes inferred from field
searches at the relevant redshifts.  Part of this difference could be
due to the modest bias towards steeper slopes as a result of the large
sizes L17 use in estimating the selection volumes (Bouwens et
al.\ 2017: see their Figure 2).

We can obtain even stronger constraints on the faint-end slope
$\alpha$ results from the HFF clusters by having at least one point on
the bright end of the LF from field searches to use as an anchor.
There is not much search volume available behind clusters to constrain
this part of the LF, and so including this information is useful.  We
therefore refit the $z\sim6$, $z\sim7$, and $z\sim8$ LF results on the
HFF clusters from L17, anchoring the fit results to the field LF
results from the same group, i.e., Finkelstein et al.\ (2015), at
$-20$ mag using their best-fit Schechter function.  We chose $-20$ mag
somewhat arbitrarily to be close enough to the knee of the LF, i.e.,
$\sim-21$ mag, while not being so bright as to be affected by
uncertainties in the assumed characteristic luminosity or form of the
bright end of the LF (Schechter vs. power-law: i.e., Bowler et
al.\ 2015).

The fits are presented in Figure~\ref{fig:livfit} as the dotted lines.
The faint-end slopes in this case are $-2.23\pm0.05$, $-2.19\pm0.08$,
and $-2.06\pm0.15$ for the $z\sim6$, $z\sim7$, and $z\sim8$ LFs,
respectively.  These values are consistent with the faint-end slope
$\alpha$ results we derive from the HFF results from L17 without a
bright anchor point.

\begin{figure}
\epsscale{0.8} \plotone{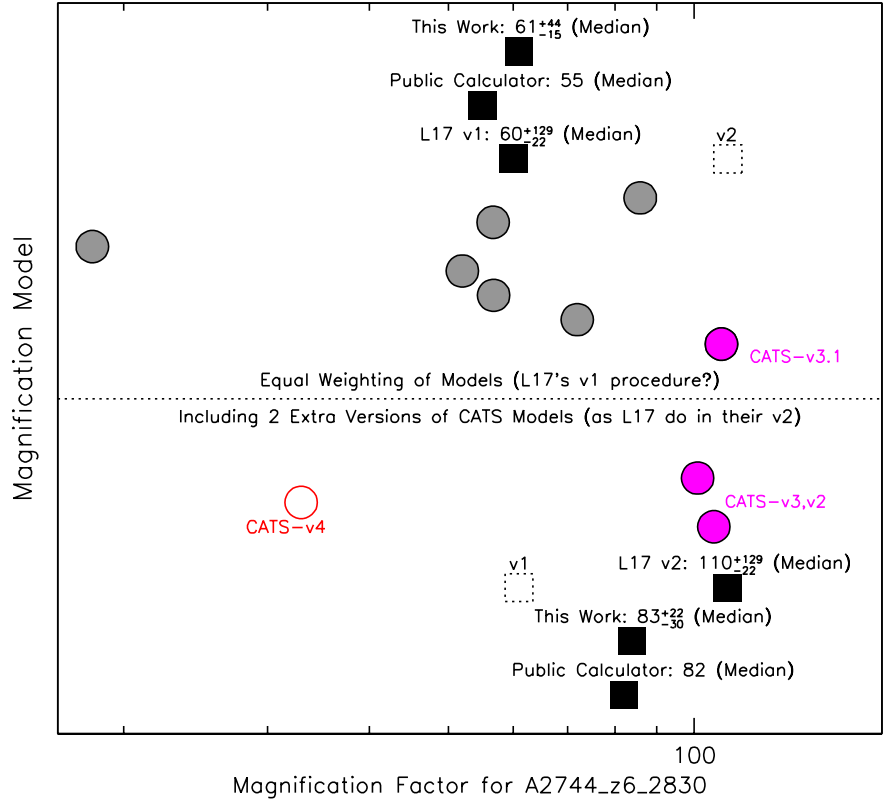}
\caption{The magnification of the faintest source A2744\_z6\_2830 in
  L17 as computed using the latest version of each of the post-HFF
  models (filled circles above the dotted line) and as computed from
  just version 2.1 and 3.0 of the CATS models (filled circles below
  the dotted line). The gray circles give the median magnification
  estimates based on all magnification models of a given type, i.e.,
  \textsc{glafic}, \textsc{Sharon/Johnson}, \textsc{Zitrin-NFW},
  \textsc{Zitrin-LTM}, \textsc{Bradac}, and \textsc{Grale}. The three
  \textsc{CATS} models given preferential weight in L17 are shown in
  magenta. The magnification estimate for A2744\_z6\_2830 from version
  4 of the CATS team (Mahler et al. 2017) is shown with the red open
  circle.  Interestingly, the latest \textsc{CATS} version 4 gives
  magnifications for A2744\_z6\_2830 that are $\sim$3.0-3.4$\times$
  lower than the earlier version 2 and 3 \textsc{CATS} models.  This
  indicates that the preferential weight given by L17 to the results
  from the v3 and v2 CATS models was not well-founded. The median
  magnification factor we compute giving equal weight to the latest
  model of a given type ($61_{-15}^{+44}$) and including two
  additional versions of the CATS models ($83_{-32}^{+20}$) are shown
  with the solid black square above and below the dotted line,
  respectively.  The median magnification factors are similarly
  estimated using the public calculator and are in excellent agreement
  with both of our own determinations.  The median magnification
  factor L17 quote (110$_{-22}^{+129}$) in the final version is
  significantly higher than what both we and the public calculator
  compute based on the public models (83 and 82, respectively). The
  high magnification reported by L17 appears to lack a clear
  justification.  By contrast, the reported magnification factor
  60$_{-22}^{+129}$ in version 1 of their paper (plotted in the upper
  panel as a black square) agrees very well with the median
  magnification factor from the public calculator and from our
  calculations. In both cases the magnifications were computed by
  weighting each model equally, suggesting this was how L17 had
  originally computed their magnification factor for the source.  The
  magnification factor L17 quote in their version 1 was an excellent
  representation of the median magnification across the models.  It is
  unclear why L17 subsequently give preferential weight to the
  \textsc{CATS} models in their version 2 instead of using the more
  appropriate values nearer the median of all models, and why they
  adopted a value that was nearly double their original magnification
  factor, and even larger than any model estimate.  All the
  indications are that a value around 60 for the magnification of
  A2744\_z6\_2830 is best justified by the v3 post-HFF
  models.\label{fig:magf_change}}
\end{figure}

\section{F.  Magnification Factor for the Faintest Source in L17}

The faintest $z\sim6$ candidate in L17 plays a key role in anchoring
their LF at the faint end, as we noted in \S6.2.  The magnification
factor they used to establish the absolute magnitude was very large
(110.0$_{-22.2}^{+129.0}$).  Given the importance of this source in
their analysis we wanted to check the high magnification quoted.  We
performed this test using the same set of magnification models that
L17 listed as being used, i.e., \textsc{GLAFIC}, Sharon/Johnson,
Zitrin-NFW, Bradac, Grale, Zitrin-LTM, and three versions of the CATS
models, including both the best model and the other models in the MCMC
chains.  The CATS model versions were the version 3.0, 3.1, and the
2.1 models (the version 2.1 gives a higher magnification estimate than
the version 2.2 model for the source).

For each model, a magnification factor can be computed using
Eq.~\ref{eq:mu} using an interpolation of the public $\gamma$ and
$\kappa$ maps and using an assumed redshift to compute the
$D_{LS}/D_S$ factor which we draw from the median redshift and
uncertainties quoted for the faintest source by L17, i.e., $z_s =
6.11_{-1.22}^{+1.05}$.  We derive the magnification factor from each
magnification map separately in the same way.  The results are shown
in Figure~\ref{fig:magf_change}.  Both the estimated magnification
factor using all models (\textit{black solid circle and horizontal
  error bar}) are plotted, as well as that estimated from each model
individually (\textit{grey circles}).  Each of our magnification
estimates matches that computed by the public calculator to typically
within 4\%, as one would expect given that the process of
computing magnifications from the public models is well defined.

The spread in the magnification factor is very large, from $<$20 to
just over 100, with the largest being the CATS version 2 and 3 models
with magnifications around 100. The median magnification estimate we
derive based on the 9 presented models is 83$_{-30}^{+22}$.  Using the
public calculator to calculate median magnification for the source
based on the full range of MCMC models available and taking the median
of the seven models, we find 82 (shown as a black solid square in
Figure 19), almost identical to our own estimate as we would expect
given that the procedure for calculating magnification factor based on
the public models is very well defined.  In contrast, L17 estimate a
magnification of 110$_{-22}^{+129}$ for the same source by purportedly
taking a median of the same models.  It is unclear why the L17's
magnification factor is higher than what we estimate based on the same
public models given that the calculation is very well defined.  In
addition, it is unclear how their estimate could be a median if it
lies higher than for all but one model or why their determination for
the \textit{lower} interquartile range on the magnification factor,
i.e., 88 ($=110-22$), lies higher than for 6 of the 9 individual
models.

One possible explanation would be if they treated the source as
extended.  While such a treatment is not particularly justified given
the essentially point-like spatial profile of the source, it turns out
this does not matter for the final result.  As the average
magnification factor would correspond to $(\int_A f(A) dA)/(\int_A
f(A)/\mu(A) dA)$ where $f(A)$ is the flux profile of a source and $A$
is the area of integration, particularly high magnification areas are
de-weighted in computing the net magnification factor (biasing the net
magnification lower not higher).  If we re-estimate the magnification
factor assuming a source size of 0.4$''$, we derive a median
magnification factor of 83 (including three of the four versions of
the CATS post-HFF models in the weighting as L17 report to do).  This
is essentially identical to the magnification factor we estimate if we
assume A2744\_z6\_2830 were a point source.

In providing context for their claimed median magnification factor of
110 in version 2 of their paper (their originally quoted median
magnification factor was 60), L17 report a large magnification factor
of 150 for A2744\_z6\_2830 from the v3 Bradac and v2 Sharon models.
These estimates, however, are significantly higher than what we find
and what is derived from the HFF public calculator.  For the Bradac v3
models, we use our own procedures and the public calculator and find
magnification factors of 52 and 56, respectively.  For the Sharon v2
models, we compute a magnification factor of 109 and 107,
respectively.  Note that these v2 models pre-date the HFF, so it is
unclear why L17 bring them into the discussion since they are not
incorporated into the computed median magnification factor.  L17's
claimed magnification factor of 150 from both models are
$\sim$3$\times$ and $\sim$1.4$\times$ larger than both our own
calculations and those from the public calculator.  Thus, L17
unfortunately appear to have likely erred, at least in v2 estimates,
in computing the magnification factor for A2744\_z6\_2830 based on the
public models.

Probably the most robust estimate for the magnification of
A2744\_z6\_2680 can be obtained by weighting each of the modeling
efforts equally (not weighting one modeling effort in excess of the
others as L17 do) and repeating the above process.  Following this
procedure, we compute 61$_{-15}^{+44}$, which is lower than our
earlier estimate including three versions of the CATS models in the
median (instead of one).  Making use of the public calculator, we
compute a median magnification of 55 (\textit{shown as a black solid
  square}). Interestingly enough, both of these estimates agree very
well with the estimate L17 provide in their version 1 for this source
($60.4_{-22.2}^{+129.0}$), suggesting that this is how L17 had
originally calculated the magnification factor for their faintest
source.  It is unclear why L17's quoted estimate approximately doubled
from version 1 to 2, since the available models appear to best support
a magnification factor of 60 for the source
(Figure~\ref{fig:magf_change}).

\end{document}